\newif\ifwordcount
\wordcountfalse

\documentclass[%
 reprint,
nofootinbib,superscriptaddress,titlepage,
 amsmath,amssymb,
prd,
]{revtex4-1}

\usepackage{color}

\def\thetaB{\boldsymbol{\theta}}

\usepackage{graphicx}
\usepackage{dcolumn}
\usepackage{bm}
\usepackage{mathtools}

\usepackage{fancyvrb}
\usepackage{hyperref}
\usepackage{graphicx}
\usepackage{caption}
\usepackage{subcaption}
\usepackage{ragged2e}
\usepackage{multirow}
\usepackage[table,xcdraw]{xcolor}
\usepackage{hhline}
\DeclareCaptionJustification{justified}{\justifying}
\begin{document}

\defcitealias{Hill2016}{H16}
\defcitealias{Ferraro2016}{F16}

\title{Constraining the Baryon Abundance with the Kinematic Sunyaev-Zel'dovich Effect: Projected-Field Detection Using  \emph{Planck}, \emph{WMAP}, and \emph{unWISE}}

\author{Aleksandra Kusiak}
\affiliation{Department of Physics, Columbia University, New York, NY, USA 10027}
\email{akk2175@columbia.edu}

\author{Boris Bolliet}
\affiliation{Department of Physics, Columbia University, New York, NY, USA 10027}

\author{Simone Ferraro}
\affiliation{Lawrence Berkeley National Laboratory, One Cyclotron Road, Berkeley, CA 94720, USA}
\affiliation{Berkeley Center for Cosmological Physics, Department of Physics, University of California, Berkeley, CA 94720, USA}

\author{J.~Colin Hill}
\affiliation{Department of Physics, Columbia University, New York, NY, USA 10027}
\affiliation{Center for Computational Astrophysics, Flatiron Institute, New York, NY, USA 10010}

\author{Alex Krolewski}
\affiliation{AMTD Fellow, Waterloo Centre for Astrophysics, University of Waterloo, Waterloo ON N2L 3G1, Canada}
\affiliation{Perimeter Institute for Theoretical Physics, 31 Caroline St. North, Waterloo, ON NL2 2Y5, Canada}
\affiliation{Berkeley Center for Cosmological Physics, Department of Physics, University of California, Berkeley, CA 94720, USA}
\affiliation{Lawrence Berkeley National Laboratory, One Cyclotron Road, Berkeley, CA 94720, USA}

\date{\today}

\begin{abstract}
The kinematic Sunyaev-Zel'dovich (kSZ) effect --- the Doppler boosting of cosmic microwave background (CMB) photons scattering off free electrons with non-zero line-of-sight velocity --- is an excellent probe of the distribution of baryons in the Universe.  In this paper, we measure the kSZ effect due to ionized gas traced by infrared-selected galaxies from the \emph{unWISE} catalog.  We employ the ``projected-field'' kSZ estimator, which does not require spectroscopic galaxy redshifts.  To suppress contributions from non-kSZ signals associated with the galaxies (e.g., dust emission and thermal SZ), this estimator requires foreground-cleaned CMB maps, which we obtain from \emph{Planck} and \emph{WMAP} data.  Using a new ``asymmetric'' estimator that combines different foreground-cleaned CMB maps to maximize the signal-to-noise, we measure the kSZ$^2$-galaxy cross-power spectrum for three subsamples of the \emph{unWISE} galaxy catalog.  These subsamples peak at mean redshifts $z \approx$ 0.6, 1.1, and 1.5, have average halo mass $\sim 1$-$5\times 10^{13}$ $h^{-1} M_{\odot}$, and in total contain over 500 million galaxies.  After marginalizing over contributions from CMB lensing, we measure the amplitude of the kSZ signal $A_{\rm kSZ^2} = 0.42 \pm 0.31 \, (stat.) \pm 0.02 \, (sys.)$, $5.02 \pm 1.01 \, (stat.) \pm 0.49 \, (sys.)$, and $8.23 \pm 3.23 \, (stat.) \pm 0.57 \, (sys.)$, for the three subsamples, where $A_{\rm kSZ^2} = 1$ corresponds to our fiducial theoretical model.  The combined statistical significance of our kSZ detection exceeds $5\sigma$.  Our theoretical model includes the first calculation of lensing magnification contributions to the kSZ$^2$-galaxy cross-power spectrum, which are significant for the $z \approx$ 1.1 and 1.5 subsamples.  We discuss possible explanations for the excess kSZ signal associated with the $z \approx 1.1$ sample, and show that foreground contamination in the CMB maps is very unlikely to be the cause.  From our measurements of $A_{\rm kSZ^2}$, we constrain the product of the baryon fraction $f_b$ and free electron fraction $f_{\rm free}$ to be $(f_b / 0.158)(f_{\rm free} / 1.0) = 0.65 \pm$ $0.24$, $2.24 \pm$ $0.25$, and $2.87 \pm$ $0.57$ at $z \approx$ 0.6, 1.1, and 1.5, respectively, consistent with a large fraction of the cosmic baryon abundance existing in an ionized state at low redshifts. 

\end{abstract}

\maketitle

\section{Introduction}
\label{sec:introduction}

The kinematic Sunyaev-Zel'dovich (kSZ) effect is the Doppler boosting of Cosmic Microwave Background (CMB) photons due to Compton scattering off free electrons moving with non-zero line-of-sight (LOS) velocity \cite{SZ_1972, SZ_1980}. It is a powerful tool to directly probe the distribution and abundance of baryons in the Universe \cite{battaglia2019probing}. The observed kSZ shift in the CMB temperature is proportional to the total mass density in electrons, and therefore it allows us to trace the electron distribution. This is crucial to probe the ionized warm–hot intergalactic medium (WHIM) that contains most of the universe’s baryons, which otherwise are difficult to detect (so-called ``missing baryons'')~\cite{Bregman2007,Ho2009,Battaglia2017}. In contrast, the thermal Sunyaev-Zel'dovich (tSZ) effect~\cite{1969Ap&SS...4..301Z} is proportional to the integrated electron pressure, and therefore primarily traces galaxy clusters and other very dense structures, where the electron temperature is high due to the depth of the gravitational potential.  The ``missing baryons'' problem has now been solved to zeroth order using multiple methods, including kSZ measurements~\cite{Hand_2012,Hill2016,kSZ_Planck2016} and fast radio burst dispersion measures~\cite{missing_baryons_FRB}, i.e., the expected cosmological abundance of baryons as inferred from the primary CMB and Big Bang nucleosynthesis (BBN) has now been confirmed to exist at low redshifts.  However, the kSZ effect still remains a powerful tool to probe the next-order question, i.e., the precise distribution of electrons in and around galaxies and clusters.  By combining kSZ measurements with tSZ and lensing mass measurements of the same tracer sample, one can determine the full thermodynamic properties of the ionized gas that provides the fuel for star formation in galaxies \cite{Battaglia2017,schaan2020act,Amodeo2020}.

The kSZ effect was first detected by the Atacama Cosmology Telescope (ACT) Collaboration in 2012~\cite{Hand_2012}.  In the past decade, multiple kSZ measurements have been performed using different methods applied to a variety of data sets (e.g.~\cite{Sayers_2013, Hill2016, Ferraro2016, kSZ_Planck2016, Soergel_2016, Schaan_2016, Sayers_2019, schaan2020act, tanimura2020direct}).  These analyses have progressively improved the kSZ detection significance, with the latest analysis of data from ACT and the Baryon Oscillation Spectroscopic Survey (BOSS) using a kSZ estimator with reconstructed velocities achieving the highest significance to date of $7.9\sigma$ \cite{schaan2020act}. 

In this work, we focus on the ``projected-field'' estimator for measuring the kSZ effect.  This technique was first suggested in \cite{Dore2004}, further developed in \cite{DeDeo}, and implemented and applied to data for the first time in Refs.~\cite{Hill2016} and~\cite{Ferraro2016} (hereafter~\citetalias{Hill2016} and~\citetalias{Ferraro2016}, respectively), in which additional theory needed to interpret the estimator was also developed.  It is based on the idea that the kSZ signal traces the overall mass distribution, and thus it can be detected by cross-correlating it with any large-scale structure (LSS) field.  Importantly, the LSS map does not need to contain 3D tracer positions, i.e., spectroscopic redshifts are not required.  Instead, only projected (2D) maps are analyzed.  The crucial step in this method is squaring the CMB temperature map (which contains kSZ) before cross-correlating it with an LSS catalog, or, more generally, measuring the $\langle T T \delta \rangle$ three-point function.  This avoids the cancellation of the kSZ signal that would occur in the two-point cross-correlation $\langle T \delta \rangle$, as the LOS velocity of the electrons sourcing the kSZ signal is equally likely to be positive or negative~\cite{Dore2004}. Since this kSZ$^2$-LSS estimator does not require redshifts for the LSS sample (apart from the overall coarse-grained $dn/dz$ distribution), it can be applied to massive photometric galaxy samples, weak lensing shear maps, or other projected LSS tracer maps. 

The kSZ$^2$-LSS estimator was first applied to data from the \emph{Planck}, \emph{Wilkinson Microwave Anisotropy Probe} (\emph{WMAP}), and \emph{Wide-field Infrared Survey Explorer} (\emph{WISE}) satellite missions~\citepalias{Hill2016}.  This analysis, which was the highest-significance kSZ measurement prior to~\cite{schaan2020act}, yielded a constraint on the product of the baryon fraction $f_b$ and the free electron fraction $f_{\rm free}$, $(f_b/0.158) (f_{\rm free}/1.0) = 1.48 \pm 0.19$ (statistical error only) at redshift $z \approx 0.4$.  The kSZ detection significance in this analysis was 3.8–4.5$\sigma$ (depending on external priors placed on the galaxy bias derived from CMB lensing cross-correlations)~\citepalias{Hill2016}. At that redshift, since hydrogen and helium are fully ionized, we can assume $f_{\rm free} \approx 1$, and thus the inferred value of $f_b$ agreed well with the values derived from primordial CMB \cite{Planck2015parameters} and BBN analyses \cite{Steigman_2007}, resolving the question of missing baryons.

In this paper, we apply the kSZ$^2$-LSS estimator to a new LSS sample, the \emph{unWISE} galaxy catalog. \emph{unWISE} was built upon five years of all-sky infrared imaging from the \emph{WISE} \cite{WISE2010} and \emph{NEOWISE} missions \cite{Mainzer11,Cutri2013,Mainzer14} at W1 (3.4 $\mu$m) and W2 (4.6 $\mu$m).  
This has yielded the deepest-ever all-sky imaging
at 3-5 $\mu$m \cite{fulldepth_neo1, fulldepth_neo2, fulldepth_neo3,Meisner19}.
The publicly available source catalog \cite{Schlafly19}\footnote{\url{https://catalog.unwise.me}}, created using the crowdsource crowded-field phtometry pipeline \cite{Schlafly18}, can be divided into 3 subsamples, blue, green and red, based upon their color and magnitude \cite{Schlafly19,Alex}, and with stars removed using
\emph{Gaia} \cite{Gaia16,Gaia18} astrometric excess noise. 

The main CMB map used here is the same as that used in \citetalias{Hill2016}, the component-separated ``local-generalized morphological component analysis'' (LGMCA \cite{Bobin_2013, Bobin_2014}) CMB map, constructed using data from \emph{Planck} and \emph{WMAP}.  We introduce several novel aspects, both in the model and in the analysis. In the theoretical prediction, we include the magnification bias contributions to the signal, which had previously been neglected, and we also perform the first calculation of the one-halo term to investigate the accuracy of our semi-perturbative model on small scales.  In the data analysis, we improve the S/N on the kSZ signal by using an ``asymmetric'' quadratic estimator (following analogous work in the CMB lensing literature~\cite{MH2018}).  In this approach, instead of simply squaring the same component-separated CMB map (as in \citetalias{Hill2016}), we multiply it with a different CMB map that has a lower noise level but potentially higher foreground contamination (the \emph{Planck} ``spectral matching independent component analysis'' (SMICA) map \cite{Planck2015_comp_sepa}). This combination reduces the noise in the final results, while remaining robust to foregrounds, due to the stringent cleaning in the LGMCA map.  Furthermore, our results are validated with several combinations of two CMB maps (LGMCA$\cdot$SMICA, SMICA-noSZ$\cdot$SMICA, LGMCA$^2$, and SMICA-noSZ$^2$; see Section~\ref{sec:validation}) which all yield similar results in our \emph{unWISE} cross-correlation analysis.  We additionally perform extensive tests for foreground contamination due to dust in the \emph{unWISE} galaxies, finding no evidence of a bias.

Our results are the most competitive up-to-date kSZ measurement with the kSZ$^2$-LSS estimator, with the overall detection significance greater than 5$\sigma$. From our measurements of the kSZ signal, we constrain the product of the baryon fraction $f_b$ and free electron fraction $f_{\rm free}$ to be $(f_b / 0.158)(f_{\rm free} / 1.0) = 0.65 \pm$ $0.24$, $2.24 \pm$ $0.25$, and $2.87 \pm$ $0.57$ (including statistical and systematic errors) for samples at redshift $z \approx$ 0.6, 1.1, and 1.5, respectively. To our knowledge, these are the highest-redshift kSZ measurements to date.  The \emph{unWISE} galaxies live in halos with characteristic halo masses $\sim 1$-$5 \times 10^{13}$ $h^{-1}$ $M_{\odot}$~\cite{Alex}, although our measurement predominantly probes the large-scale regime beyond the virial radius of these objects.  Since $f_{\rm free}$ is close to unity at these redshifts (nearly all baryons are ionized), our measurements thus show that there are no ``missing baryons'' on large scales at low-$z$.

The remainder of this paper is organized as follows. In Section \ref{sec:theory}, we describe the kSZ$^2$-LSS estimator and our theoretical model of the harmonic-space cross-correlation of a galaxy catalog with the real-space product of two blackbody CMB maps. In Section \ref{sec:data}, we describe the data used in the analysis: foreground-cleaned CMB maps and the \emph{unWISE} galaxy catalog. In Section \ref{sec:analysis}, we discuss the foreground cleaning methods and null tests on the CMB maps to exclude the possibility of dust contamination, which might bias our measurement, and present the results of the main (LGMCA$\cdot$SMICA)$\times$\emph{unWISE} cross-correlation. In Section \ref{sec:interpretation} we discuss the fit of the measured data points to our predicted theoretical model from Section \ref{sec:theory}. Section \ref{sec:validation} addresses the additional validation of our results performed with different combinations of three CMB maps, and fits these measurements to the same theoretical model. In Section \ref{sec:discussion} we discuss the kSZ measurements and possible sources of the anomalously high amplitude of the kSZ signal in one of the galaxy samples. In Appendix \ref{sec:append:const_alpha} we present a different cleaning method of the CMB, which agrees very well with our original cleaning, and in Appendix~\ref{sec:append:dust} we assess the dust level in the LGMCA map. Both appendices demonstrate that the LGMCA map contains negligible thermal dust contamination from the \emph{unWISE} galaxies compared to our measured kSZ signal, and it can be safely assumed that the foreground contamination is not a source of the detected signal.  Appendix~\ref{app:posteriors} contains the full posteriors for our parameter fits.  In Appendix \ref{s:hm}, we present a first calculation of the one-halo term, to explore the validity of our semi-perturbative model on small scales.  Appendix~\ref{sec:append_valid_plots} contains additional validation plots for our analysis pipeline. 

Our fiducial theoretical model assumes a $\Lambda$CDM cosmology with \emph{Planck} 2018 best-fit parameters (the first column in Table 1 of Ref.~\cite{Planck2018}).  The cosmic baryon abundance in this model is $\Omega_b / \Omega_m = 0.158$.  All error bars quoted in this work are 1$\sigma$.

\section{Theory}
\label{sec:theory}

In this section we describe the kSZ$^2$-LSS estimator along with the predicted contributions to the cross-correlation of the product of two CMB maps and a galaxy catalog. 

\subsection{The kSZ$^2$-LSS estimator}\label{ss:kszest}
The kSZ$^2$-LSS estimator is based on the idea of cross-correlating the squared kSZ signal extracted from a frequency-cleaned CMB map with a projected LSS tracer map.  Here, we focus on the projected galaxy overdensity, but the estimator can also be applied to a tracer map of weak lensing convergence, thermal SZ, or other projected fields.  By squaring the kSZ signal, the estimator avoids the cancellation that would otherwise occur when averaging over LOS velocities that are equally likely to be positive or negative.

The fractional CMB temperature shift in a direction $\boldsymbol{ \hat{n} }$ due to the kSZ effect, $\Theta^{{\rm kSZ}}(\boldsymbol{ \hat{n} })$, is
\begin{equation}
\begin{multlined}
 \Theta^{{\rm kSZ}}(\boldsymbol{ \hat{n} }) = -  \int_{0}^{\eta_{re}} d\eta \,  g(\eta) \, \boldsymbol{p_e} \cdot \boldsymbol{ \hat{n} }  \\
 = - \sigma_T \int_{0}^{\eta_{re}} \frac{d \eta}{1+z} e^{-\tau} n_e(\boldsymbol{ \hat{n} }, \eta) \boldsymbol{v_e} \cdot \boldsymbol{ \hat{n} } ,
 \label{eq.kSZdef}
\end{multlined}
\end{equation}
where $\eta(z)$ is the comoving distance to redshift $z$,  $\eta_{re}$ is the comoving distance to the end of hydrogen reionization, $g(\eta) = e^{-\tau} d\tau / d\eta$ is the  visibility function, $\tau$ is the optical depth to Thomson scattering, $n_e$ is the free electron number density, $\boldsymbol{p_e} = (1+ \delta_e) \boldsymbol{v_e}$ is the electron momentum, and $\boldsymbol{v_e}$ is the peculiar electron velocity.

The projected galaxy overdensity $\delta_g(\boldsymbol{ \hat{n} })$, our tracer of the three-dimensional matter overdensity, is
\begin{equation}
    \delta_g(\boldsymbol{ \hat{n} }) = \int_{0}^{\eta_{\rm max}} d\eta W_g(\eta) \delta_m(\eta \boldsymbol{ \hat{n} }, \eta),
    \label{eq.delta_g}
\end{equation}
where $\eta_{\rm max}$ is the maximum comoving distance of the galaxy sample, $\delta_m = (\rho_m - \bar{\rho}_m )/ \bar{\rho}_m$ is the matter overdensity, $\rho_m$ is the matter density, and $W_g(\eta) = b(\eta) p_s(\eta)$ is the projection kernel, $b (\eta)$ is the linear galaxy bias (for which we allow a redshift dependence), and $p_s(\eta)$ is the distribution of the galaxies in comoving distance (normalized to have unit integral). 

In order to extract the kSZ signal from a blackbody CMB temperature map, we apply a Wiener filter $F(\ell)$  in harmonic space, which is defined as 
\begin{equation}
    F(\ell)=C_{\ell}^{{\rm kSZ}}/C_{\ell}^{\rm tot},
    \label{eq.filter}
\end{equation}
where $C_{\ell}^{{\rm kSZ}}$ is the theoretical kSZ power spectrum (obtained from hydrodynamic simulations and semi-analytic models~\cite{Battaglia_2010, Battaglia_2013} as in~\citetalias{Hill2016}) and $C_{\ell}^{\rm tot}$ is the total power spectrum, including the CMB, kSZ, and integrated Sachs-Wolfe (ISW) contributions, noise, and residual foregrounds.  The result of Eq.~\ref{eq.filter} is then normalized to have a maximum value of unity.  This filter downweights low-$\ell$ modes that are dominated by the primary CMB and high-$\ell$ modes that are dominated by noise.

\begin{figure}
\includegraphics[width=1.\columnwidth]{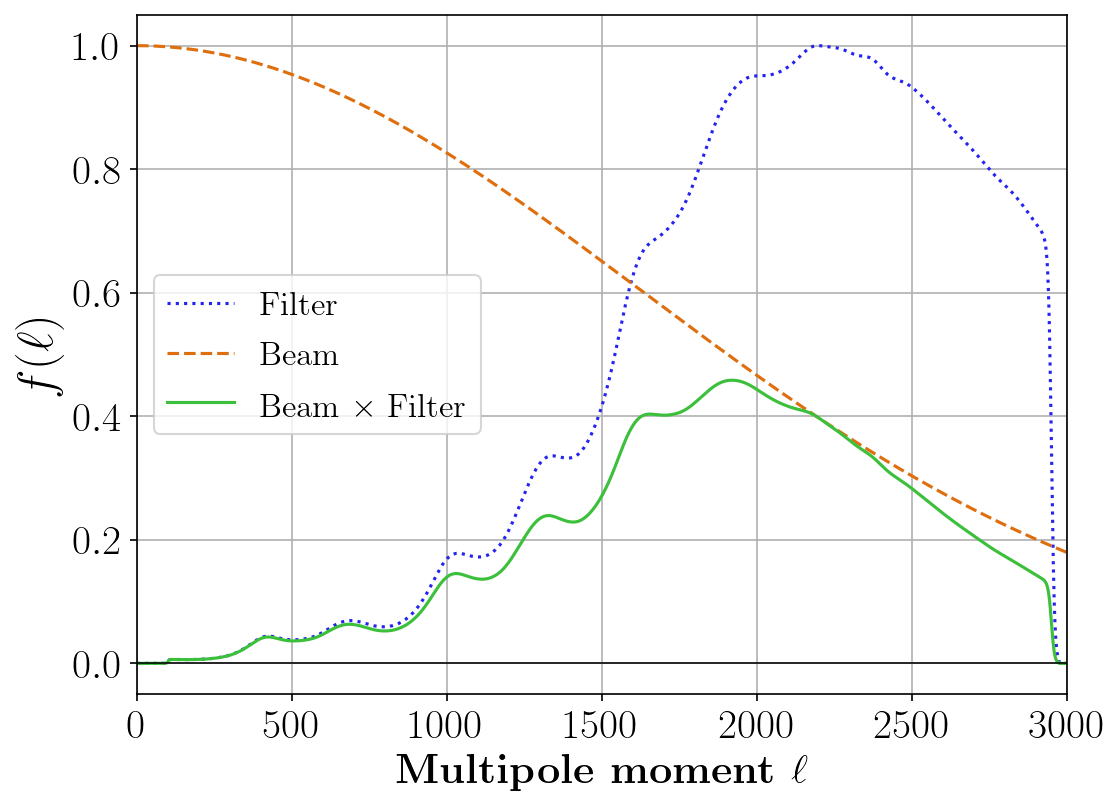}
    \caption{The filter (used to extract the kSZ signal) and beam functions (to account for the finite resolution of the telescope), along with their product versus multipole moment used in the main analysis. See Eqs. \ref{eq.filter}, \ref{eq.filterbeam}, and \ref{eq.beam} for more details. }
    \label{Filter}
\end{figure}

Furthermore, the CMB maps are observed with a finite telescope beam $b(\ell)$, which we treat as an additional factor in the overall filter as in~\cite{Hill2016,Ferraro2016}. Then, the filtered fractional temperature shift $\Theta_f(\ell)$ (in harmonic space) is related to the true CMB anisotropy $\Theta(\ell)$ as 
\begin{equation}
    \Theta_f(\ell) = F(\ell) b(\ell ) \Theta(\ell) \equiv f(\ell) \Theta(\ell),
    \label{eq.filterbeam}
\end{equation}
where the defined function $f(\ell)$ is the product of the filter function $F(\ell)$ and beam window function $b(\ell)$, which is
\begin{equation}
b\left(\ell\right)=\exp\left\{ -\frac{1}{2}\ell\left(\ell+1\right)\hat{\sigma}^{2}\right\},
\label{eq.beam}
\end{equation}
with $\hat{\sigma}=\mathrm{FWHM}/\sqrt{8\ln2}$, where $\mathrm{FWHM}$ is the full width at half maximum of the telescope beam in radians.  For all maps used in our analysis, the FWHM = 5 arcmin.  The filter, beam functions, and their product are shown in Fig.~\ref{Filter}.  The theoretical calculations in this work are specific to this choice of filter and beam; results for higher-resolution experiments were computed in~\citetalias{Ferraro2016}.  Hereafter, we will drop the subscript ``$f$'' on CMB temperature (and kSZ) fields, which should be understood to be filtered unless otherwise specified.

A direct cross-correlation of $\Theta^{{\rm kSZ}}$ and $\delta_g$ is expected to vanish because of the LOS velocity symmetry of the kSZ signal, which is as likely to be positive as it is to be negative \cite{Dore2004,Hill2016}. Therefore, following the method from \cite{Dore2004,DeDeo,Hill2016,Ferraro2016}, we first square the filtered kSZ fractional CMB temperature shift $\Theta^{{\rm kSZ}}$ in real space, and then cross-correlate it with the galaxy overdensity $\delta_g$ \cite{Dore2004, DeDeo}. In the Limber approximation \cite{Limber, LimberExtended}, this yields
\begin{equation}
    C_{\ell}^{{\rm kSZ}^2 \times \delta_g} = \frac{1}{c^2} \int_{0}^{\eta_{\rm max}} \frac{d\eta}{\eta^2} W_g(\eta)g^2(\eta) \mathcal{T} (k= \frac{\ell}{\eta}, \eta),
\label{eq.ClkSZ2delta}
\end{equation}
where $\mathcal{T} (k, \eta)$ is the \emph{triangle power spectrum}~\cite{Dore2004}:
\begin{equation}
    \mathcal{T} (k, \eta) = \int \frac{d^2 \boldsymbol{q}}{(2\pi^2)} f(q\eta)f(|\boldsymbol{k}+\boldsymbol{q}|\eta) B_{\delta_g {p_{\boldsymbol{\hat{n}}} p_{\boldsymbol{\hat{n}}}}}(\boldsymbol{k}, \boldsymbol{q}, \boldsymbol{-k-q}).
\end{equation} 
The triangle power spectrum $\mathcal{T} (k, \eta)$ is an integral over the hybrid bispectrum $B_{\delta_g {p_{ \boldsymbol{\hat{n}}} p_{\boldsymbol{\hat{n}}}}}$ of one density contrast and two LOS electron momenta, $p_e(\hat{\bf n})$.  Following previous work~\cite{Dore2004,DeDeo,Hill2016,Ferraro2016}, we approximate the hybrid bispectrum as the product of the velocity dispersion and the non-linear matter bispectrum: 
\begin{equation}
B_{\delta_g {p_{ \boldsymbol{\hat{n}}} p_{\boldsymbol{\hat{n}}}}} = \frac{1}{3}v_\mathrm{rms}^2 B_\mathrm{m}^\mathrm{NL} \,.
\label{eq:bispec}
\end{equation}
This approximation is equivalent to the statement that the only strongly non-Gaussian field in the five-point function $\langle \delta {\bf v} \delta {\bf v} \delta \rangle$ is the density fluctuation $\delta$, and thus the only non-zero connected term amongst the Wick contractions of this five-point function is that in Eq.~\ref{eq:bispec}.  The validity of this approach was first assessed analytically in Ref.~\cite{DeDeo}, who showed that Eq.~\ref{eq:bispec} was indeed the dominant term; this approximation was then confirmed in detail via comparison to cosmological hydrodynamics simulations in~\citetalias{Ferraro2016}.  Here, we follow~\cite{Hill2016,Ferraro2016} in modeling the non-linear matter bispectrum $B_\mathrm{m}^\mathrm{NL}$ using a fitting function derived from dark-matter-only $N$-body simulations~\cite{Gil_Mar_n_2012}.  Implicitly, we thus assume that the free electron distribution traces that of the dark matter on the scales probed by our measurements.  Appendix~\ref{s:hm} discusses this issue in further detail.  We compute the velocity dispersion $v_\mathrm{rms}^2$ using the Halofit prescription for the non-linear matter power spectrum, which increases the overall signal by $\approx 20$\% as compared to a linear-theory calculation.  The velocity dispersion is discussed further in Section~\ref{sec:discussion} and Appendix~\ref{s:hm}.

We note that the non-linear matter bispectrum (or, more directly relevant here, the galaxy-electron-electron bispectrum) can also be computed within the halo model of LSS.  In Appendix~\ref{s:hm}, we outline the main aspects of the halo model computation for the one-halo term, which we compute as a cross-check of the fitting-function-based model on small scales, where it may not be expected to be highly precise, and where uncertainties in the galaxy-halo connection may lead to large uncertainties in the projected-field kSZ signal.

Since the visibility function $g(\eta) \propto f_b f_{\rm free}$, we have $C_{\ell}^{{\rm kSZ}^2 \times \delta_g} \propto f_b^2 f_{\rm free}^2$ \cite{Hill2016,Ferraro2016}.  Thus, the kSZ${}^2$-LSS cross-correlation is a direct probe of ionized gas in galaxies, groups, and clusters.  Alternatively, on large scales for which $f_b$ is expected to be equal to the cosmic baryon abundance and $f_{\rm free}$ is close to unity, the kSZ${}^2$-LSS cross-correlation could be used to probe dark energy, neutrino masses, modified gravity, and other novel physics through their effects on the large-scale velocity dispersion in Eq.~\ref{eq:bispec}~(e.g.,~\cite{DeDeo}).

\subsection{CMB lensing contribution}

Gravitational lensing generates non-zero correlations between quadratic combinations of Fourier modes in CMB temperature and polarization maps~(see, e.g.,~\cite{Lewis-Challinor2006} for a review of CMB lensing).  To lowest order in the CMB lensing potential $\psi_{\rm CMB}$, the correlation of a squared CMB temperature map with the galaxy overdensity field depends directly on the CMB lensing -- galaxy cross-correlation, i.e., $\langle T^2 g\rangle \propto C_{\ell}^{\psi_{\rm CMB} \delta_g}$.  Since the Wiener filter in Eq.~\ref{eq.filter} does not completely remove all primary CMB fluctuations, we must account for the lensing contribution in our analysis --- indeed, it turns out to be the largest term in the overall signal.\footnote{Similarly, the kSZ${}^2$-LSS cross-correlation leads to a contribution in quadratic-estimator-reconstructed CMB lensing cross-correlation measurements that will require mitigation in upcoming surveys~\cite{FerraroHill2018}.}  
Note that the lensing contribution is proportional to the overall amplitude of the $\delta_g$ field, which we will denote $b_g$ (the amplitude of the linear galaxy bias), and thus its presence actually allows the degeneracy between $b_g$ and $A_{\rm kSZ^2}$ in the kSZ signal to be broken in our data analysis~\citepalias{Hill2016}.

To lowest order in the lensing potential, the CMB lensing contribution to the squared temperature -- galaxy cross-correlation is~\citepalias{Ferraro2016}
\begin{equation}
\begin{multlined}
 \Delta C_\ell^{T^2 \times \delta_g} \approx -2 \frac{\ell\ C_\ell^{\psi_{\rm CMB}\delta_g}}{(2\pi)^2} \int_{0}^\infty dL'L'^{2}f(L') C_{L'}^{TT} \\ \int_{0}^{2\pi} d\phi\ f(| \pmb{L'} + \pmb{\ell} |) \cos\phi, 
 \label{eq.lens}
\end{multlined}
\end{equation}
where $C_\ell^{\psi_{\rm CMB} \delta_g}$ is the cross-power spectrum between the gravitational lensing potential and the galaxy overdensity, $C_{L'}^{TT}$ is the unlensed primary CMB temperature power spectrum, $f(\ell)$ is defined in Eq.~\ref{eq.filterbeam}, and $\phi$ is the angle between the 2D wavevectors $\pmb{L'}$ and $\pmb{\ell}$.  This analytic expression for the lensing contribution was compared to numerical simulations in Refs.~\cite{Ferraro2016, Hill2016} and was found to be extremely accurate (agreement at $< 3$\% precision up to $\ell=3000$).  Note that Eq.~\ref{eq.lens} neglects the ISW contribution to $T^2$, but this is orders of magnitude below the lensing term on the scales in our analysis~\citepalias{Ferraro2016}.

To obtain $C_\ell^{{\psi_{\rm CMB}}{\delta_g}}$, we start with the lensing convergence $\kappa(\boldsymbol{\theta})$, which is a weighted projection of the three-dimensional matter overdensity $\delta_m$ along the LOS,
\begin{equation}
\kappa(\boldsymbol{\theta}) = \int_0^{\infty} dz W(z) \delta_m(\eta(z)\thetaB, z),
\label{eq.lensingkernel}
\end{equation}
where the kernel $W(z)$ indicates the lensing strength at redshift $z$ for sources with a redshift distribution $dn(z_s)/dz_s$. For a flat universe,
\begin{eqnarray}
W(z) &=& \frac{3}{2}\Omega_{m} H_0^2 \frac{(1+z)}{H(z)} \frac{\eta(z)}{c} \int_z^{\infty} dz_s \frac{dn}{dz_s} \frac{\eta(z_s) - \eta(z)}{\eta(z_s)},\nonumber\\
\label{eq:wkappag}
\end{eqnarray}
where $\Omega_{m}$ is the matter density as a fraction of the critical density at $z=0$, $H(z)$ is the Hubble constant at redshift $z$, with a present-day value $H_0$, $c$ is the speed of light, and $z_s$ is the source redshift. Note that $\int_0^{\infty} dz_s dn/dz_s = 1$.

For CMB lensing, there is only one source plane at the last scattering surface $z_\star = 1090$. Using $dn(z_s)/dz_s=\delta_D(z_s-z_\star)$, where $\delta_D$ is the Dirac delta function, the CMB lensing kernel is
\begin{eqnarray}
W^{\kappa_{\rm CMB}}(z) &=&  \frac{3}{2}\Omega_{m}H_0^2  \frac{(1+z)}{H(z)} \frac{\eta(z)}{c} \frac{\eta(z_\star)-\eta(z)}{\eta(z_\star)}.\label{eq:kappacmb}
\end{eqnarray}

In the expression for the angular galaxy number density field $\delta_g$, defined in Eq. \ref{eq.delta_g}, we can write the projection kernel $W_g(\eta)$ explicitly, and thus obtain
\begin{equation}
\delta_g(\boldsymbol{\theta}) = \int_0^{z_{\rm max}} dz \, b(z) \frac{dn}{dz} \delta_m(\eta(z)\thetaB, z) \,, 
\label{eq.deltag}
\end{equation}
where $b(z)$ is the galaxy bias, which we allow to depend on redshift (c.f. Eq.~\ref{eq.delta_g}), with an overall amplitude $b_g$.

We can use these results to obtain the auto- and cross-power spectra of these fields as LOS integrals of the three-dimensional matter power spectrum.  In the Limber approximation~\cite{LimberExtended,Limber}, the CMB lensing -- angular galaxy number density cross-correlation is
\begin{equation}
\label{eqn:theory_eqn}
C_\ell^{{\kappa_{\rm CMB}}{\delta_g}} = \int_0^\infty \frac{dz}{c} \frac{H(z)}{\eta^2(z)} W^{\kappa_{\rm CMB}}(z) \, b(z) \frac{dn}{dz} P\left(k, z \right),
\end{equation}
where $P\left(k, z \right)$ is the matter power spectrum evaluated at wavenumber $k = (\ell + 1/2)/\eta(z)$ at redshift $z$. We compute the matter power spectrum with CLASS \cite{CLASS}, using the Halofit fitting function  \cite{Takahashi_2012} to compute non-linear corrections.  Our models for $b(z)\frac{dn}{dz}$ and $\frac{dn}{dz}$ for the \emph{unWISE} galaxy samples are discussed in Section~\ref{ss:gc} (see Fig.~\ref{dn/dz}).

Finally, note that the lensing convergence $\kappa$ is related to the lensing potential $\psi$ by $\kappa(\boldsymbol{\theta}) =
-\nabla^2 \psi(\boldsymbol{\theta}) / 2$ where $\nabla^2$ is the two-dimensional Laplacian on the sky, which yields $\kappa_{\ell}
= \ell(\ell+1) \psi_{\ell}/2$ in multipole space.  Thus
\begin{equation}
C_\ell^{{\psi_{\rm CMB}}{\delta_g}} = \frac{2}{\ell(\ell+1)} C_\ell^{{\kappa_{\rm CMB}{\delta_g}}} \,,
\label{eq.lensing_deltag}
\end{equation}
which directly enters Eq.~\ref{eq.lens}.

\subsection{Magnification bias contributions}

The observed galaxy number density fluctuation field, $\delta_g$, is the sum of the contribution from intrinsic (physical) fluctuations in the galaxy number density and the contribution from lensing magnification fluctuations coupling to the galaxy field.  The latter contribution due to magnification bias can be non-negligible for galaxy samples with luminosity functions that are steep at the faint end, near the threshold for detection.  This is quantified by $s \equiv d \log_{10} N/dm$, the logarithmic slope of the galaxy number counts as a function of apparent magnitude $m$ near the magnitude limit of the survey.

The contribution due to magnification bias in the observed galaxy number density field, $\mu_g$, can be written as a LOS integral:
\begin{equation}
\mu_{g}(\boldsymbol{\theta}) = (5s-2) \int_0^{\infty} dz W(z) \delta(\eta(z)\thetaB, z) \,,
\label{eq.magbias}
\end{equation}
where $W(z)$ is the lensing kernel defined in Eq.~\ref{eq:wkappag}.  The total observed angular number density fluctuation field is then $\delta_g^{\rm obs} = \delta_g + \mu_g$, where $\delta_g$ is given by Eq.~\ref{eq.deltag} and $\mu_g$ is given above.

The galaxy magnification bias field has a non-zero correlation with both the CMB lensing field and the kSZ${}^2$ field.  We account for both in our analysis, improving upon Refs.~\cite{Hill2016,Ferraro2016}, which neglected these contributions.  As will be seen below, the magnification bias contributions are relatively small for the lowest-redshift (blue) sample in our analysis, which is similar to the sample used in~\cite{Hill2016,Ferraro2016}; thus, the conclusions of that work are unlikely to be strongly biased due to neglecting these terms.  However, for the higher-redshift (green and red) samples that we analyze, the magnification bias terms are large and must be included to obtain unbiased results.

Analogous to Eq.~\ref{eq.lens}, the lowest-order expression for the contribution of the CMB lensing -- magnification bias cross-correlation to our measured statistic is
\begin{equation}
\begin{multlined}
 \Delta C_\ell^{T^2 \times \mu_g} \approx -2\frac{\ell\ C_\ell^{\psi_{\rm CMB}\mu_g}}{(2\pi)^2} \int_{0}^\infty dL'L'^{2}f(L') C_{L'}^{TT} \\ \int_{0}^{2\pi} d\phi\ f(| \pmb{L'} + \pmb{\ell} |) \cos\phi \,.
 \label{eq.lens_magn}
\end{multlined}
\end{equation}
Analogous to Eq.~\ref{eqn:theory_eqn}, the cross-power spectrum between the CMB lensing and lensing magnification bias fields is
\begin{equation}
C_\ell^{{\kappa_{\rm CMB}}{\mu_g}} = (5s-2) \int_0^\infty \frac{dz}{c} \frac{H(z)}{\eta^2(z)} W^{\kappa_{\rm CMB}}(z) W(z) P\left(k, z \right),
\end{equation}
where $W(z)$ is the  lensing kernel of Eq. \eqref{eq:wkappag} and $W^{\kappa_{\rm CMB}}(z)$ is the CMB lensing kernel of Eq. \eqref{eq:kappacmb}.

The magnification bias field also correlates with the squared kSZ signal.  Following the same approach used above to obtain Eq.~\ref{eq.ClkSZ2delta}, the cross-correlation of $(\Theta^{{{\rm kSZ}}})^2$ with the magnification bias field $\mu_g$ (Eq.~\ref{eq.magbias}) is given by
\begin{equation}
C_\ell^{{{\rm kSZ}^2} \times {\mu_g}} = (5s-2) \frac{1}{c^2} \int_{0}^{\eta_{\rm max}} \frac{d\eta}{\eta^2} W(\eta)g^2(\eta) \tau (k= \frac{\ell}{\eta}, \eta) \,. 
\label{eq.ClkSZ2mu}
\end{equation}
This kSZ$^2$-magnification bias term is the final component of the theoretical model that we fit to the data, in addition to the terms in Eqs.~\ref{eq.ClkSZ2delta},~\ref{eq.lens}, and~\ref{eq.lens_magn}.  Note that Eq.~\ref{eq.ClkSZ2mu} describes the correlation between the (squared) kSZ field and the total matter density field, as traced by lensing --- the galaxy bias does not appear.  Thus a measurement of this term directly probes the manner in which free electrons trace the underlying matter density field, with no need to model the relation between galaxies and matter.

\section{Data} 
\label{sec:data}

\subsection{CMB Maps}

We use several foreground-cleaned blackbody-component CMB temperature maps in this analysis.  The primary map that we focus on, as was used in~\citetalias{Hill2016}, is the ``local-generalized morphological component analysis'' (LGMCA) map, which was constructed from \emph{WMAP} 9-year \cite{WMAP9} and {\emph{Planck-PR2}} \cite{Planck2015_comp_sepa} data.  The LGMCA method extracts the blackbody CMB signal (including kSZ and ISW) from other sky components by relying on the sparse distribution of non-CMB foregrounds in the wavelet domain \cite{LGMCA, Bobin_2013, Bobin_2014}. We note that the tSZ signal is explicitly removed in the LGMCA map using a constraint, by taking advantage of the known frequency spectrum of the tSZ effect.  While other foregrounds are not explicitly nulled, the dust contamination in the LGMCA map is minimal, as will be shown below.

We also make use of the \emph{Planck} ``spectral matching independent component analysis'' (SMICA) map, which is constructed from a linear combination of the nine pre-processed \emph{Planck} frequency channels from 30 to 857 GHz \cite{Planck2015_comp_sepa,PlanckComponentSeparation}.  The SMICA algorithm is quite different from that of LGMCA, and is applied in the harmonic rather than wavelet domain, thus providing a useful test for our results.  In addition, we consider the ``SMICA-noSZ'' map, which is similar to SMICA but imposes an explicit spectral constraint to null the tSZ signal, as described above for the LGMCA map.  All of the CMB maps described here have an angular resolution of FWHM = 5 arcmin.

Other CMB maps which we use are the LGMCA noise map \cite{Bobin_2013, Bobin_2014} (constructed from the half-difference of splits of the \emph{Planck} and \emph{WMAP} data) for the null tests (see Section \ref{sec:analysis}), \emph{Planck} 545 GHz (FWHM = 4.63 arcmin) and 857 GHz maps (FWHM = 4.83 arcmin) \citep{Planck2015_comp_sepa} for the explicit dust removal (cleaning procedure described in Section~\ref{sec:analysis}) and dust null tests.

Our analysis employs an ``asymmetric'' technique to construct the $T^2$ field, in which we multiply two different CMB maps, one of which has more robust foreground cleaning but lower S/N, and the other of which has higher S/N but slightly higher foreground contamination.  The motivation for doing this is to improve the overall S/N of our measurement --- by only using a fully-foreground-nulled map in one ``leg'' of the $T^2$ field, we can decrease the noise penalty associated with this foreground cleaning on our measurement (see, e.g.,~\cite{Abylkairov2020} for a recent exploration of such tradeoffs in CMB foreground cleaning).  This method remains unbiased due to the fact that the full foreground deprojection --- in our case, the tSZ removal --- has been applied to one of the $T$ maps, and thus even if it is present in the other $T$ map there will still be zero correlation in the final result.  This is analogous to recent developments in foreground-immune CMB lensing reconstruction quadratic estimators, which employ one foreground-deprojected $T$ map and one $T$ map in which only total variance minimization has been applied~\cite{MH2018,Darwish2021}.  In our fiducial analysis, the LGMCA map serves as the fully-foreground-deprojected $T$ map and the SMICA map serves as the $T$ map with slightly lower noise in the asymmetric combination.  We explicitly confirm that this choice increases the S/N of our final kSZ measurements by $\approx 10$\%, as compared to using the LGMCA map in both legs of the $T^2$ combination. 

To optimally extract the kSZ signal from the component-separated CMB maps, we use a Wiener filter (nearly identical to the filter used in \citetalias{Hill2016}) $F_{\ell}=C_{\ell}^{{\rm kSZ}}/C_{\ell}^{\rm tot}$. As described in Section \ref{sec:theory}, the numerator, $C_{\ell}^{{\rm kSZ}}$, is the theoretical kSZ power spectrum and the denominator, $C_{\ell}^{\rm tot}$, is set to $C_{\ell}^{\rm LGMCA}$, the auto-power spectrum of our main CMB map (LGMCA). The filter is set to zero for $\ell <100$ (to remove any contribution from the ISW effect) and $\ell>3000$ (to remove the noise-dominated modes), multiplied by a hyperbolic tangent to ensure smoothness at the boundaries, and normalized to a maximum value of unity (at $\ell \approx 2200$). The filter function $F(\ell)$ is shown in Fig. \ref{Filter}. The beam window function of the LGMCA maps $b(\ell)$ is modeled as a Gaussian with FWHM = 5 arcmin, and is also shown on this plot, along with the product of the beam and filter functions, $f(\ell)$ (see Equations~\ref{eq.filterbeam} and~\ref{eq.beam}). To ensure consistency in the filtered kSZ and lensing signals that enter our estimator, we also filter the SMICA map (and SMICA-noSZ) using the same $F(\ell)$ and beam $b(\ell)$ (the SMICA maps also have FWHM = 5 arcmin).

\subsection{Galaxy Catalogs}
\label{ss:gc}

Our LSS tracer is the \emph{unWISE} galaxy catalog, which is constructed using data from the Wide-Field Infrared Survey Explorer (\emph{WISE}) mission, and additionally includes the post-hibernation \emph{NEOWISE} mission \cite{Schlafly19,Alex}. The \emph{unWISE} catalog provides a sample of more than 500 million galaxies with redshift $0 < z < 2$ across the full sky, and is the largest full-sky galaxy catalog currently available.\footnote{The recent Dark Energy Survey DR2 data release contains a similar number of galaxies ($\approx$ 543 million), but on a $5000$ deg${}^2$ footprint rather than the full sky~\cite{DES_DR2}.} It is further divided into 3 subsamples using cuts on infrared galaxy color and magnitude, as detailed below: \emph{blue}, \emph{green}, and \emph{red} at mean redshifts $\approx 0.6, 1.1,$ and $1.5$, respectively~\cite{Alex}. 

\begin{figure}[htbp!]
    \centering
    \includegraphics[width=.45\textwidth]{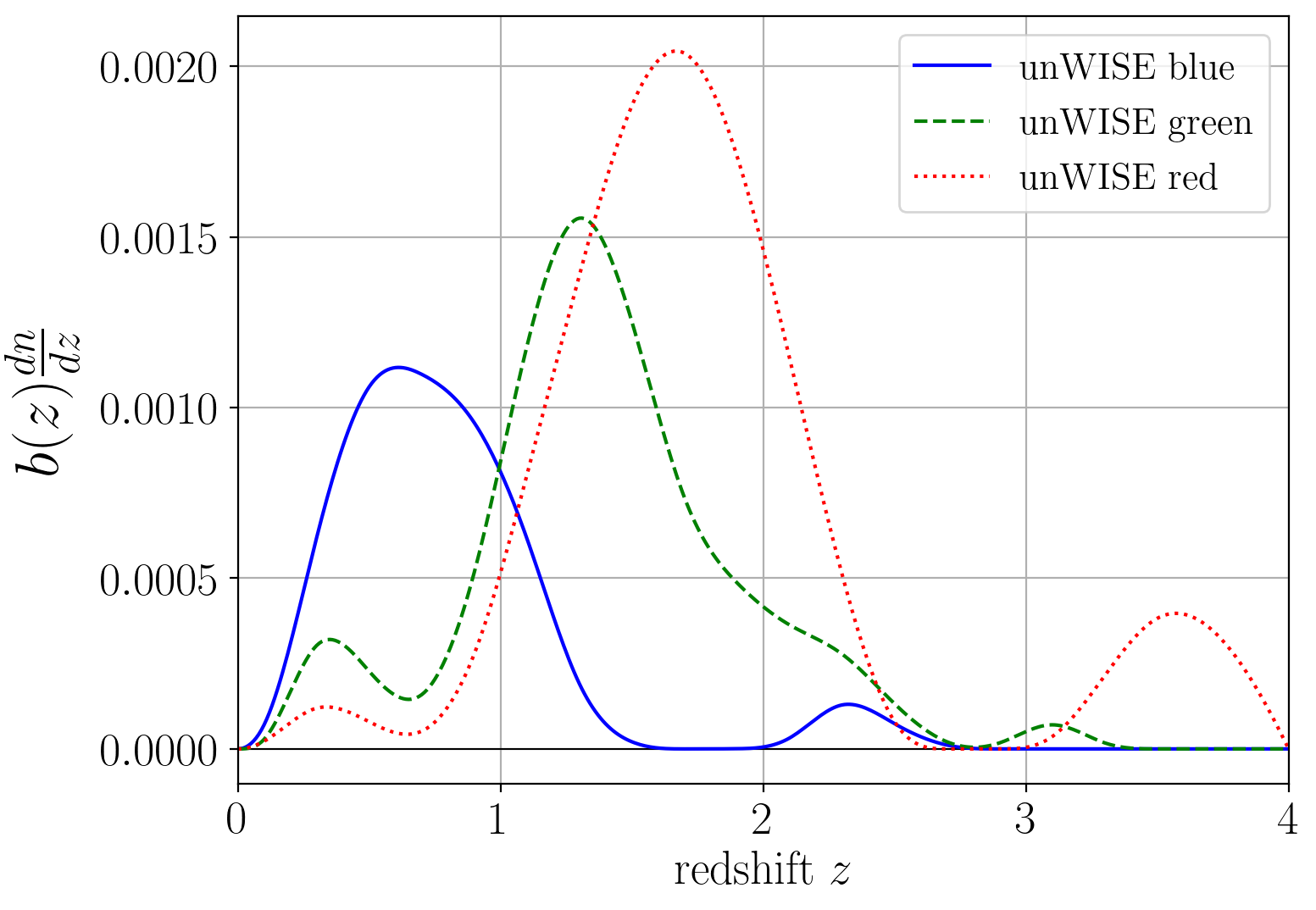}
    \includegraphics[width=.44\textwidth]{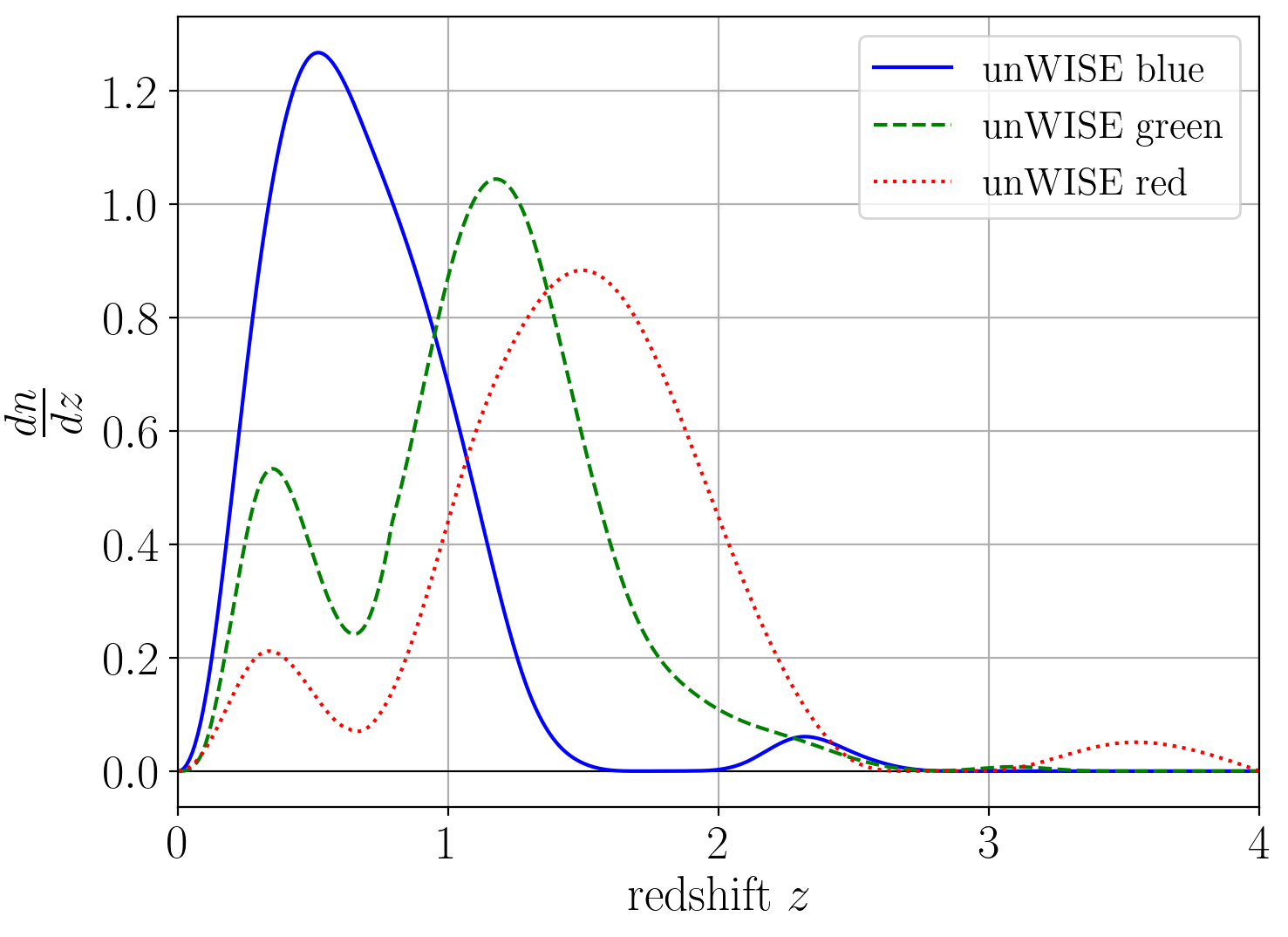}
    \caption{Top: Bias-weighted redshift distributions $b(z)dn/dz$ (obtained by cross-correlation with \emph{BOSS} and \emph{eBOSS} surveys) for the blue (solid line), green (dashed line), and red (dotted line) \emph{unWISE} samples, as measured in \citep{Alex}. 
    Bottom: Normalized statistical redshift distribution of the \emph{unWISE} galaxies for the blue (solid line), green (dashed line), and red (dotted line) obtained by dividing the curves from the top panel by a redshift dependent bias $b(z)$ function, and normalizing. See Table \ref{table:1} for mean redshifts and other properties associated with these galaxy samples.} 
    \label{dn/dz}
\end{figure}

The \emph{WISE} mission mapped the entire sky at 3.4, 4.6, 12, and 22 $\mu$m (W1, W2, W3, and W4) with angular resolution of $6.1''$, $6.4''$, $6.5''$, and $12''$, respectively \cite{Wright_2010, Alex}. The \emph{unWISE} galaxies are selected from the \emph{unWISE} catalog based on cuts on the W1 and W2 magnitudes.  Further, stars are removed by cross-matching the catalog with \emph{Gaia}; each of the samples is required to be either undetected or not point-like in \emph{Gaia}~\cite{Alex,Gaia16, Gaia18}.  Table~\ref{table:1} gives a summary of the cuts made to construct the three \emph{unWISE} subsamples. Their important properties are summarized in Table~\ref{table:1a}, including redshift distribution (mean redshift $\bar{z}$, and redshift width $\delta_z$), number density $\bar{n}$, and faint-end logarithmic slope of the luminosity function $s = d \log_{10} N/dm$. 
The corresponding average halo masses of the samples, as inferred from their halo biases, are $\sim 1$-$5 \times 10^{13}$ $h^{-1}$ $M_{\odot}$~\cite{Alex}.
The redshift distribution of the galaxies, $b(z)dn/dz$, obtained for each sample in Ref.~\cite{Alex} from cross-correlation with large-area spectroscopic surveys (\emph{BOSS} galaxies and quasars and \emph{eBOSS} quasars) is shown in Fig.~\ref{dn/dz} (the method is described in detail in \cite{Newman_2008, McQuinn_2013}). The $dn/dz$ curves in Fig.~\ref{dn/dz} are obtained by dividing the cross-correlation $b(z)dn/dz$ measured in~\cite{Alex} by the best-fit bias model for each sample ($b(z)=0.8 + 1.2z$ for blue, $b(z)={\rm max}(1.6z^2,1)$ for green, and $b(z)={\rm max}(2z^{1.5},1)$ for red) and normalizing \cite{Alex}.

    \begin{table}[h!]
    \begin{tabular}{| c| c |c|c|c| } 
    \hline
    \emph{unWISE} & W2 $<$ & W2 $>$ & W1 - W2 $>$ x &  W1 - W2 $<$ x  \\ 
    \hline\hline
    blue & 15.5 & 16.7 & & $\frac{(17 - W2)}{4}+0.3$ \\
    green & 15.5 & 16.7 & $\frac{(17 - W2)}{4}+0.3$ & $\frac{(17 - W2)}{4}+0.8$\\ 
    red & 15.5 & 16.2 & $\frac{(17 - W2)}{4}+0.8$ &  \\ 
    \hline
    \end{tabular}
    \caption{Cuts made on infrared color and magnitude in the \emph{WISE} data to construct the \emph{unWISE} catalogs  (see \cite{Schlafly19,Alex} for further details).}
    \label{table:1}
    \end{table}
    
    \begin{table}[h!]
    \begin{tabular}{| c| c |c|c|c| } 
    \hline
    \emph{unWISE} & $\bar{z}$ & $\delta_{z}$ & $\bar{n}$ & s \\ 
    \hline\hline
    blue & 0.6 & 0.3 & 3409 & 0.455\\
    green & 1.1& 0.4 & 1846 & 0.648\\ 
    red &  1.5 & 0.4 & 144 & 0.842\\ 
    \hline
    \end{tabular}
    \caption{Important properties of each \emph{unWISE} sample: $\bar{z}$, mean of redshift from $dn/dz$, see Section~\ref{sec:data}; $\delta_z$, width of the redshift distribution (as measured by matching to objects with photometric redshifts on the \emph{COSMOS} field \citep{Laigle_2016}); $\bar{n}$, the number density per deg$^2$; and $s$, the faint-end slope of the luminosity function $s = d \log_{10} N/dm$. See \cite{Schlafly19,Alex} for further details.}
    \label{table:1a}
    \end{table}

To remove stellar contamination, we apply a Galactic mask, based on the \emph{Planck} CMB lensing mask, which also masks point sources, bright stars (cross-matched with \emph{CatWISE} catalog~\cite{CatWISE_paper}), galaxies (\emph{LSLGA}\footnote{https://github.com/moustakas/LSLGA.} catalog), and planetary nebulae, as described in \cite{Alex}. The \emph{Planck} lensing mask is apodized, while the stellar, large galaxy, planetary nebulae and area-lost masks for the \emph{unWISE} galaxy map are not, which leaves a sky fraction of $f_{\rm sky} = 0.586$.

\section{Analysis}
\label{sec:analysis}

\begin{figure*}[htbp!]
    \includegraphics[width=0.45\textwidth]{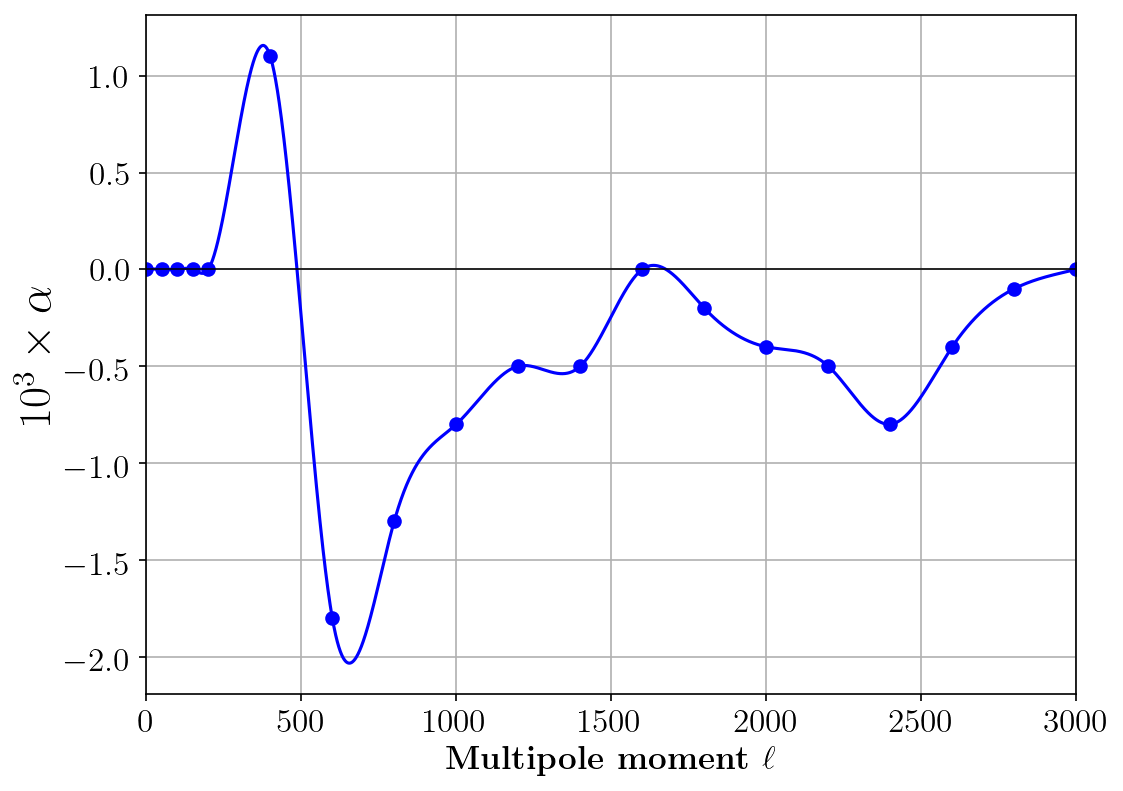}
    \includegraphics[width=0.45\textwidth]{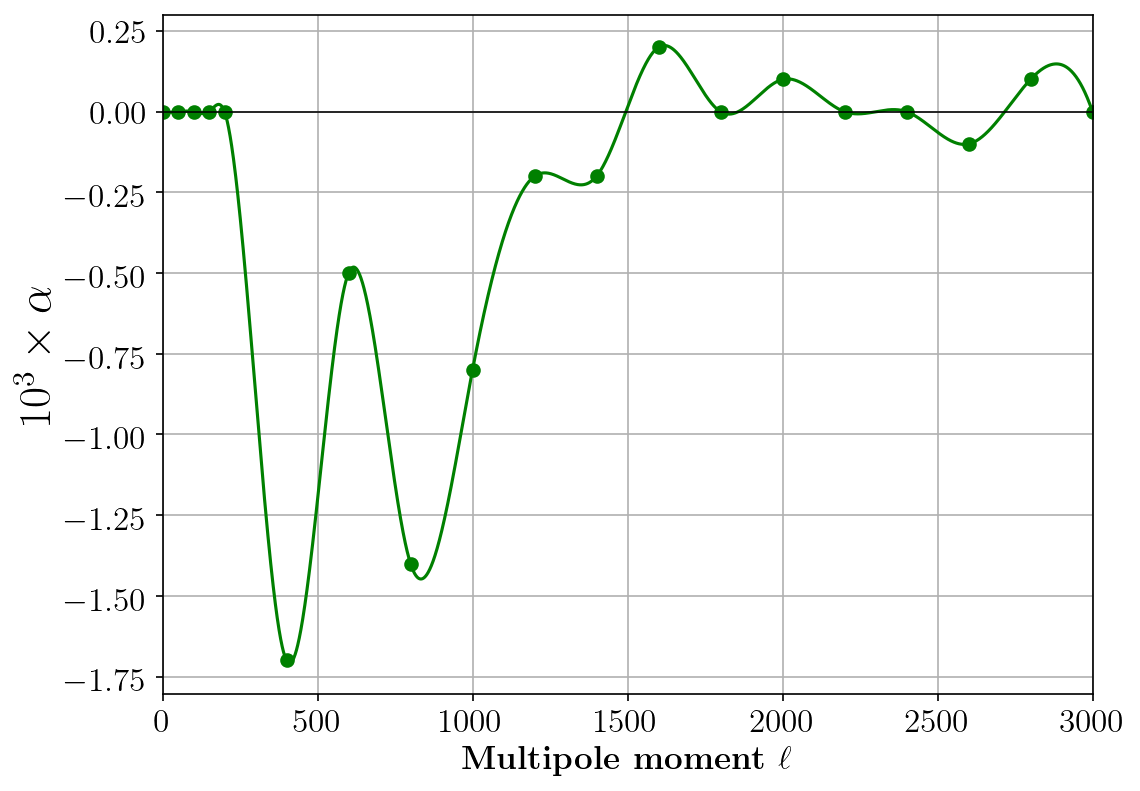}
    \includegraphics[width=0.45\textwidth]{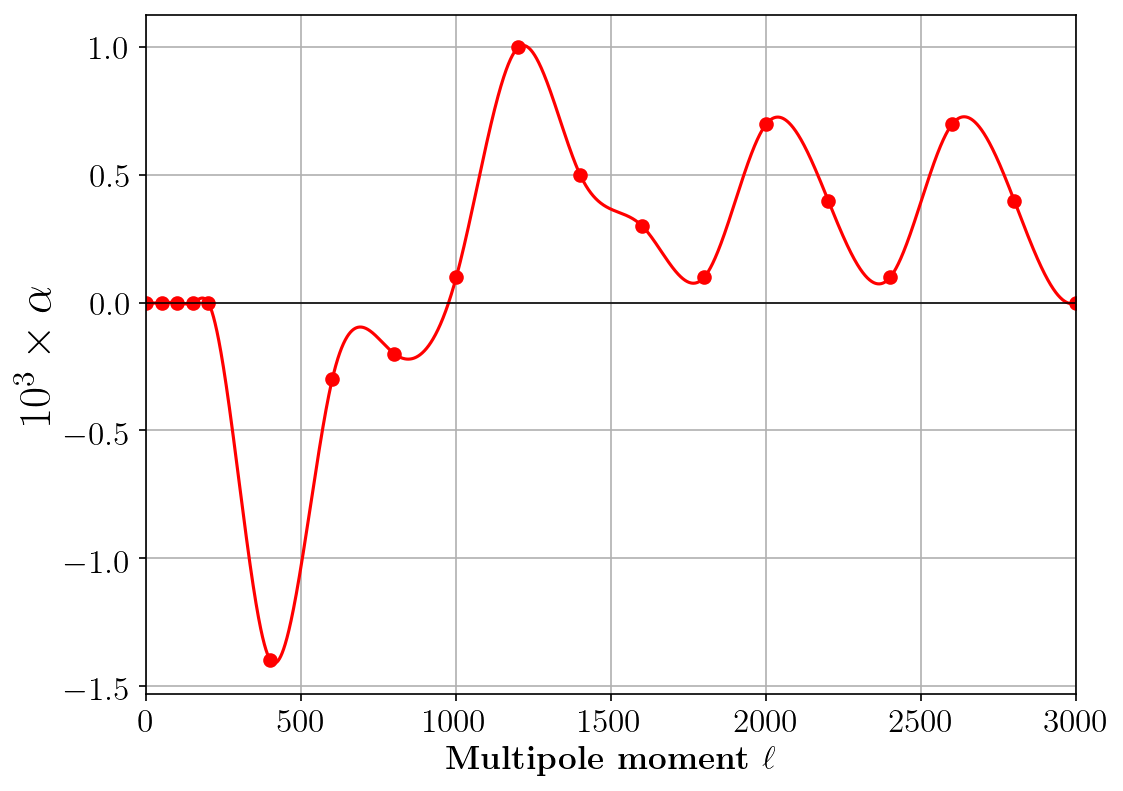}
    \caption{Multipole-dependent $\alpha$ values that null the cross-correlations of $\delta_{g}$ and $((1 + \alpha)T_{\rm LGMCA} - \alpha T_{\rm dust})$ for each of the \emph{unWISE} maps. The $\alpha$ values are found for each of the 13 multipole bins, and then interpolated (using a \emph{CubicSpline}) over the entire $\ell$ range used in the analysis.} 
    \label{fig_alphas_interpolated}
\end{figure*}

\subsection{Foreground cleaning}
\label{subsec:cleaning}

Our kSZ${}^2$-LSS estimator can potentially be biased by contributions from residual foreground contamination in the cleaned CMB temperature map.  As described above, we employ tSZ-deprojected CMB maps (in at least one leg of the estimator), which allows us to robustly avoid any bias from tSZ signal associated with the \emph{unWISE} galaxies that could have leaked through the component separation process.  However, residual thermal dust emission that is present in the CMB maps at the locations of the \emph{unWISE} galaxies could potentially affect our analysis, since the dust SED is not perfectly known a priori (unlike tSZ), and thus it cannot be exactly removed.  Note that our analysis is not biased by residual dust emission in the CMB map that is uncorrelated with the \emph{unWISE} objects (e.g., from the Milky Way or from galaxies beyond the redshift range of the \emph{unWISE} samples), but only with residual emission that would be detected by our cross-correlation method.

In order to remove possible \emph{unWISE}-correlated dust contamination from the LGMCA CMB map that might not have been removed in the original foreground cleaning~\citep{Bobin_2013, Bobin_2014}, we use a cleaning technique similar to the one from \citetalias{Hill2016}.  We construct a linear combination of the filtered $T_{\rm LGMCA}$ map and a filtered dust map $T_{\rm dust}$ such that it has zero cross-correlation with the \emph{unWISE} galaxies, i.e., such that any dust associated with the \emph{unWISE} galaxies is removed.  In our analysis, the dust removal procedure is performed bin-by-bin in multipole space.  Note that~\citetalias{Hill2016} employed a scale-independent cleaning procedure, which we also consider in Appendix~\ref{sec:append:const_alpha}, finding very similar results to those obtained in our scale-dependent approach.  Our procedure, which we will often refer to as ``$\alpha$-cleaning'', works as follows: for each of the \emph{unWISE} subsamples, we look for a value $\alpha$ that nulls the cross-correlation of the \emph{unWISE} galaxy overdensity $\delta_{g}$ and the linear-combination map $((1 + \alpha)T_{\rm LGMCA} - \alpha T_{\rm dust})$, where $T_{\rm dust}$ is a map dominated by dust emission (note that our linear combination ensures that even if the dust map contains a small amount of CMB signal, the linear-combination map remains an unbiased CMB map).  Our fiducial dust map is the \emph{Planck} 545 GHz map.\footnote{We convert the 545 and 857 GHz maps from MJy/sr to CMB temperature units using the conversion factors provided by \emph{Planck}~\cite{2014A&A...571A...9P}.}  We perform this operation independently for each of the 13 linearly-spaced multipole bins from $\ell = 300$ to $\ell=2900$ with width $\Delta\ell=200$; we look for an $\ell$-dependent $\alpha_{\rm min}$ that gives exactly null when cross-correlating $\delta_{g}$ and $((1 + \alpha)T_{\rm LGMCA} - \alpha T_{\rm dust})$.  After determining the $\alpha_{\rm min}$ values for the central $\ell$ value in each bin, we interpolate them over the entire range of multipoles (using the \emph{CubicSpline} routine).  This gives us effectively a smooth $\alpha_{\rm min}$-filter, which we use to construct
\begin{equation}
T_{\rm clean} = (1 + \alpha_{\rm min})T_{\rm LGMCA} - \alpha_{\rm min} T_{\rm dust} \,,
\label{eq.Tclean}
\end{equation}
where $\alpha_{\rm min}$ is $\ell$-dependent and the operations are performed in harmonic space. The results of this procedure (the $\alpha_{\rm min}$ values and the interpolating function) are shown in Fig.~\ref{fig_alphas_interpolated}. 

Even after performing the $\alpha$-based dust cleaning, we want to ensure that our detected $C_{\ell}^{{\rm kSZ}^2 \times \delta_{g}}$ signal is not due to other effects.  When a product of two CMB maps is cross-correlated with a galaxy map, the possible sources of non-zero measurements are: kSZ (our intended signal); tSZ, which we eliminate by always using the tSZ-nulled LGMCA map (or SMICA-noSZ) map in at least one leg of our estimator, such that tSZ will not be picked up in the cross-correlation $(T_{\rm LGMCA} T_{\rm SMICA}) \times \delta_g^{unWISE}$; CMB lensing contributions, which we account for in our theoretical prediction (see Sections \ref{sec:theory} and \ref{sec:interpretation}); radio source contamination, which has been tested in \citetalias{Hill2016} to be fully negligible; the dust contribution whose mitigation is described above; and the ISW effect, which we avoid by removing all modes at $\ell < 100$ with the filter shown in Fig.~\ref{Filter}.

\begin{figure*}[t]
    \centering
    \includegraphics[width=0.45\textwidth]{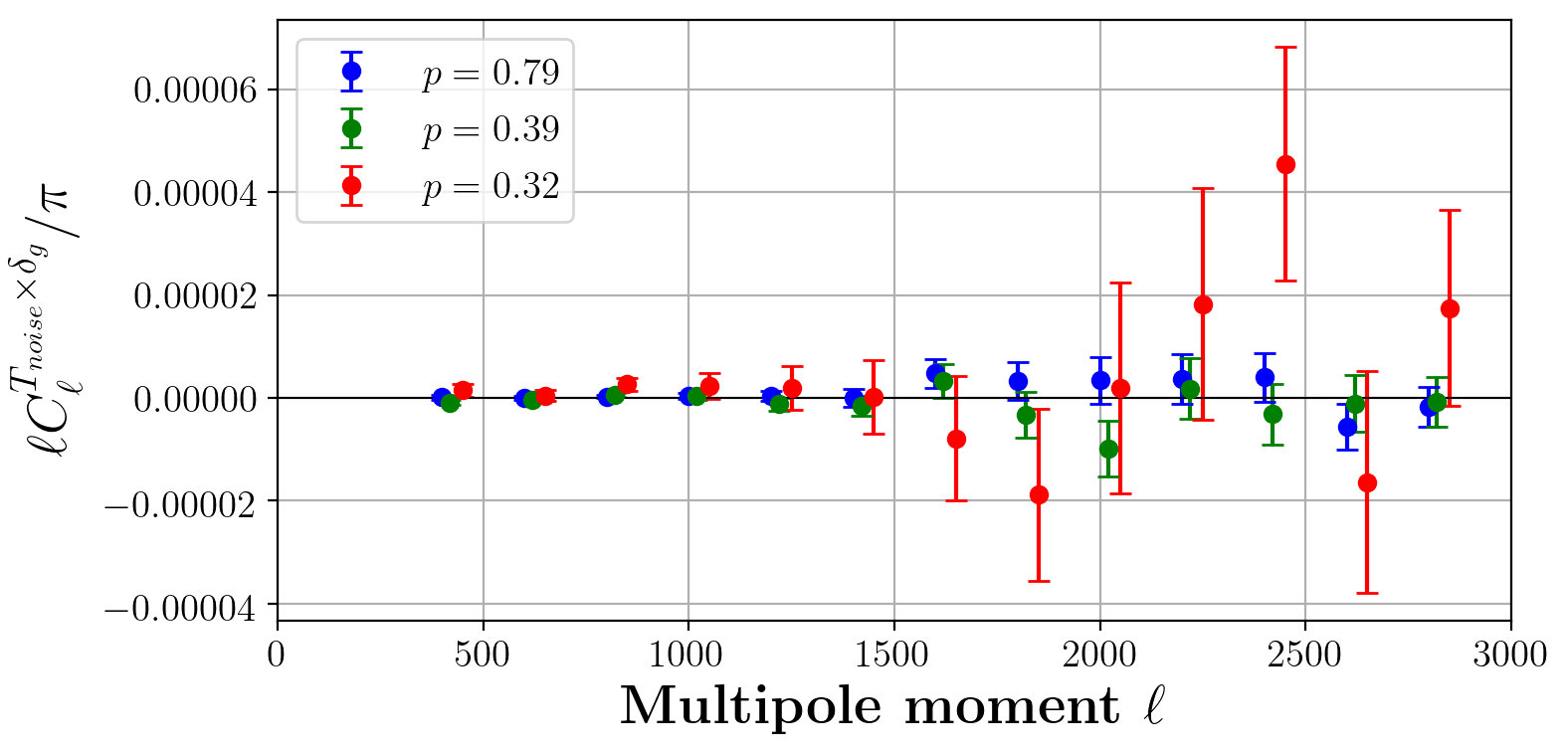}
    \includegraphics[width=0.45\textwidth]{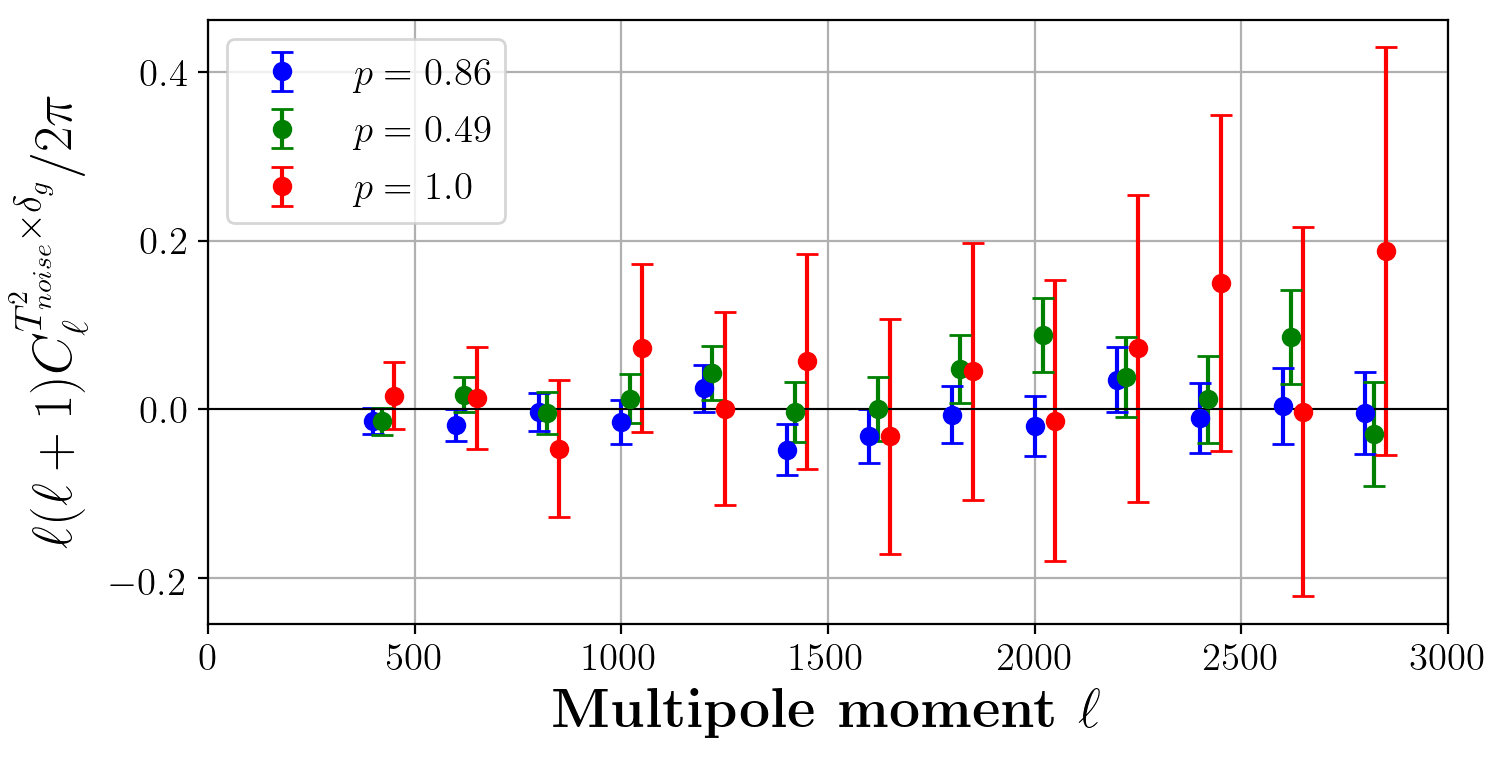}
    \includegraphics[width=0.45\textwidth]{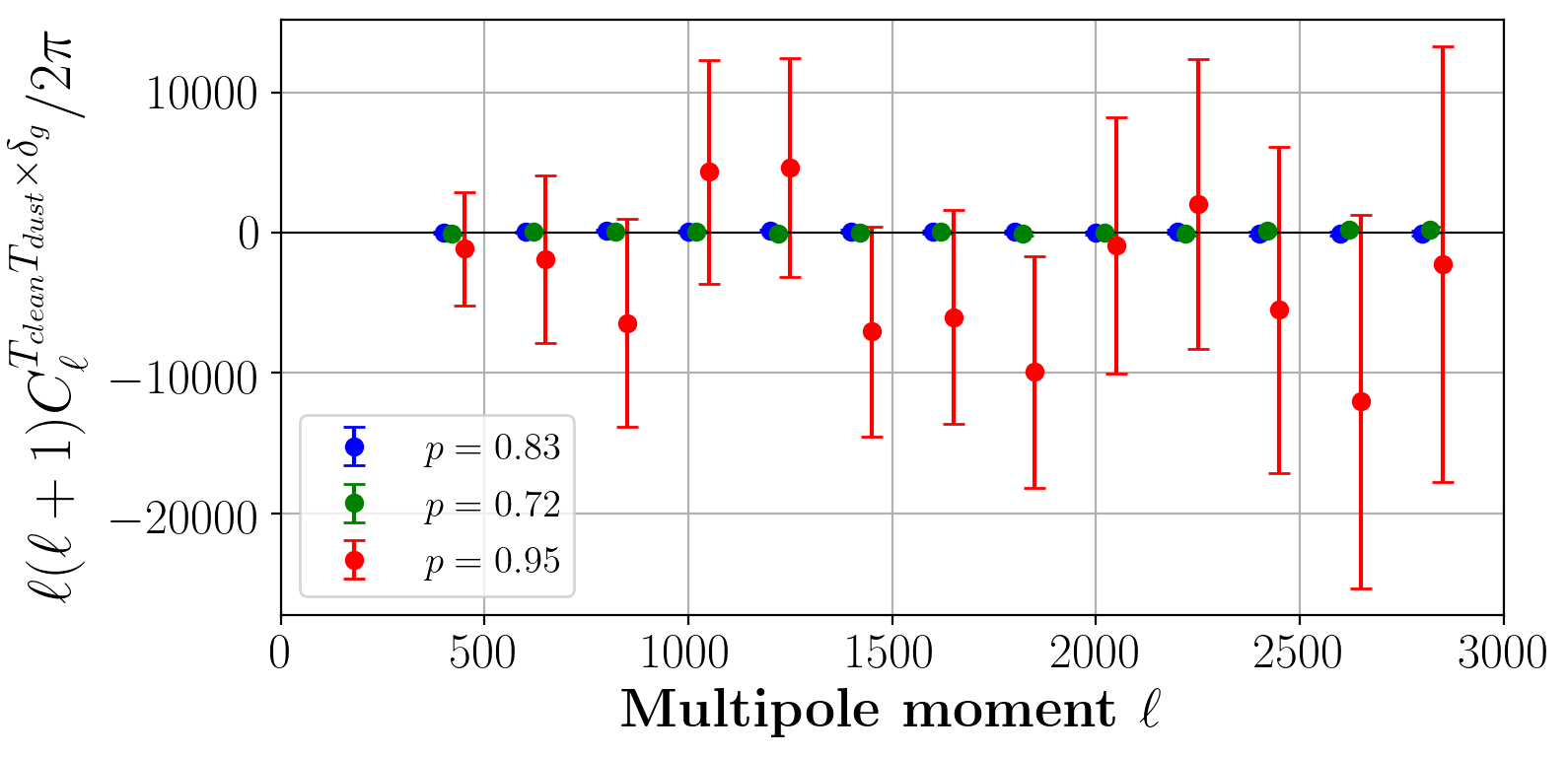}
    \includegraphics[width=0.45\textwidth]{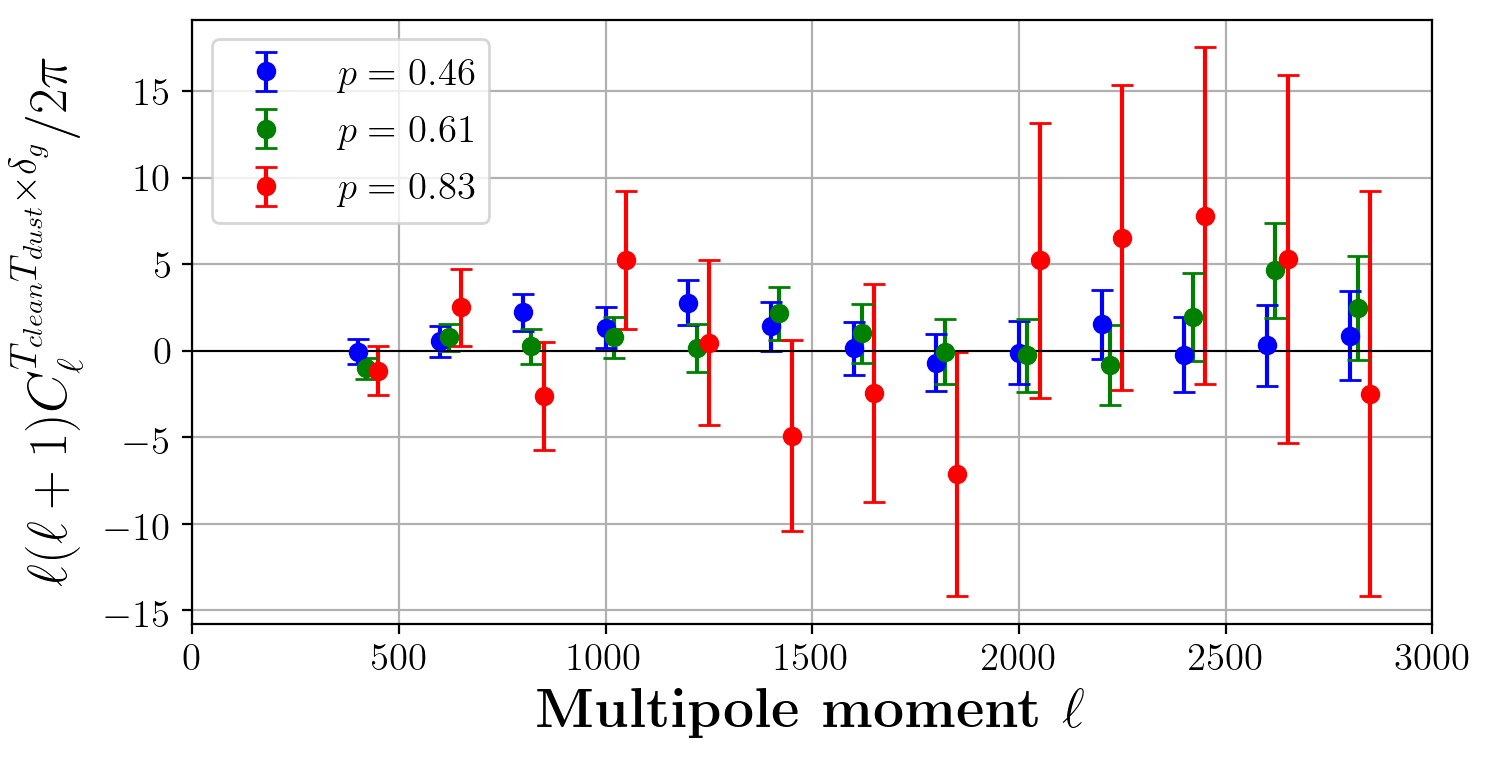}
    \caption{Null tests on the LGMCA maps (noise maps in top panels and the $\alpha$-cleaned maps used in our main data analysis in the bottom panels). All are color coded based on the \emph{unWISE} colors: blue, red, and green, and include probabilities-to-exceed from fitting them to null. The green points are offset by $\ell=20$, and the red by $\ell=50$ with respect to the true multipole moment values, for visual purposes. Top left: Cross-correlation between $T_{\rm noise}$ and unWISE, where $T_{\rm noise}$ is the LGMCA$_{\rm noise}$ map. Top right: Cross-correlation between $T_{\rm noise}^2$ and unWISE. Bottom left: Cross-correlation of $(T_{\rm clean}T_{\rm dust})$ and \emph{unWISE}, where $T_{\rm dust}$ is the \emph{Planck} 857 GHz map. Bottom right: Cross-correlation of $(T_{\rm clean}T_{\rm dust})$ and \emph{unWISE}, where $T_{\rm dust}$ is the \emph{Planck} 545 GHz map.
    All tests are consistent with null, which provides robust evidence that our $\alpha$-cleaned LGMCA map contains negligible thermal dust contamination from the \emph{unWISE} galaxies.} 
    \label{fig_tests}
\end{figure*}

Although we implement the explicit cleaning procedure described above, we apply several tests to ensure the results are not artificial or due to dust contamination, and that our $\alpha$-cleaning was indeed effective. Firstly, we consider cross-correlating the LGMCA noise map ($T_{\rm noise}$) with the \emph{unWISE} $\delta_{g}$ maps, where one would expect null (Fig.~\ref{fig_tests} top left). A similar test is done with $T_{\rm noise}^2$ (Fig.~\ref{fig_tests} top right). Both are consistent with null. 

The next test we perform is the cross-correlation $(T_{\rm clean}T_{\rm dust}) \times \delta_{g}$, which would yield a strong signal if $T_{\rm clean}$ contained even a small amount of \emph{unWISE}-correlated dust emission, due to the brightness of thermal dust at high frequencies~\citepalias{Hill2016}. This particular test is performed twice with two different dust maps: \emph{Planck} 545 GHz and \emph{Planck} 857 GHz, where both maps are heavily dust-dominated.  In~\citetalias{Hill2016}, this dust null test was the closest to failing, with probability-to-exceed $p=0.02$ (for 13 degrees of freedom) when fitted to null, although this was dominated by a single data point \citepalias{Hill2016}.  Thus it is of interest to see whether our analysis using larger galaxy samples (and hence higher S/N) may see stronger evidence of a failure of this null test, if indeed the $\alpha$-cleaned LGMCA map does contain some residual dust.  However, we find that this test is fully consistent with null.  Using \emph{Planck} 545 GHz as $T_{\rm dust}$, we obtain $p= $ 0.46, 0.61, and 0.83 for \emph{unWISE} blue, green, and red, respectively, for 13 degrees of freedom, when fitting to null.  This provides very strong evidence against the presence of dust contamination in our kSZ measurements.  The tests (with probabilities-to-exceed $p$) are summarized in Table \ref{table_tests_pvalues}, and shown in Fig. \ref{fig_tests}.

An additional natural test to perform might seem to be $T_{\rm clean} \times \delta_{g}$.  However, since $T_{\rm clean}$ is defined as the linear combination of $T_{\rm LGMCA}$ and $T_{\rm dust}$ that yields zero when cross-correlated with $\delta_g$ (see Eq.~\ref{eq.Tclean}), this test is meaningless as it is guaranteed to pass.  Put more simply, the construction of $T_{\rm clean}$ already removes any mean dust residual associated with the \emph{unWISE} galaxies; the $(T_{\rm clean}T_{\rm dust}) \times \delta_{g}$ test is meaningful and informative because it is sensitive to any potential contribution arising from fluctuations in the dust residuals around the zero mean level.

Even though all of our tests show that the dust present in the LGMCA map is effectively negligible, we still want to assess its approximate level, if possible.  We do this by measuring $T_{\rm clean}^2 \times \delta_g$ for the \emph{unWISE} samples and then rescaling the measurements to the primary CMB frequencies using a fiducial thermal dust SED.  This procedure is described in detail in Appendix \ref{sec:append:dust}, and further demonstrates that we can safely assume the dust is not a source of the measured signal in our kSZ analysis.

Moreover, we also ensure that our results are not contaminated by Galactic dust, which could in principle affect the \emph{unWISE} selection function, and is also present in the mm-wave sky via its thermal emission. Firstly, we note that the CMB maps are filtered to remove the low multipoles (the Wiener filter peaks at $\ell \approx 2000$), where Galactic dust is most important. However, since this is not a stringent test of a potential Galactic dust bias, we re-run our kSZ$^2$-galaxy pipeline (described in detail below in Section~\ref{subsec:kSZ_measure}) with a \emph{Planck} mask with $f_{sky}=0.4$0 combined with the original \emph{unWISE} mask ($f_{sky}=60$). The results are consistent within the statistical error bars with our fiducial measurements, which indicates that Galactic dust contamination in our measurement is statistically negligible.

        \begin{table}[h!]
        \begin{tabular}{ |c|c|c|c|c|c|c| } 
        \hline
         \emph{unWISE} $\times$ & $T_{noise}$ & $T_{noise}^2$ & $T_{\rm clean}T_{\rm dust, 857}$ & $T_{\rm clean}T_{\rm dust, 545}$ \\ 
        \hline\hline
        blue & 0.79 & 0.86 & 0.83 & 0.46\\
        green & 0.39 & 0.49 & 0.72 & 0.61\\ 
        red & 0.32 & 1.0 & 0.95 & 0.83 \\ 
        \hline
        \end{tabular}
        \caption{Testing the dust contamination in the $\alpha$-cleaned LGMCA map. Probability-to-exceed $p$ (for 13 degrees of freedom) for fitting the cross-correlation between the \emph{unWISE} galaxy maps (blue, green, and red) and the CMB temperature maps ($T_{noise}$, $T_{noise}^2$, and $T_{\rm clean}T_{\rm dust}$) to null. $T_{\rm clean}$ is the LGMCA$_{\rm clean}$ map, $T_{\rm noise}$ is the LGMCA$_{\rm noise}$ map, and $T_{\rm dust}$ is either the \emph{Planck} 857 GHz or \emph{Planck} 545 GHz map.}
        \label{table_tests_pvalues}
        \end{table}

\subsection{kSZ measurement}
\label{subsec:kSZ_measure}

\begin{figure*}[htbp!]
    \centering
    \includegraphics[width=0.45\textwidth]{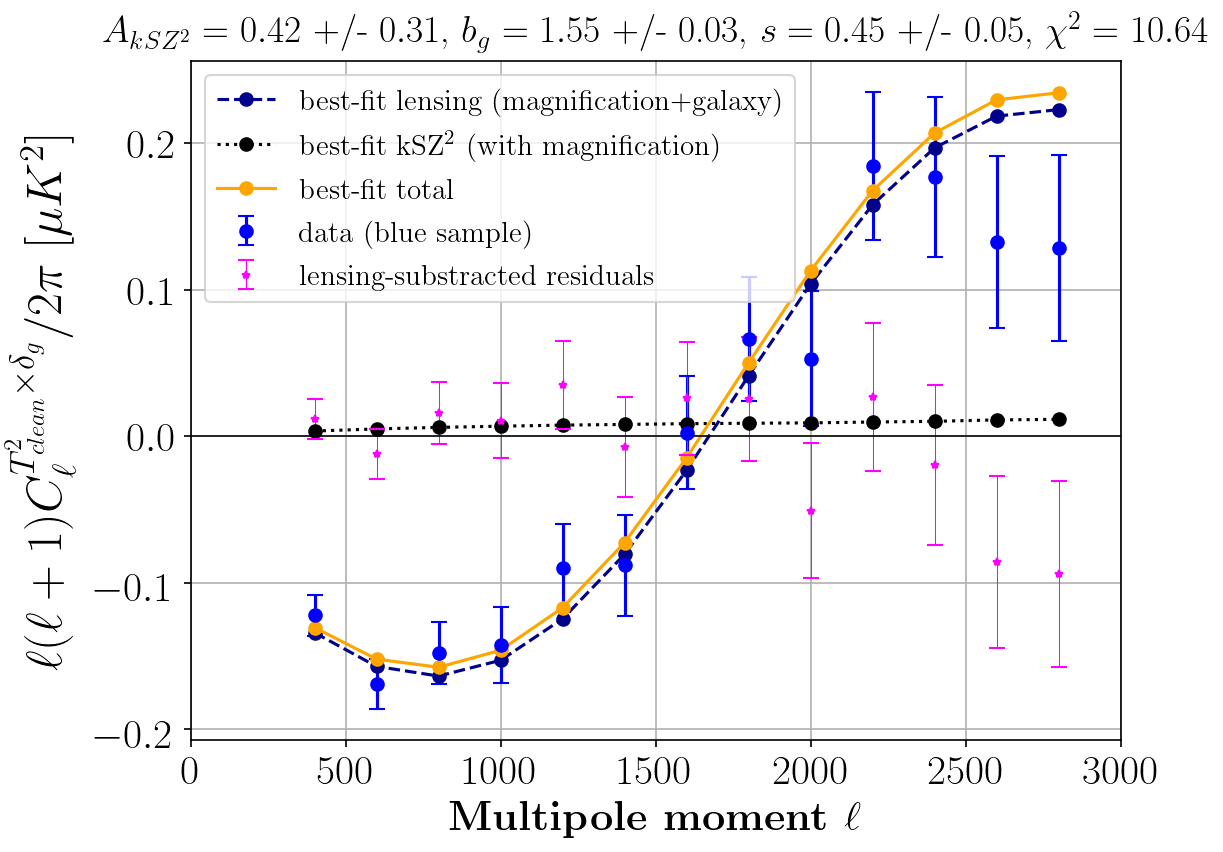}
    \includegraphics[width=0.45\textwidth]{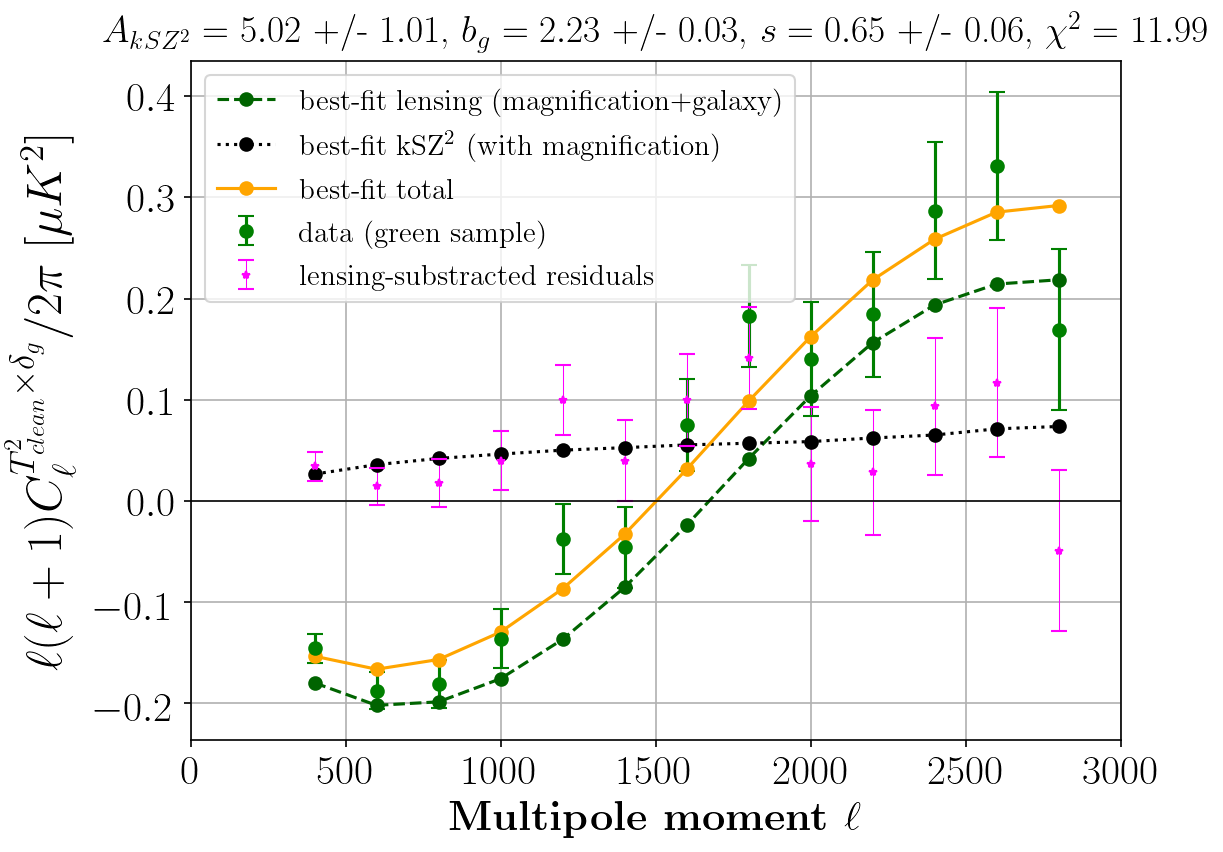}
    \includegraphics[width=0.45\textwidth]{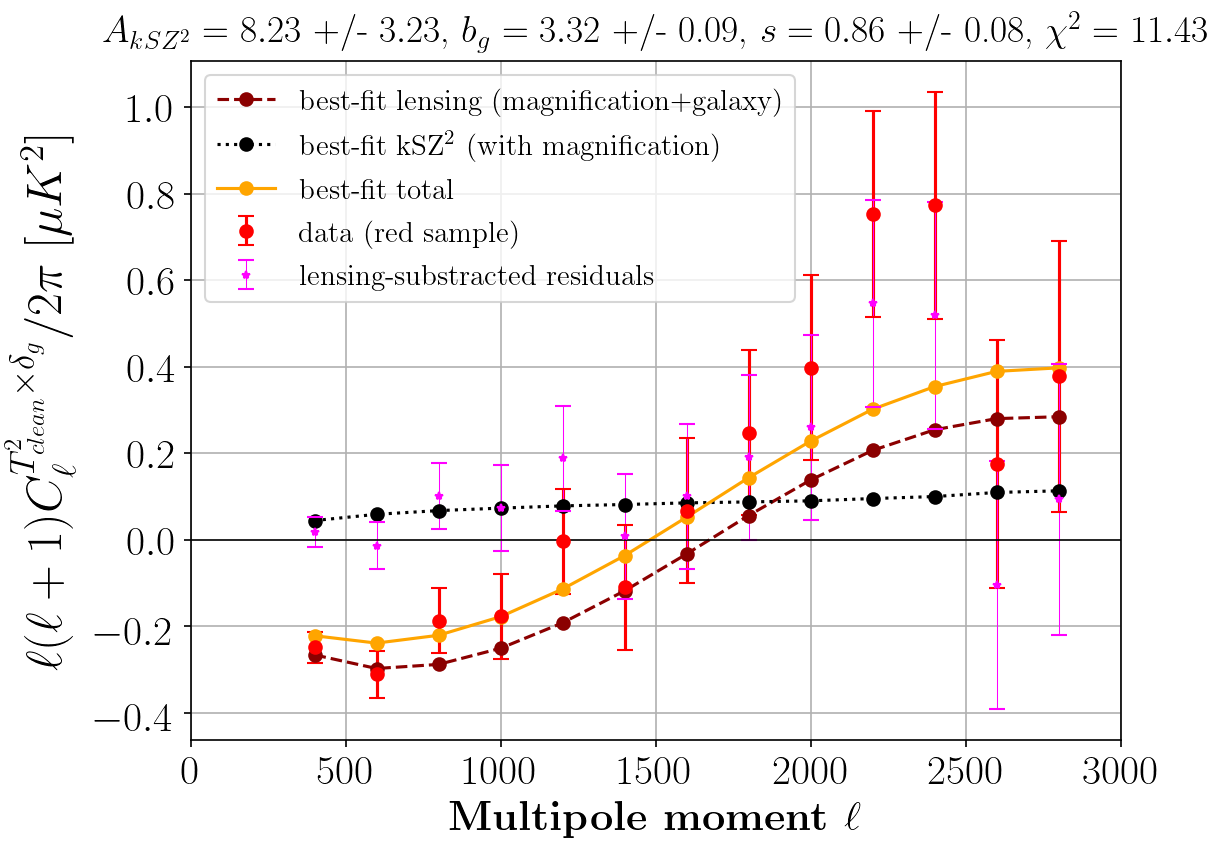}
    \caption{Cross-power spectra of the real-space product of the cleaned, filtered LGMCA map and the filtered SMICA map with each of the \emph{unWISE} galaxy density maps: blue, green, and red (data points in respective colors). The thin dashed curves show the best-fit CMB lensing contribution (including the lensing-galaxy and lensing-magnification bias terms), the black dotted shows the best-fit kSZ contribution (including the kSZ-galaxy and kSZ-magnification bias terms), and the pink stars show the lensing-subtracted residuals for illustration. The yellow solid curves in each plot show the total best-fit curves, which are the sum of best-fit lensing and best-fit kSZ contributions. The best-fit values for each of the free parameters in the theory model (the kSZ$^2$ amplitude $A_{{\rm kSZ}^2}$, the galaxy bias $b_g$, and the magnification response $s$) are presented in the plot titles. Our fiducial model assumes $A_{{\rm kSZ}^2} = 1$. The kSZ signal is detected at 1.4$\sigma$, 5.0$\sigma$, and 2.5$\sigma$ significance, respectively, for the three \emph{unWISE} subsamples. } 
    \label{fig:1}
\end{figure*}

Finally, after ensuring that our cleaned CMB map is not contaminated by dust, we can apply the $C_{\ell}^{{\rm kSZ}^2 \times \delta_{g}}$ estimator to the product of LGMCA$_{\rm clean}$ and SMICA, and the \emph{unWISE} $\delta_{g}$, which provides higher S/N than LGMCA$_{\rm clean}^2 \times \delta_g$ while remaining robust to foregrounds.  We multiply the filtered, $\alpha$-cleaned LGMCA map with the filtered SMICA map in real space and cross-correlate the product with each of the masked \emph{unWISE} galaxy maps, using the \texttt{Namaster} software package~\cite{namaster,Namaster_github}. The results are shown in Fig.~\ref{fig:1}, where we bin the cross-power spectrum signal into 13 multipole bins from $\ell = 300$ to $\ell=2900$ with width $\Delta\ell=200$. The covariance matrices of the measurements are estimated in the Gaussian approximation, again using the \texttt{Namaster} code.  As expected for a harmonic-space measurement with wide multipole bins, the covariance matrices are close to diagonal, but we use the full covariance matrices in the likelihood analysis in Section~\ref{sec:interpretation}.  The error bars shown in all plots are the square root of the diagonal elements of the covariance matrices.  We describe the interpretation of these measurements in the next section.

As a robustness check, we repeat the analysis of the original \emph{WISE} catalog from \citetalias{Hill2016} using our pipeline (and using the sky mask from that work).  We reproduce the results of \citetalias{Hill2016} essentially perfectly, including both the central values and the uncertainties (we find a $\lesssim 10$\% difference in the error bar values, which arises from the different cleaning methods used).

\section{Theoretical interpretation}
\label{sec:interpretation}

We fit the measured cross-power spectra using the model described in Section \ref{sec:theory}. The model is a sum of the ${\rm kSZ}^2$ signal, $C_{\ell}^{{\rm kSZ}^2 \times \delta_g}$; kSZ magnification bias contribution, $C_{\ell}^{{\rm kSZ}^2 \times \mu_g}$; the CMB gravitational-weak lensing contribution, $C_{\ell}^{T^2 \times \delta_g}$; and the CMB lensing—magnification bias term, $C_{\ell}^{T^2 \times \mu_g}$. We allow a free amplitude for each term: $A_{{\rm kSZ}^2}b_g$ for the kSZ$^2$ term, $A_{{\rm kSZ}^2}(5s-2)$ for the kSZ magnification bias, $b_g$ for the CMB gravitational lensing, and $(5s-2)$ for the CMB lensing-magnification bias, where  $A_{{\rm kSZ}^2}$ is the amplitude of the kSZ$^2$ signal, $b_g$  is the galaxy bias (an overall amplitude of the $\delta_g$ field in Eq. \ref{eq.delta_g}), and $s$ is the magnification response. The total model is then:

\begin{equation}
\begin{multlined}
C_{\ell}^{T^2 \times \delta_g} = A_{{\rm kSZ}^2} b_g C_{\ell}^{{\rm kSZ}^2 \times \delta_g} 
+ A_{{\rm kSZ}^2}(5s-2) C_{\ell}^{{\rm kSZ}^2 \times \mu_g} \\ + b_g \Delta C_{\ell}^{T^2 \times \delta_g} + (5s-2) \Delta C_{\ell}^{T^2 \times \mu_g}
\end{multlined}
\label{eq.model_tot}
\end{equation}

\begin{figure*}[htbp!]
    \centering
    \includegraphics[width=0.45\textwidth]{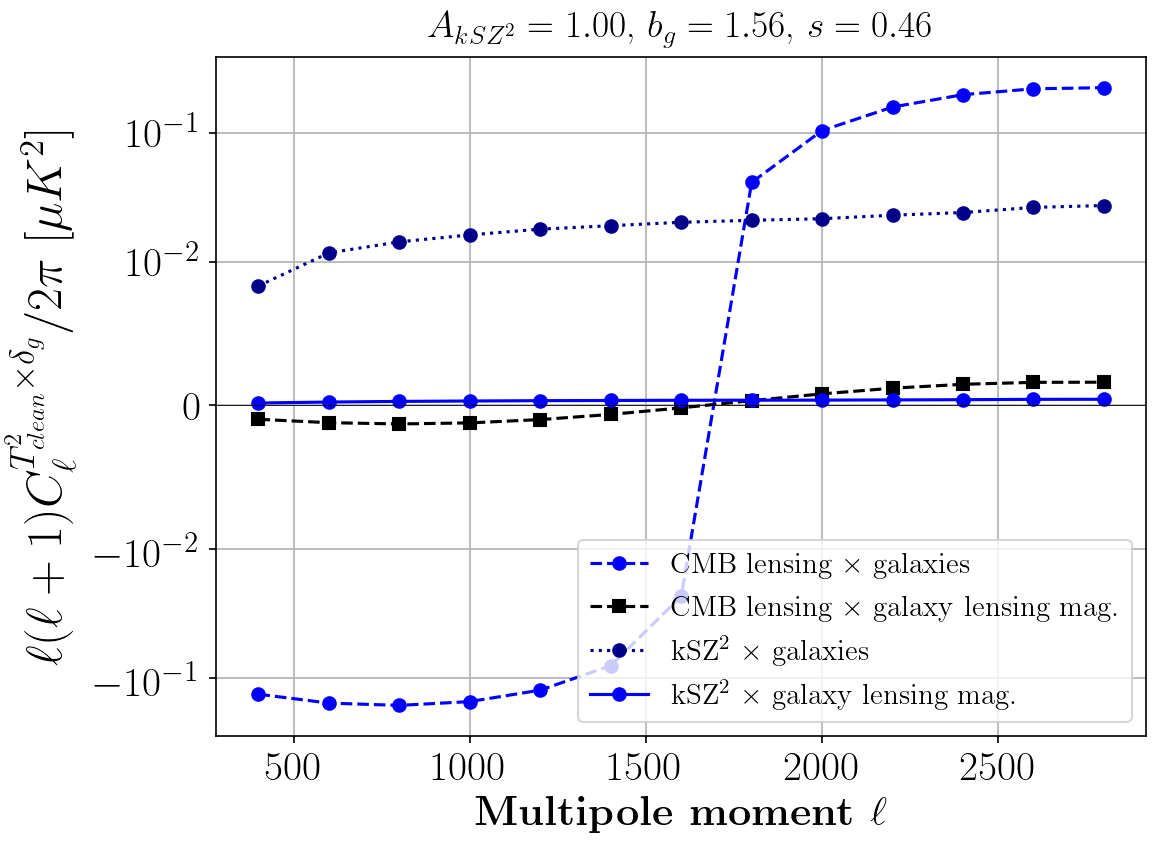}
    \includegraphics[width=0.45\textwidth]{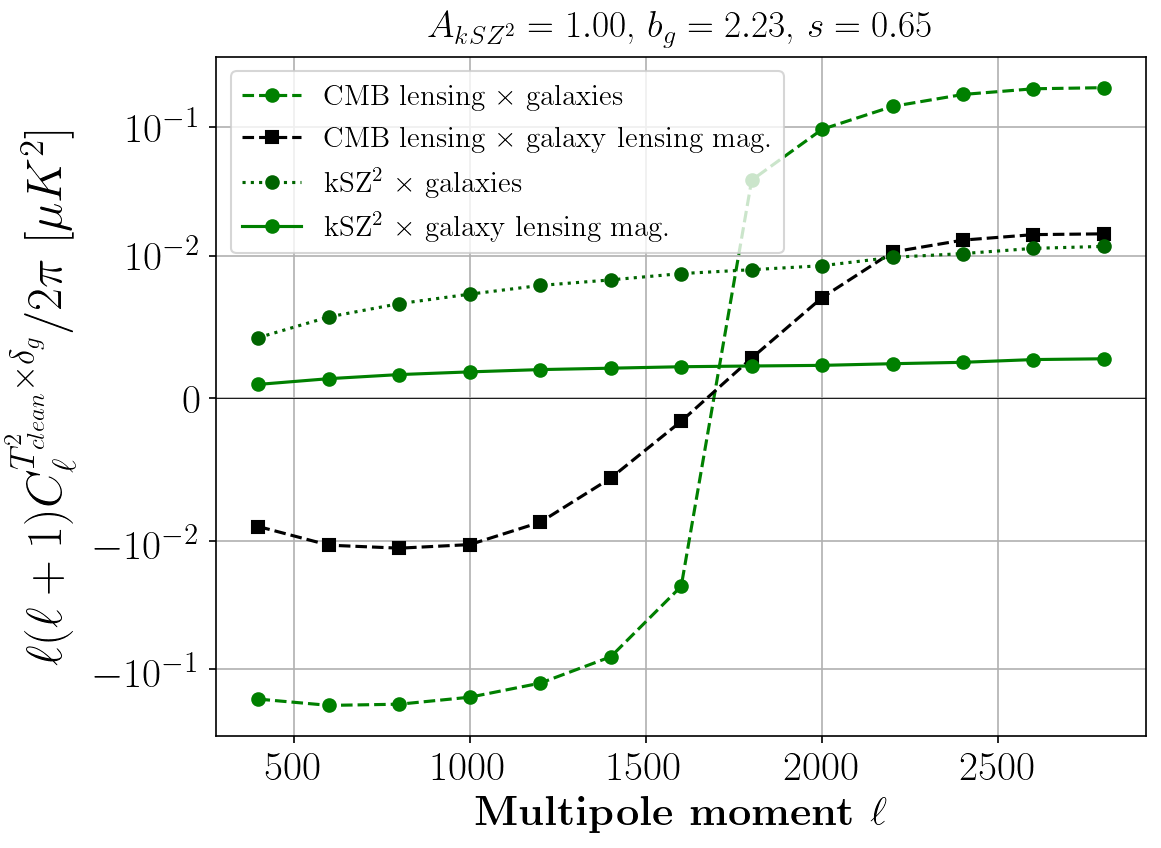}
    \includegraphics[width=0.45\textwidth]{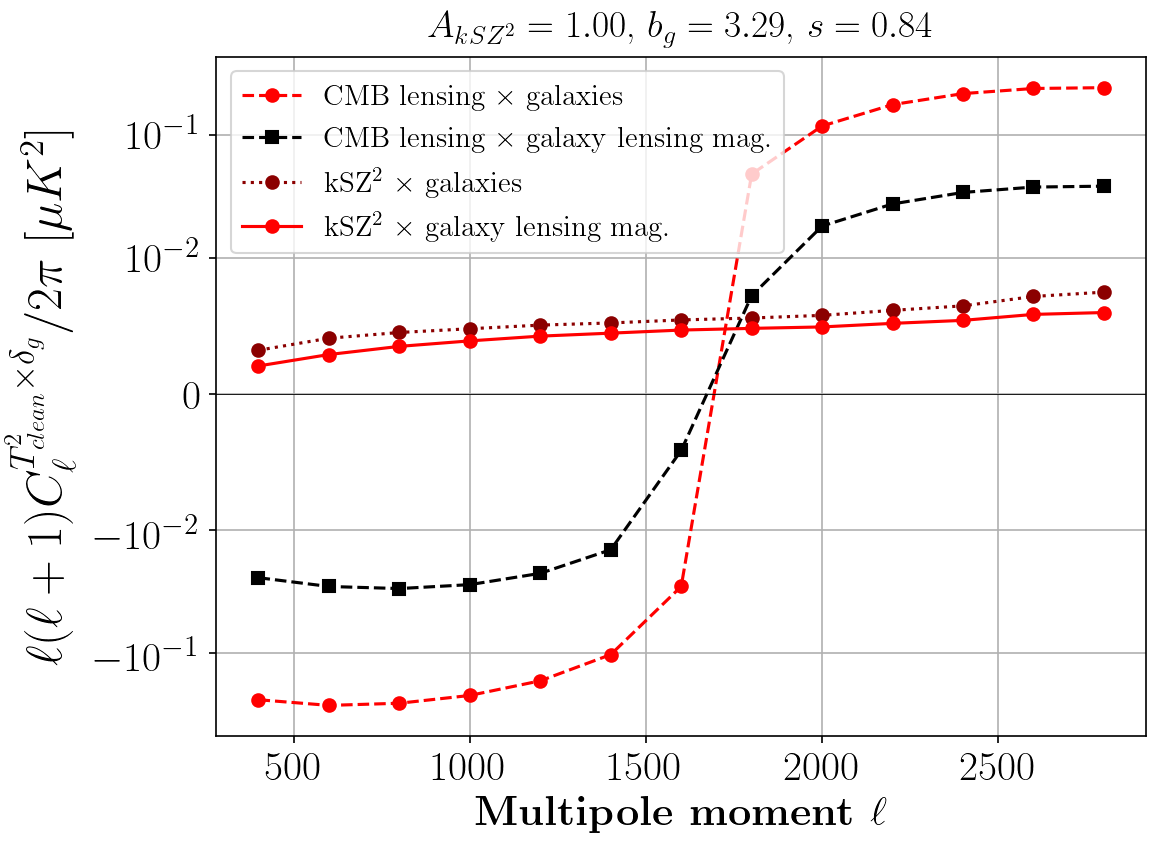}
    \caption{Fiducial model prediction for each of the \emph{unWISE} samples (color-coded). Each contribution (${\rm kSZ}^2$-galaxy cross-correlation, kSZ${}^2$-magnification bias cross-correlation, CMB lensing-galaxy cross-correlation, and CMB lensing-magnification bias cross-correlation) is plotted separately. The fiducial model assumes the kSZ$^2$ amplitude $A_{{\rm kSZ}^2}$ = 1, and the values for the galaxy bias $b_g$ and the magnification response $s$ to be equal to the the values measured in Ref.~\cite{Alex}. These values are given in the plot titles for ease of reference.  The curves are slightly non-smooth because we apply the same binning used in the data analysis to the theory curves here.  Note the logarithmic scale on the vertical axis, as compared to the linear scale in Fig.~\ref{fig:1}.}
    \label{Fiducial}
\end{figure*}

The fiducial model (where the cosmic baryon abundance is taken to be $\Omega_b / \Omega_m = 0.158$) assumes that the kSZ$^2$ amplitude is equal to unity, $A_{{\rm kSZ}^2} = 1$, and is shown in Fig. \ref{Fiducial}. In these plots, we assume the values derived in \cite{Alex} for the galaxy bias $b_g$ and for the magnification response $s$. They are summarized in Table \ref{param_values}.  We note that our model includes the first calculation of the kSZ${}^2$ $\times$ magnification bias contribution and the CMB lensing $\times$ magnification bias contribution to the $T^2 \times \delta_g$ estimator (these terms were neglected in~\citetalias{Hill2016} and~\citetalias{Ferraro2016}).  For the blue sample, which is most similar to the original \emph{WISE} sample used in~\citetalias{Hill2016} and~\citetalias{Ferraro2016}, these terms are negligible, which indicates that the results in those papers were not biased by neglecting these terms.  However, for the green and red samples, the magnification bias terms are non-negligible.  For the red sample in particular, the fiducial kSZ${}^2$ $\times$ magnification bias term is nearly equal to the fiducial kSZ${}^2$ $\times$ galaxy term.

\begin{table}[h!]
    \centering
    \begin{tabular}{|c|c|c|c|c|}
    \hline
         \emph{unWISE} & $b_g$ & $\sigma_{b_g}$ & $s$ & $\sigma_{s}$ \\
        \hline\hline
        blue & 1.56 & 0.0276  & 0.455 & 0.046\\
        green & 2.23 & 0.0352 & 0.648 &  0.065\\
        red & 3.29 & 0.0352 & 0.842  & 0.084\\
    \hline
    \end{tabular}
    \caption{Prior values for the galaxy bias $b_g$ and the magnification response $s$ for each of the \emph{unWISE} samples, taken from \cite{Alex}. We take $\sigma_{s}$ to be $0.1s$. }
    \label{param_values}
\end{table}

To fit the (LGMCA$_{\rm clean}\cdot$SMICA) $\times$ \emph{unWISE} measurements using the theory model in Eq.~\ref{eq.model_tot}, we assume a multivariate Gaussian likelihood for the data with three free parameters, $A_{{\rm kSZ}^2}$, $b_g$, and $s$. We adopt a flat prior on $A_{{\rm kSZ}^2} > 0$, and Gaussian priors on $b_g$ and $s$, so that $b_g$ is within $1\sigma$ from the values in Table \ref{param_values} derived in \cite{Alex},\footnote{Specifically, we use the $b_g$ results obtained in~\cite{Alex} from the \emph{unWISE} -- CMB lensing cross-correlation (rather than the \emph{unWISE} auto-correlation), since this is precisely the same quantity that appears in our measurement.} and the prior on $s$ conservatively has 10\% width (since the $s$ values are well-determined in Ref.~\citep{Alex}). The fit is done using the Python Monte Carlo Markov Chain (MCMC) package \texttt{emcee} \cite{emcee_paper}. The full posterior distributions for the parameters are shown in Appendix \ref{app:posteriors}, where we also present an analysis without imposing external priors on $b_g$, finding consistent results (with slightly larger error bars).

The best-fit values for $A_{{\rm kSZ}^2}$, $b_g$, and $s$ obtained from the MCMC procedure, along with the best-fit $\chi^2$ values for each \emph{unWISE} sample are shown in Table \ref{best-fit}, and the best-fit theory curves are plotted as yellow solid curves in Fig. \ref{fig:1}. The best-fit CMB lensing contribution (the sum of the cross-correlation with both galaxy overdensity and magnification bias) is plotted in dashed \emph{unWISE}-color-coded curves, and the best-fit kSZ$^2$ terms (with the magnification bias term included) are plotted in black dotted curves.  Overall we find that the theory model provides a good fit to the data, but with an anomalously high kSZ amplitude for the green subsample.  These results are interpreted and discussed further in Section \ref{sec:discussion}.

\begin{table}[h!]
    \centering
    \begin{tabular}{|c||c|c|c||c|}
    \hline
         $unWISE$ & $A_{{\rm kSZ}^2}$ & $b_g$ &  $s$ & $\chi^2$ \\
        \hline\hline
        blue & 0.42 $\pm$ 0.31 & 1.55 $\pm$ 0.03  & 0.45 $\pm$ 0.05 & 10.64\\
        \hline
        green & 5.02 $\pm$ 1.01 & 2.23 $\pm$ 0.03 &  0.65 $\pm$ 0.06 &  11.99\\
        \hline
        red & 8.23 $\pm$ 3.23 & 3.32 $\pm$ 0.09 & 0.86 $\pm$ 0.08 & 11.43\\
    \hline
    \end{tabular}
    \caption{Best-fit values for the model parameters (amplitude of the kSZ$^2$ signal $A_{{\rm kSZ}^2}$, galaxy bias $b_g$, and magnification response $s$), obtained by fitting the (LGMCA$_{\rm clean}\cdot$SMICA) $\times$ \emph{unWISE} data with the theory model in Eq.~\ref{eq.model_tot} for each \emph{unWISE} sample, along with the minimum $\chi^2$ values of the fit (for 10 degrees of freedom).} 
    \label{best-fit}
\end{table}

\section{Additional validation}
\label{sec:validation}

To validate the results for the (LGMCA$_{\rm clean}\cdot$SMICA) $\times$ \emph{unWISE} measurement and to ensure they are not artificial or due to foreground contamination, we repeat the analysis with different CMB maps.  We consider the following map combinations: (SMICA-noSZ$_{\rm clean}\cdot$SMICA) $\times$ \emph{unWISE}; LGMCA$_{\rm clean}^2$ $\times$ \emph{unWISE}; and SMICA-noSZ$_{\rm clean}^2$ $\times$ \emph{unWISE}.  In all cases, we utilize at least one map that is fully tSZ-deprojected and has been cleaned with the $\alpha$-cleaning procedure described in Sec.~\ref{sec:analysis}, which ensures robustness against both tSZ and dust foregrounds.  The results of all analyses agree with those of our main analysis presented earlier, (LGMCA$_{\rm clean}\cdot$SMICA) $\times$ \emph{unWISE}, and with each other within the uncertainties (see Figs.~\ref{smicas_all_comparison} and~\ref{smica_lgmca_comparison}).  As a further test, we also consider the cross-power spectrum of SMICA$_{\rm clean}^2$ with \emph{unWISE}, which in principle contains a non-zero bias from tSZ residuals in the SMICA map (since it is not a tSZ-deprojected map); nevertheless, we find that it gives fairly close results to those from our main analysis, indicating that any tSZ bias was already quite small, even before tSZ deprojection was applied in the LGMCA and SMICA-noSZ maps.

We briefly describe each of these analyses in the following subsections.  The associated plots can be found in Appendix~\ref{sec:append_valid_plots}.

\subsection{ (SMICA-noSZ$\cdot$SMICA) $\times$ \emph{unWISE} analysis}

Firstly, we consider an analysis using the \emph{Planck} SMICA-noSZ map as the main CMB map instead of the LGMCA map.  Like LGMCA, the SMICA-noSZ map includes tSZ deprojection in the component separation process, ensuring no tSZ residuals are present in this map. The SMICA-noSZ map is $\alpha$-cleaned, filtered, multiplied in real space with the original SMICA map, and the product is then cross-correlated with each of the \emph{unWISE} samples. The $\ell$-dependent $\alpha$-cleaning procedure for SMICA-noSZ is performed with the \emph{Planck} 545 GHz dust map.  We also apply the dust null tests described in Section~\ref{subsec:cleaning} on the clean SMICA-noSZ map with both the \emph{Planck} 545 GHz and 857 GHz dust maps (Fig. \ref{fig_tests_smicanosz}). They are both consistent with null, showing no evidence of bias. 


The (SMICA-noSZ$_{\rm clean} \cdot SMICA)$ $\times$ \emph{unWISE} measurements (Fig. \ref{fits_smicanosz_smica}) are then fit with the same theory model as in the main analysis. The results are very similar to those of our fiducial analysis in Section~\ref{sec:interpretation}, which ensures that the main results are not artificial.  Again, the theory model describes the data well, but the inferred value of $A_{{\rm kSZ}^2}$ for the green sample is anomalously high.

\subsection{LGMCA$^2 \times$ \emph{unWISE} analysis}

The analysis is also conducted as in \citetalias{Hill2016}, where we simply square the filtered, $\alpha$-cleaned LGMCA map (the same map used as the first leg of the estimator in our main analysis), and then cross-correlate it with each of the \emph{unWISE} samples.  The results are shown in Fig.~\ref{LGMCA^2} in the Appendix \ref{sec:append_valid_plots}.  They are consistent with those of our main analysis, which further validates our measurements.  Also, the comparison of these LGMCA$_{\rm clean}^2$ results with our ``asymmetric'' (LGMCA$_{\rm clean} \cdot$ SMICA) results clearly demonstrates that the asymmetric approach is not biased by dust (despite only cleaning one leg of the estimator), as expected due to its construction.

\subsection{SMICA-noSZ$^2 \times$ \emph{unWISE} analysis}

We also conduct a similar analysis for the SMICA-noSZ map.  We square the filtered, $\alpha$-cleaned SMICA-noSZ map, and then cross-correlate it with the \emph{unWISE} samples.  The results are shown in in the Appendix in Fig.~\ref{SMICAnoSZ^2}, and they are again consistent with our fiducial analysis and with those in the previous subsections.  This analysis thus also further validates our results.

\subsection{Comparison of the different combinations}

A comparison of the results of all combinations is shown in Figs.~\ref{smicas_all_comparison} (SMICA-noSZ) and \ref{smica_lgmca_comparison} (LGMCA), along with the cross-correlation of the original SMICA map (filtered, $\alpha$-cleaned, and squared) with the \emph{unWISE} data.  SMICA-noSZ or LGMCA in these plots are the clean maps (using the $\alpha$-cleaning procedure).

The original SMICA map, in contrast to SMICA-noSZ or LGMCA, includes residuals due to tSZ contamination, which can bias our estimator; this is why we do not use it as the primary map in our analysis.  However, for the same reason, the SMICA map also has lower noise than SMICA-noSZ or LGMCA, as the explicit deprojection of the tSZ signal in the latter maps leads to the loss of a degree of freedom in the variance-minimization procedures used in component separation.\footnote{A direct comparison of the noise power spectra of the SMICA and SMICA-noSZ maps can be found in Fig.~D.4 of Ref.~\cite{PlanckComponentSeparation}, showing that the former indeed has lower noise than the latter.}  This is why it is advantageous for us to build the asymmetric combination (LGMCA$_{\rm clean}\cdot$SMICA) or (SMICA-noSZ$_{\rm clean}\cdot$SMICA), which have lower noise than LGMCA$_{\rm clean}^2$ or SMICA-noSZ$_{\rm clean}^2$, respectively.  Moreover, these asymmetric combinations remain robust against tSZ contamination because the tSZ signal is explicitly nulled in the first leg of the combination (LGMCA or SMICA-noSZ).

The primary takeaway from Figs.~\ref{smicas_all_comparison} and \ref{smica_lgmca_comparison} is that all of the measurements agree within the statistical uncertainties.  However, it is also noticeable that the SMICA$^2 \times$ \emph{unWISE} points show some differences compared to the other measurements, which is very likely due to the tSZ residuals in the SMICA map.  This justifies the need for our use of the LGMCA or SMICA-noSZ maps as the primary map in our analysis.  Overall, the consistency of the measured cross-correlation of the \emph{unWISE} maps with all of the quadratic combinations considered here, including (LGMCA$_{\rm clean}\cdot$SMICA), (SMICA-noSZ$_{\rm clean}\cdot$SMICA), LGMCA$_{\rm clean}^2$, and SMICA-noSZ$_{\rm clean}^2$, clearly demonstrates the robustness of our measurements.

\subsection{{Shifting the Redshift Distribution}}
\label{subsec:shift_dndz}

We also test the impact of varying the \emph{unWISE} redshift distributions on our final results. As shown in Fig.~\ref{dn/dz}, we use both the redshift distributions $dn/dz$ and bias-weighted redshift distributions $b(z)dn/dz$ for each sample to obtain the theory components in our model. Note that $dn/dz$ is used to compute the kSZ$^2$-galaxy and lensing-galaxy theory terms, while $b(z)dn/dz$ is used in the kSZ$^2$-magnification and lensing-magnification terms.
The bias-weighted redshift distributions are obtained by cross-correlating the \emph{unWISE} catalogs with the spectroscopic BOSS and eBOSS surveys. The redshift distributions $dn/dz$ are on the other hand obtained by dividing $b(z)dn/dz$ by the redshift-dependent bias functions $b(z)$ for each sample (see \cite{Alex} for more details). Since the $b(z)dn/dz$ are known to a better precision than $dn/dz$, we consider a 3\% shift in the bias-weighted distribution and a 10\% shift in $dn/dz$.  We emphasize that these should be viewed as fairly extreme shifts, particularly in $b(z) dn/dz$.

We then re-run our parameter inference pipeline for these shifted theoretical templates.  The shift changes the amplitude of the kSZ signal $A_{kSZ^2}$ for the blue sample to $0.38 \pm 0.28$, for the green sample to $4.05 \pm 0.93$, and for the red sample to $7.09 \pm 3.10$. Even though these values are within the statistical error bars of our original results, and the considered case is quite extreme, the shift in $A_{kSZ^2}$ is non-negligible.  Thus, we use the results obtained from shifting $dn/dz$ and $b(z)dn/dz$ to estimate the associated systematic error in our final quoted results.  We consider the shift in the central values to represent a $2\sigma$ systematic error (as the shifted templates are quite extreme), and thus quote half of these shifts as a representative $1\sigma$ systematic error.

\section{Discussion and Future Directions}
\label{sec:discussion}

Our measurements are well-described by the model in Eq.~\ref{eq.model_tot}.
From the best-fit $\chi^2$ values, for 10 degrees of freedom, the probabilities-to-exceed are $p=$ 0.39, 0.29, and 0.33 for the blue, green, and red samples, respectively. These galaxies have average halo masses $\sim 1$-$5\times 10^{13}$ $h^{-1} M_{\odot}$~\cite{Alex}.

The measured kSZ$^2$ amplitudes, $A_{{\rm kSZ}^2}$, are $ = 0.42 \pm 0.31$, $5.02 \pm 1.01$, and $8.23 \pm 3.23$ for the cross-correlation with the \emph{unWISE} blue, green, and red samples, respectively. Including the systematic error from shifting the $dn/dz$ and $b(z)dn/dz$ distributions (see Subsection \ref{subsec:shift_dndz}), the results are $A_{\rm kSZ^2} = 0.42 \pm 0.31 \, (stat.) \pm 0.02 \, (sys.)$, $5.02 \pm 1.01 \, (stat.) \pm 0.49 \, (sys.)$, and $8.23 \pm 3.23 \, (stat.) \pm 0.57 \, (sys.)$. To our knowledge, these are the highest-redshift kSZ measurements to date. Including both the statistical and systematic errors, these results are within 1.9$\sigma$, 3.6$\sigma$, and 2.2$\sigma$ of our fiducial model, respectively, for which $A_{{\rm kSZ}^2}=1$ (Fig. \ref{Fiducial}).  We discuss the tension between our fiducial model and the inferred kSZ amplitude for the green sample in detail below.

The statistical significance of our $A_{{\rm kSZ}^2}$ detections for the blue, green, and red samples are 1.4$\sigma$, 5.0$\sigma$, and 2.5$\sigma$, respectively.  If the measurements were all uncorrelated, the overall detection significance would be $5.8\sigma$; however, since the samples have non-negligible overlap in redshift (see Fig.~\ref{dn/dz}) and we use the same CMB data in all three measurements (hence with the same noise), the measurements will be somewhat correlated.  It is beyond the scope of our current analysis to estimate the covariance of the three measurements; however, it is clear that the correlation will not be 100\%, and thus the overall detection significance of the kSZ signal across the three samples is greater than 5$\sigma$.  Our measurement is the most sensitive to date of the kSZ signal using the kSZ$^2$-LSS estimator.

Using the inferred kSZ amplitudes, we can constrain the product of the fraction of baryonic matter $f_b$ and the fraction of free electrons $f_{\rm free}$ for each \emph{unWISE} sample.  We find $f_{\rm free}$ to be $(f_b / 0.158)(f_{\rm free} / 1.0) = 0.65 \pm$ $0.24$, $2.24 \pm$ $0.25$, and $2.87 \pm$ $0.57$ at redshifts $z \approx $ 0.6, 1.1, and 1.5, respectively.  At these redshifts, we expect $f_{\rm free} \approx 0.8-0.9$, since all baryons are in an ionized form except for the fraction locked up in stars or the small pockets of neutral hydrogen gas.  Thus, as found in \citetalias{Hill2016}, our measurements confirm that the expected abundance of baryons based on BBN and primary CMB measurements is indeed present at these redshifts (i.e., there are no ``missing baryons'' on the relatively large scales probed by our data).

While the data are well-described by the theory model, and the blue and red $A_{{\rm kSZ}^2}$ values are within 1.9$\sigma$ and 2.2$\sigma$ of the fiducial prediction, the green $A_{{\rm kSZ}^2}$ is anomalously high.  As we have robustly verified that our measurements are not contaminated by dust (see Sections~\ref{sec:analysis} and~\ref{sec:validation} and Appendices~\ref{sec:append:const_alpha} and~\ref{sec:append:dust}), 
we consider other possible explanations for the high $A_{{\rm kSZ}^2}$ amplitude of the green sample.  Here, we emphasize that the detection significance of the kSZ signal in our measurements would not be affected by a change in the theoretical model that solely rescales the signal by an overall factor (as the best-fit $A_{\rm kSZ^2}$ value and its error bar would both be changed by the same factor).

First, we note that our fiducial theoretical model may be less accurate on small scales, particularly at high redshifts, as it relies on a fitting function for the non-linear matter bispectrum that is calibrated from $N$-body simulations (see Section~\ref{sec:theory}), which may begin to break down in these regimes.  It also assumes that the full signal simply scales as the linear galaxy bias, which is expected to break down on small scales where the one-halo term dominates.  To investigate these assumptions, we perform a first calculation of the one-halo term for the kSZ${}^2$-LSS estimator in Appendix~\ref{s:hm}.  We generally find consistency with the fitting function-based model, although the halo model calculation makes clear that the signal depends strongly on the assumed shape of the electron distribution on small scales (where our fiducial calculation assumes it to trace the dark matter) --- see Fig.~\ref{fig:halo-model-cvir}.  In future work, a full halo model will be used to interpret the measurements presented in this paper and derive constraints on the shape and amplitude of the electron profiles of the \emph{unWISE} galaxies.

As a further test of potential issues with our model on small scales, we consider fitting the cross-power spectra as a function of $\ell_{\rm max}$ (in our fiducial analysis we have $\ell_{\rm max}=2900$). Since the one-halo term is only significant at high $\ell$ (small scales), we would expect the kSZ$^2$ amplitude $A_{{\rm kSZ}^2}$ to approach our fiducial model value of unity as we decrease the maximum value of $\ell$, if indeed our model for the small-scale signal is inaccurate. The trend that we observe is qualitatively consistent with this expectation, although the changes in $A_{\rm kSZ^2}$ are not dramatic, and the error bars on $A_{\rm kSZ^2}$ are somewhat too large to draw a firm conclusion.  As an example, for the green sample, when including only the first seven data points ($\ell_{\rm max}=1700$), we obtain $A_{\rm kSZ^2}=4.82\pm1.19$, and for the first five data points ($\ell_{\rm max}=1300$) $A_{\rm kSZ^2}=4.52 \pm 1.31$. This may be some evidence that a more accurate model of the one-halo term could explain our high green $A_{{\rm kSZ}^2}$ measurement, but the trend is only suggestive at present, as these shifts are $< 1\sigma$.

Second, in our model we make the assumption that the hybrid bispectrum (Eq. \ref{eq:bispec}) can be factored into the product of the 3D velocity dispersion $v_{\rm rms}$ (computed with the Halofit non-linear prescription) and the non-linear matter bispectrum $B^{\rm NL}_m$.  Our calculation of $v_{\rm rms}$ is a slight modification of that in~\citetalias{Hill2016} and~\citetalias{Ferraro2016}, in which linear theory was used; the inclusion of non-linear power increases the predicted kSZ signal by roughly 20\%.  Additional refinement of the non-linear corrections is unlikely to change this value significantly, as the rms velocity is dominated by linear fluctuations. However, on small scales we might expect the presence of highly non-perturbative effects due to galaxy formation physics, such as velocity outflows in the ionized gas driven by feedback events (e.g., accretion onto supermassive black holes).  Recent analyses of the Illustris-TNG simulations indicate that these outflows can reach velocities up to $\approx 2000$ km/s (i.e., several times larger than $v_{\rm rms}$) out to scales as large as 300 kpc~\cite{Nelson_2019, Pillepich_2019}.  Moreover, these outflows are expected to be larger at higher redshifts~\cite{Nelson_2019}, which is consistent with the trend in our results from the low-redshift blue sample to the progressively higher-redshift green and red samples.  Since our measured cross-power spectrum is an integral over the full bispectrum that is particularly sensitive to squeezed triangles (see Eq.~\ref{eq.Clintegral}), seemingly small-scale effects like gas outflows can change the signal on surprisingly large scales.  Thus, our measurements could be an intriguing sign of feedback activity at high redshifts.  In this case, the use of perturbation theory to approximate the hybrid bispectrum would be invalid, and a model of the outflow velocities would be needed to interpret the signal.  We plan to explore this possibility using hydrodynamical simulations in future work.

In conclusion, the excess signal in one of the \emph{unWISE} samples can potentially be explained by physical effects that have not been accounted for in our analysis, and we have robustly shown that it is not due to foreground contamination.  We will investigate these explanations in detail in future work using the halo model and hydrodynamical simulations.

The future of measurements with the kSZ${}^2$-LSS estimator is bright.  As shown in~\citetalias{Ferraro2016}, the S/N of this estimator grows rapidly as smaller scales can be probed, which is now possible with multifrequency high-resolution CMB maps from ACT~\cite{Aiola2020,Naess2020} and the South Pole Telescope~\cite{SPT3G}, as well as future maps from the Simons Observatory~\cite{SO2019} and CMB-S4~\cite{CMBS4DSR}.  Upcoming data will yield kSZ detections with S/N an order of magnitude (or more) higher than that found here (see forecasts in~\citetalias{Ferraro2016}).  Crucial to this approach, however, will be the construction of robust component-separated blackbody CMB temperature maps from high-resolution ground-based data (see~\cite{ACT} for a first effort in this direction).  Moreover, in our analysis, we have considered galaxies as tracers of the matter overdensity field, but the method can be extended to other tracers, including quasars, weak lensing shear maps, or 21 cm fluctuations~\cite{Dore2004,DeDeo,MaHelgasonKomatsu2018,LaPlante2020}.

\begin{acknowledgements}

We thank Lehman Garrison and David Spergel for useful conversations. We are thankful to Mathew Madhavacheril for help with his publicly available halo model code \verb|hmvec|\footnote{\href{https://github.com/simonsobs/hmvec}{https://github.com/simonsobs/hmvec}} used to validate our numerical halo model computations carried out with \verb|class_sz|.

Some of the results in this paper have been derived using the healpy and HEALPix packages \cite{healpy_paper1, healpy_paper2}.  SF is supported by the Physics Division of Lawrence Berkeley National Laboratory. JCH thanks the Simons Foundation for support. AGK thanks the AMTD Foundation for support.

\end{acknowledgements}

\appendix
\label{Appendic}

\section{Constant $\alpha$ cleaning for LGMCA}
\label{sec:append:const_alpha}
To remove possible \emph{unWISE}-correlated dust contamination from the LGMCA map, we use the $\ell$-dependent $\alpha$ cleaning technique, which is described in detail in Section~\ref{sec:analysis}.  However, this procedure can also be done for a constant (not $\ell$-dependent) $\alpha$. As before, we cross-correlate the \emph{unWISE} galaxy overdensity $\delta_{g}$ with $((1 + \alpha)T_{\rm LGMCA} - \alpha T_{\rm dust})$, where $T_{\rm dust}$ is a dust-dominated map; as in the main analysis, we use the \emph{Planck} 545 GHz map as our fiducial choice.  Now, we look for an $\alpha_{\rm min}$ which minimizes the $\chi^2$ when fitting all 13 binned $D_{\ell}$ values to null (in the main analysis, we look for an $\alpha_{\rm min}$ which is closest to zero when cross-correlating $\delta_{g}$ with $((1 + \alpha)T_{\rm LGMCA} - \alpha T_{\rm dust})$ in each bin separately, and then interpolate these $\alpha$ values within the range $\ell = 300$ to $\ell=2900$).  In the scale-independent approach here, we find the $\alpha_{\rm min}$ values are -0.00040 $\pm$ 0.00001, -0.00007 $\pm$ 0.00001, 0.00037 $\pm$ 0.00001 for \emph{unWISE} blue, green, and red, respectively. Then, a new map $T_{\rm clean} =(1+\alpha_{\rm min})T_{\rm LGMCA}-\alpha_{\rm min} T_{\rm dust}$ is constructed separately for each of the \emph{unWISE} maps, as in Eq.~\ref{eq.Tclean}, but now with a trivial scale-independent operation instead of a multipole-dependent operation. 

\begin{figure*}[htbp!]
\includegraphics[width=0.45\textwidth]{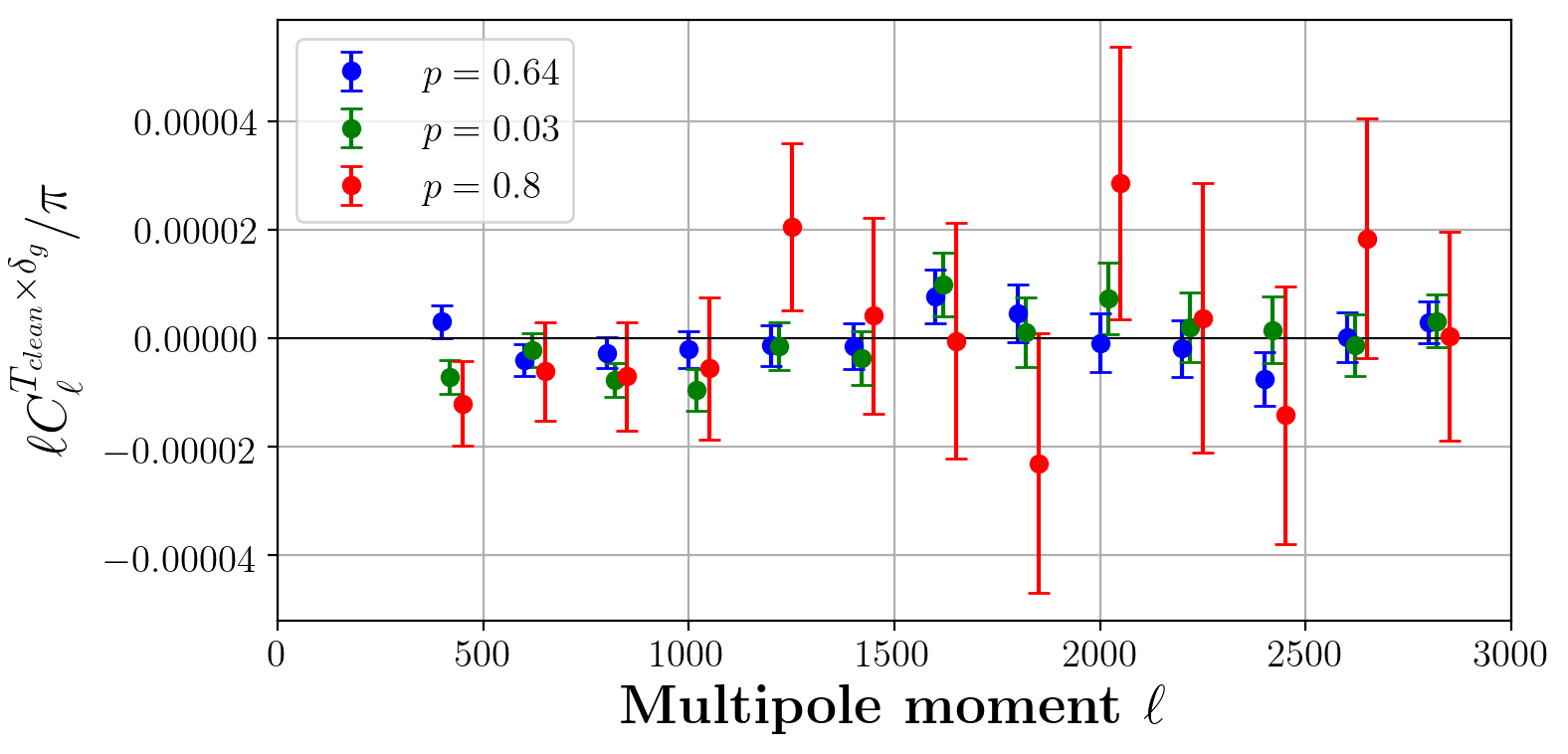}
\includegraphics[width=0.45\textwidth]{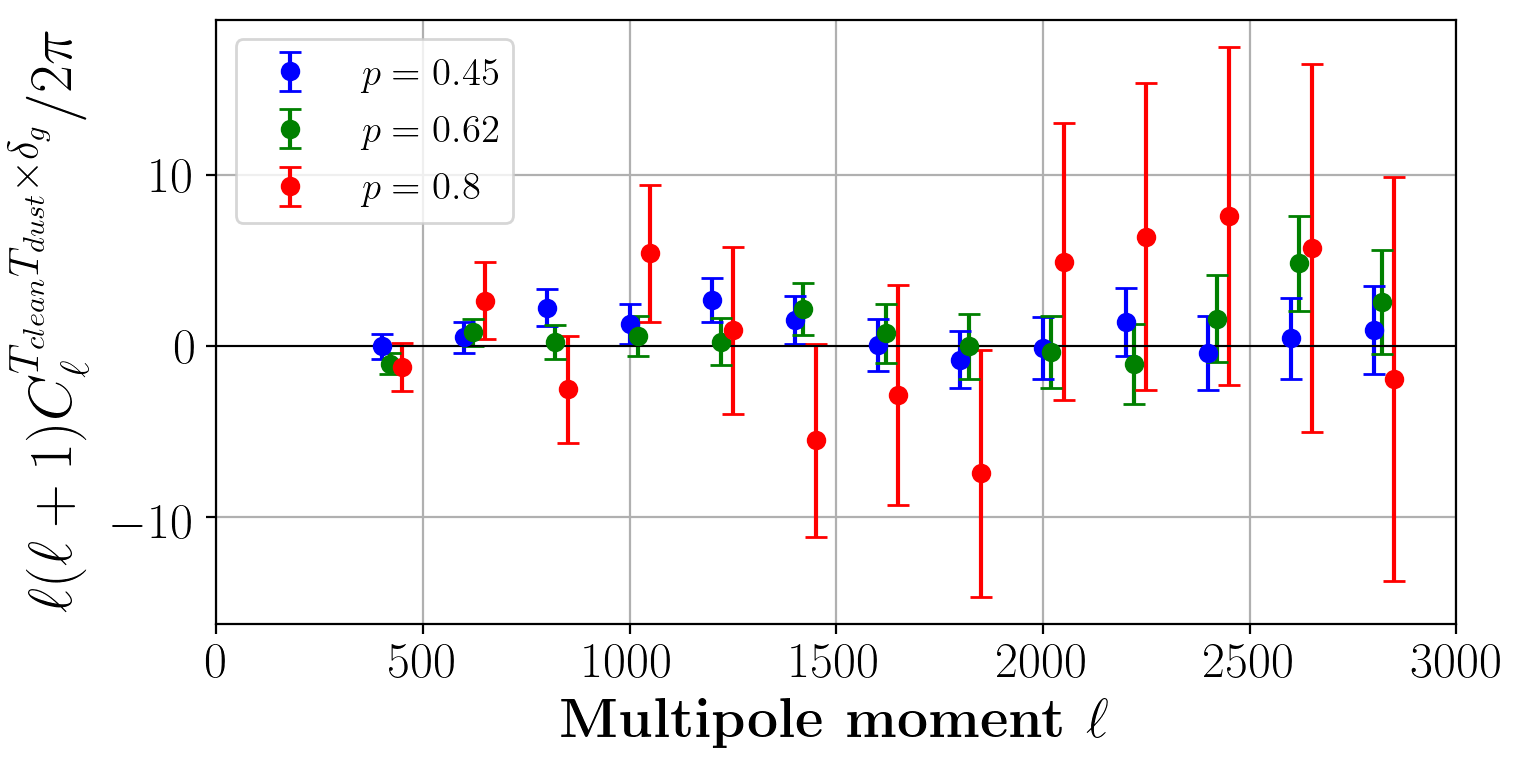}
\caption{Null tests on the LGMCA$_{\rm clean}$ map for the constant $\alpha$ cleaning. Both plots are color coded based on the \emph{unWISE} colors: blue, red, and green, and include their respective $p$-values (for 13 degrees of freedom) in the legend when fitting them to null. Green is offset by $\ell=20$, and red by $\ell=50$ with respect to the true multipole moment values, for visual purposes. Left: Cross-correlation between $T_{\rm clean}$ and \emph{unWISE}. Right: Cross-correlation between $T_{\rm clean}T_{\rm dust}$ and unWISE, where $T_{\rm dust}$ is \emph{Planck} 545 GHz map. } 
\label{fig_const_alpha_tests}
\end{figure*}

As in Section~\ref{sec:analysis}, we perform null tests ($T_{\rm clean} \times \delta_g$, $T_{\rm clean}T_{\rm dust} \times \delta_g $,) with the $T_{\rm clean}$ maps. The constant $\alpha$-cleaning results are summarized below; Fig.~\ref{fig_const_alpha_tests} and Table \ref{table_const_alpha_pvalues} show the null tests performed, along with the probabilities-to-exceed for each test and \emph{unWISE} color in the table. Note that we do not perform the $T_{noise}$ and $T_{noise}^2$ tests since the LGMCA noise map does not change here.

In the analysis in~\citetalias{Hill2016} (where the constant $\alpha$-cleaning method was also used), the dust null test $(T_{\rm clean}T_{\rm dust}) \times \delta_g$ was only marginally consistent with null, with $p=0.02$.  However, we find that it is consistent with null ($p= $ 0.45, 0.62, and 0.82 for \emph{unWISE} blue, green, and red, respectively) for the constant-$\alpha$ cleaning here, as for the $\ell$-dependent $\alpha$-cleaning in Section~\ref{sec:analysis}.  However, the null test is marginal for the cross-correlation between the \emph{unWISE} green sample and $T_{\rm clean}$, with $p=0.03$.  This test is semi-meaningful here since the nulling procedure to determine $\alpha$ is not done $\ell$-by-$\ell$, but rather for all bins simultaneously.  We take this as a clear sign that the $\ell$-dependent approach used in our main analysis is preferable due to its robust ability to handle residuals that have non-negligible changes in sign (which are responsible for the marginal issue here --- see the green points in Fig.~\ref{fig_const_alpha_tests} left panel). 

Overall we conclude that both $\alpha$-cleaning approaches (scale-dependent or scale-independent) yield maps with no evidence of \emph{unWISE}-correlated dust contamination, but we adopt the $\ell$-dependent cleaning approach in our main analysis due to its greater robustness.

        \begin{table}[h!]
        \begin{tabular}{ |c|c|c|c| } 
        \hline
         \emph{unWISE} $\times$ & $T_{\rm clean}$ & $T_{\rm clean}T_{\rm dust}$ \\ 
        \hline\hline
        blue & 0.64 & 0.45\\
        green & 0.03 & 0.62\\ 
        red & 0.80 & 0.80\\ 
        \hline
        \end{tabular}
        \caption{Cross-correlation between the \emph{unWISE} galaxy maps (blue, green and red), and the CMB temperature maps ($T_{\rm clean}$, $T_{\rm clean}T_{\rm dust}$) with probabilities-to-exceed (for 13 degrees of freedom) to test the dust contamination, where $T_{\rm dust}$ is \emph{Planck} 545 GHz map, and $T_{\rm clean}$ is the LGMCA$_{\rm clean}$ map, cleaned using the constant $\alpha$ method. }
        \label{table_const_alpha_pvalues}
        \end{table}

\section{Dust assessment in the LGMCA map}
\label{sec:append:dust}

Although we performed foreground cleaning and found no evidence of bias, we decided to further investigate the possible dust contamination in the LGMCA map, and assess its level.
To estimate the amount of dust in this CMB map, we cross-correlate the squared \emph{Planck} 545 GHz with \emph{unWISE}, and then rescale it down to the \emph{Planck} 143 GHz frequency. The \emph{Planck} 545 GHz map is dominated by dust, so it is justified to neglect any other components at this particular frequency channel. Then, however, we have to rescale this measurement down to the lower frequencies that dominate our LGMCA kSZ measurement, i.e., to 143 GHz. 
Using a standard modified blackbody dust SED (e.g., Eq. 6 in \cite{ACT}) with temperature $T_{\rm dust}$ = 24 K and spectral index $\beta$ = 1.40, 1.33, 1.27 for the blue, green, and red samples, respectively, we find the rescaling factor from 545 GHz to 143 GHz to be 328.5 for maps in CMB temperature units, and from 545 GHz to 217 GHz to be 116.6. As a test of this rescaling factor, we measure the ratio of \emph{Planck} 545 GHz $\times unWISE$ to \emph{Planck} 217 GHz $\times unWISE$ measurements, both of which are dust-dominated (since the tSZ signal vanishes at 217 GHz).  This gives a ratio in the range of 115-120 for our $\ell$ range (which avoids any ISW contributions), which validates the SED rescaling model that we adopt.

\begin{figure*}[htbp!]
\includegraphics[width=0.45\textwidth]{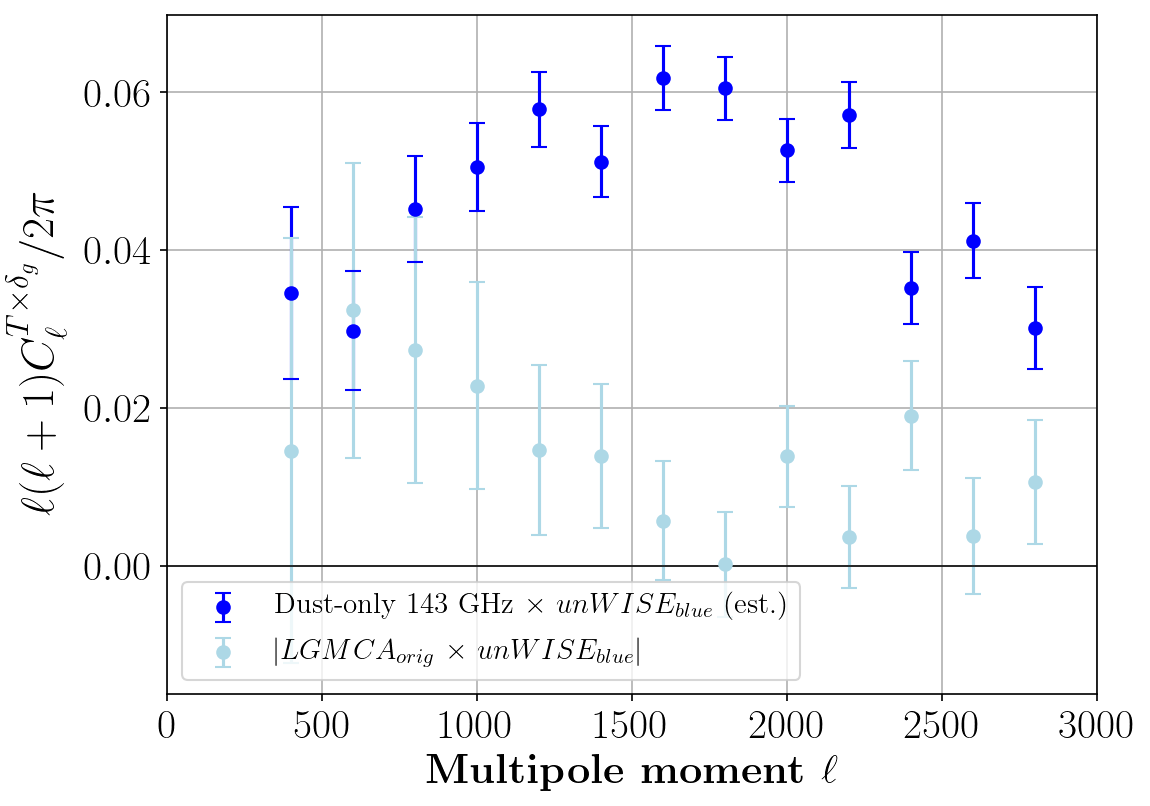}
\includegraphics[width=0.45\textwidth]{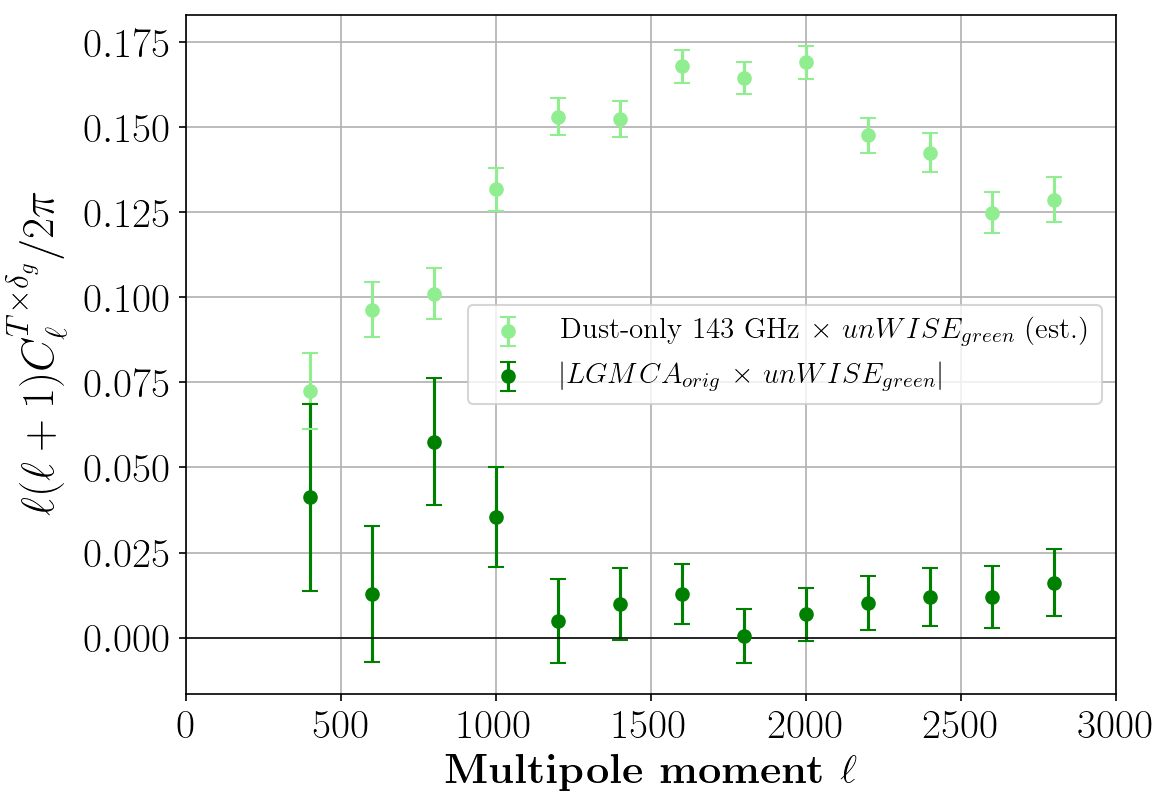}
\includegraphics[width=0.45\textwidth]{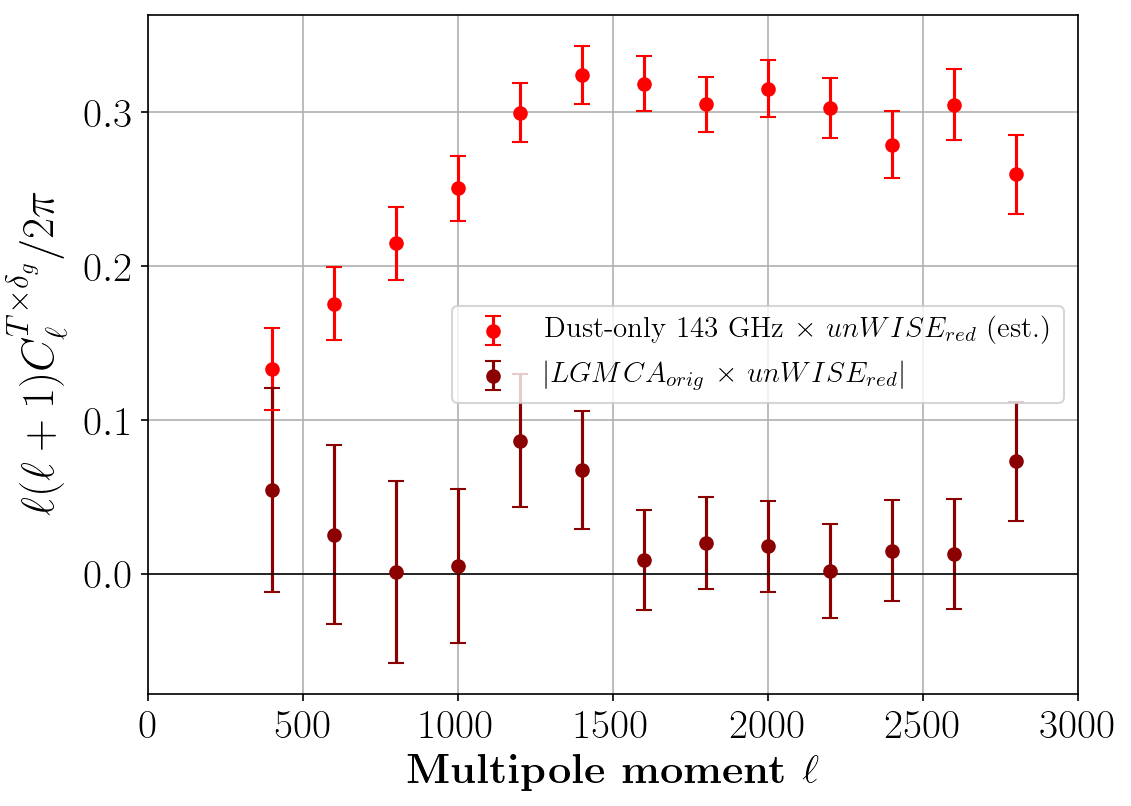}
\caption{To assess the ``additional cleaning factor'' described in Appendix~\ref{sec:append:dust}, we show a comparison of the absolute value of $D_{\ell}$ for LGMCA $\times$ \emph{unWISE} and $D_{\ell}$ for an estimate of the dust-only contribution to 143 GHz $\times$ \emph{unWISE}, where the latter has been obtained by rescaling from \emph{Planck} 217 GHz $\times$ \emph{unWISE} using a dust SED determined from 217 and 545 GHz data (see the text for details). We show these quantities for each of the \emph{unWISE} maps (color-coded).}
\label{add_cleaning_factor}
\end{figure*}

In this dust assessment, we want to include an ``additional cleaning factor'' that approximately captures how much the original LGMCA cleaning method has reduced the dust in the map compared to its 143 GHz value.  We cannot directly measure the dust contribution at 143 GHz (or, e.g., 100 GHz) by cross-correlating \emph{unWISE} with the \emph{Planck} 143 GHz map, because this cross-correlation also receives a non-negligible contribution from the tSZ effect.  Thus, instead we take the \emph{unWISE} $\times$ \emph{Planck} 217 GHz measurement and rescale it to 143 GHz using our dust SED model determined from the 545 and 217 GHz measurements described above.  The rescaling factor from 217 to 143 GHz in our SED model is 2.8.  In Fig. \ref{add_cleaning_factor}, we show this estimate for the \emph{unWISE} $\times$ 143 GHz dust-only contribution, and compare it to the absolute value of LGMCA $\times$ \emph{unWISE}.  The ratio of these two measurements gives us an estimate of the $\ell$-dependent ``additional cleaning factor'' by which the \emph{unWISE}-correlated dust in the LGMCA map has been reduced below its value at 143 GHz.  These additional cleaning factor values are given in Table \ref{table_clean_factor}.  Note that this additional cleaning factor will enter squared in a $T^2 \times unWISE$ measurement.

\begin{table}[h!]
\begin{tabular}{ |c|c|c|c|}
\hline
$\ell$ & blue & green & red  \\
\hline
400 & 2 & 2 & 2\\
600 & 1 &  8 & 7\\
800 & 2 &  2 & 203\\
1000 & 2 & 4 & 50\\
1200 & 4 & 32 & 3\\
1400 & 4 & 16 & 5\\
1600 & 11 & 13 & 35\\ 
1800 & 312 & 301 & 15\\
2000 & 4 & 24 & 18\\
2200 & 16 & 15 & 167\\
2400 & 2 & 12 & 19\\
2600 & 11 & 10 & 24\\
2800 & 3 & 8 & 4\\
\hline
\end{tabular}
\caption{The ``additional cleaning factors'' used to assess how much the original LGMCA cleaning method has reduced the dust compared to its value at 143 GHz. The ``additional cleaning factor'' for each of the \emph{unWISE} samples is estimated from the ratio of $\lvert {\rm LGMCA} \times unWISE \rvert$ to our estimate of the dust-only contribution to 143 GHz $\times$ \emph{unWISE}, which we obtain by measuring \emph{Planck} 217 GHz $\times$ \emph{unWISE} and rescaling it with our dust SED model to 143 GHz.}  
\label{table_clean_factor}
\end{table}

\begin{figure}
\includegraphics[width=0.5\textwidth]{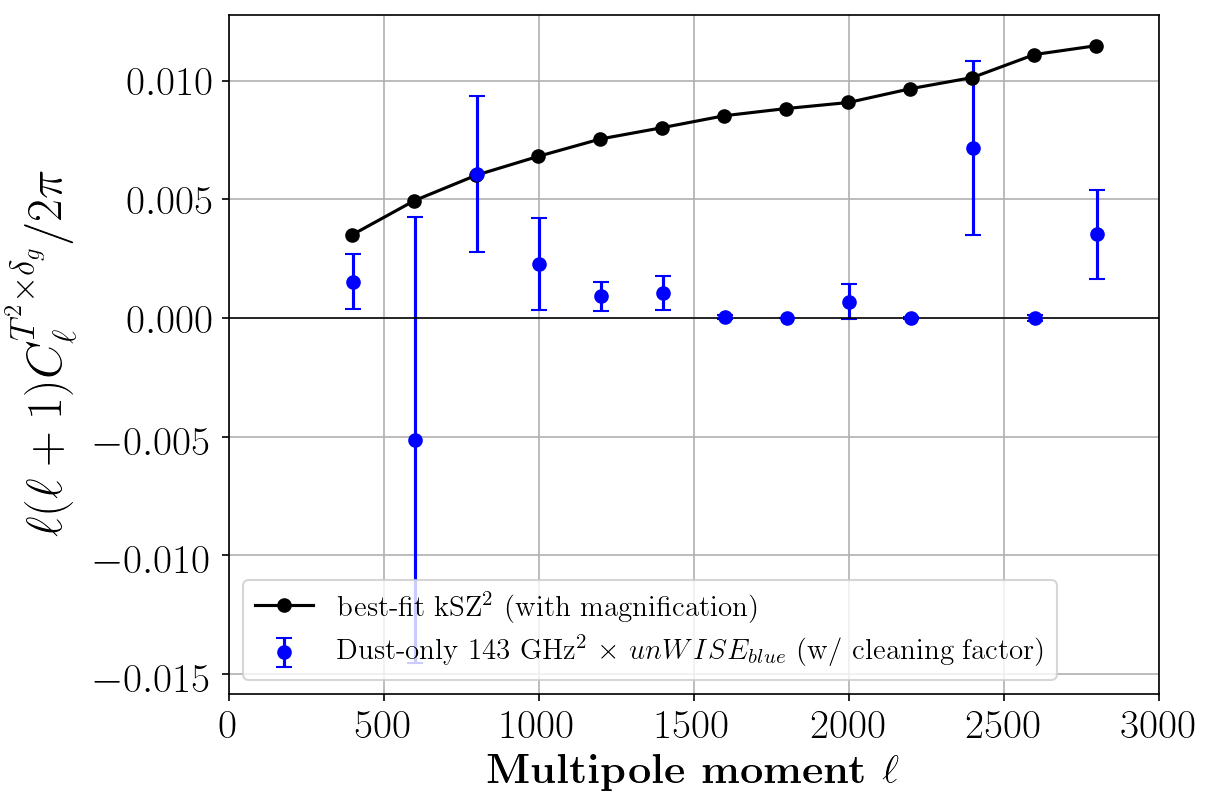}
\includegraphics[width=0.5\textwidth]{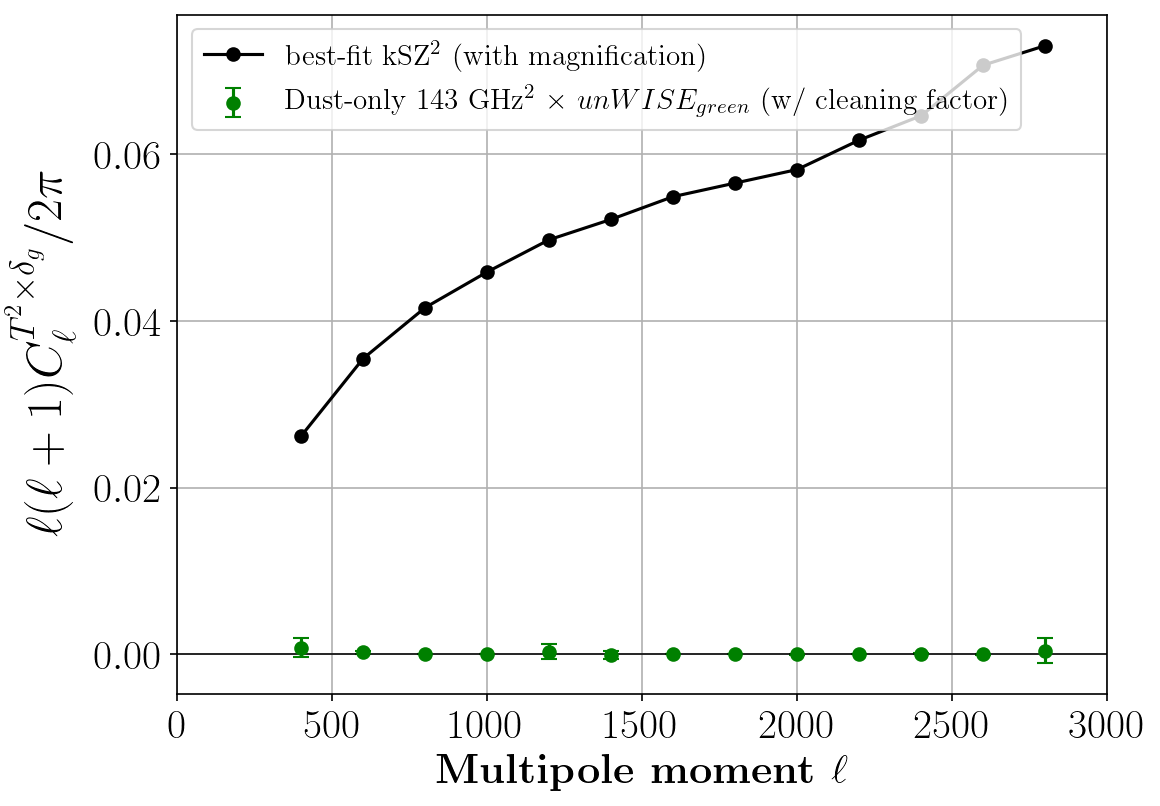}
\includegraphics[width=0.5\textwidth]{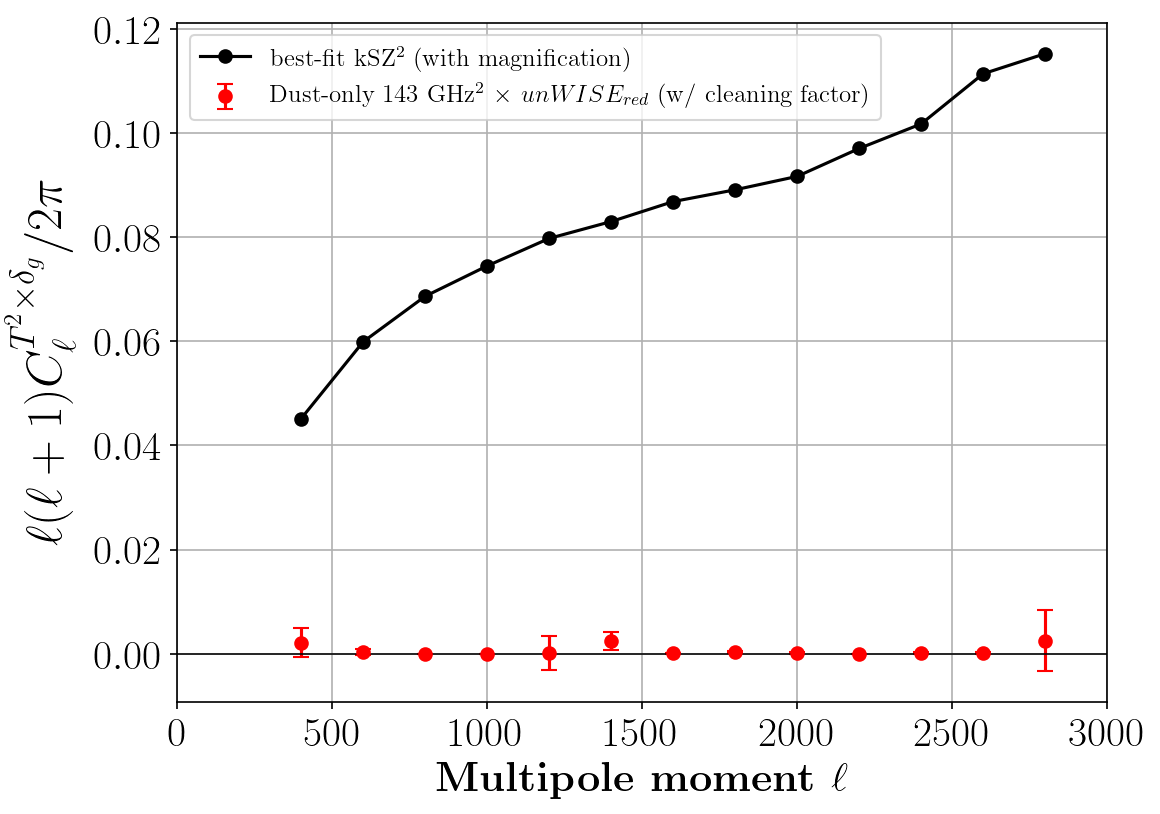}
\caption{ A comparison of the dust contamination present in the LGMCA map versus the best-fit kSZ signal from our analysis for each of the \emph{unWISE} maps. The dust is assessed by rescaling (\emph{Planck} 545 GHz)${}^2 \times unWISE$ down to 143 GHz and dividing by the ``additional cleaning factors'' from Table~\ref{table_clean_factor} squared. This demonstrates that any \emph{unWISE}-correlated dust that is present in the LGMCA maps is negligible, and we can safely assume it is not a source of our detected signal.} 
\label{dust_kSZ}
\end{figure}

Finally, our (\emph{Planck} 545 GHz$)^2$ $\times$ \emph{unWISE} measurement rescaled to 143 GHz frequency (so divided by $328.5^2$), and then further divided by the additional $\ell$-dependent cleaning factor (also squared), are shown in Fig.~\ref{dust_kSZ}, and compared with our best-fit kSZ signal. This demonstrates that the dust contamination, even in the original (non-$\alpha$-cleaned) LGMCA map is indeed negligible.

In addition, note that the additional cleaning factor was obtained for the original LGMCA map (before the implementation of any $\alpha$-cleaning method), so the dust contamination will be even more negligible for the LGMCA$_{\rm clean}$ (i.e., $T_{\rm clean}$, see Eq.~\ref{eq.Tclean}) map used in our kSZ analysis. Therefore we can safely state that residual \emph{unWISE}-correlated thermal dust emission present in the LGMCA map yields a negligible bias to the measured kSZ signal in this work.

\section{Posteriors of the model parameters}
\label{app:posteriors}

\begin{figure*}[htbp!]
\includegraphics[width=0.45\textwidth]{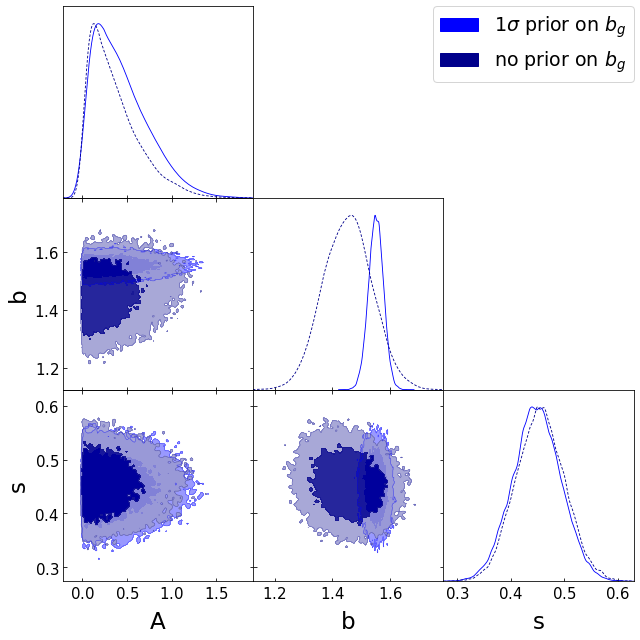}
\includegraphics[width=0.45\textwidth]{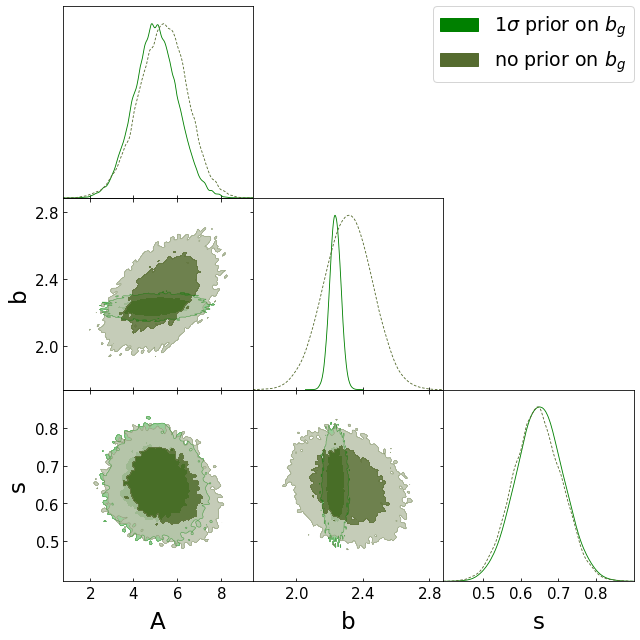}
\includegraphics[width=0.45\textwidth]{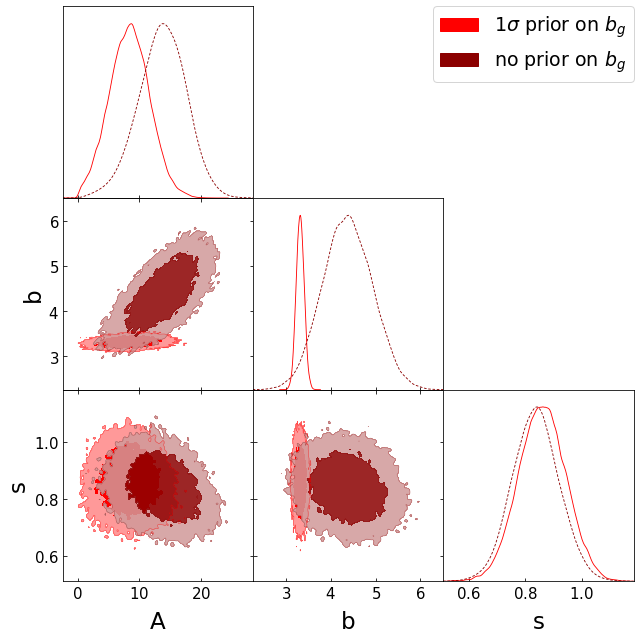}
\caption{The 2D and 1D marginalized posterior distributions of the $A_{{\rm kSZ}^2}$, $b_g$, and $s$ parameters for the fit to the cross-power spectra of the product of filtered LGMCA$_{\rm clean}$ and SMICA with each of the \emph{unWISE} galaxy maps (solid lines). This analysis imposes external priors on $b_g$ and $s$ derived from Ref.~\cite{Alex}.  The $b_g$ priors are Gaussians with mean and 1$\sigma$ width corresponding to the values in Table~\ref{param_values}, while the $s$ priors are Gaussians centered at the values in Table~\ref{param_values} with a fractional standard deviation of 10\%.  The dashed curves are the posteriors of the same model fit, but without the external priors on $b_g$.  Clockwise from top-left: \emph{unWISE} blue, green, and red.}
\label{posteriors}
\end{figure*}

In this appendix, we present the full posterior distributions for the fit parameters: the amplitude of the kSZ$^2$ signal, $A_{{\rm kSZ}^2}$; the galaxy bias, $b_g$; and the magnification response, $s$, for each of the \emph{unWISE} samples (solid, color-coded lines in Fig. \ref{posteriors}). They are obtained by fitting the model in Eq.~\ref{eq.model_tot} to the data points in Fig.~\ref{fig:1} with a Gaussian likelihood function using the Python \texttt{emcee} package.  External priors on $b_g$ and $s$ are imposed in this analysis, as described in Section~\ref{sec:interpretation}.  In particular, we impose a Gaussian prior on $b_g$ for each sample with standard deviation equal to the error bars found in Ref.~\cite{Alex}, and a Gaussian prior on $s$ with fractional width of $10\%$.

We also perform the fit without imposing external priors on the galaxy bias $b_g$, as a cross-check of our main results.  The posterior distributions from this MCMC run are shown in Fig.~\ref{posteriors} in dashed lines. They are consistent with our main analysis, which confirms that our data and model yield $b_g$ values consistent with those found in Ref.~\cite{Alex}.  This verifies that our use of the priors based on that work is statistically sound, i.e., we are not imposing an inconsistent assumption on our data.

\section{The 1-halo term for kSZ with projected fields}\label{s:hm}

To build confidence and further understanding of our estimator for $C_{\ell}^{{\rm kSZ}^2 \times \delta_g}$, we present the first computation of the 1-halo term within the halo model formalism. For details about the halo model and its assumptions we refer the reader to the review by Cooray \& Sheth \cite{Cooray:2002dia}.  Our goal here is to illustrate the sensitivity of  $C_{\ell}^{{\rm kSZ}^2 \times \delta_g}$ to the Halo Occupation Distribution (HOD), i.e., to the distribution of galaxies within halos, and to the properties of the electron gas (velocity and profile). We leave the presentation of a detailed analysis, including the 2-halo and 3-halo terms, for a future paper. 

The projected field power spectrum $C_{\ell}^{{\rm kSZ}^2 \times \delta_g}$ is an integral of a bispectrum over a range of scales, modulated by a  Wiener filter and beam function  to account for the resolution of the CMB maps (see Section \ref{ss:kszest}). The bispectrum contains two copies of the electron momentum field and one copy of the galaxy overdensity field. An expression for the 1-halo term of the bispectrum is given in Eq.\ B60 of \cite{Coulton:2017crj}. The main feature of this formulation is that the product of the electron momentum fields is  written as the product  the electron density fields times the cluster velocity dispersion. The cluster velocity dispersion can then be approximated with the RMS of the matter velocity field, 
 \begin{equation}
     v_{\mathrm{rms}}^{2}\left(z\right)=\frac{1}{2\pi^{2}}\int\mathrm{d}k\left[f\left(z\right)a\left(z\right)H\left(z\right)\right]^{2}P\left(k,z\right),
\label{eq:vrms2}
 \end{equation}
and factored out of the mass integral \citep[see, e.g.\ Section V D of][]{Hill_2018}. In this equation, $P(k,z)$ is the matter power spectrum and $f$ is the logarithmic derivative of the growth function. In our fiducial model we use the matter power spectrum including non-linear corrections computed with Halofit \cite{Takahashi_2012}. Meanwhile, the electron density field multiplied by the Thomson cross-section can be interpreted as  an optical depth profile and referred to as a  $\tau$-profile. 

More explicitly, the 1-halo term is computed as
\begin{equation}
    C_{\ell}^{{\rm kSZ}^2 \times \delta_g,\mathrm{1h}} =\int\frac{d^{2}\ell'}{\left(2\pi\right)^{2}}f\left(|\vec{\ell}-\vec{\ell}'|\right)f\left(\ell^{\prime}\right)B_{|\vec{\ell}-\vec{\ell}'|\ell'\ell}^{{\rm kSZ}^{2}\times \delta_g,\mathrm{1h}}, \label{eq.Clintegral}
\end{equation}
where the function $f$ is the product of a filter and beam, as in Eq.~\ref{eq.filterbeam}. 
The 1-halo term of the bispectrum at multipoles $(\ell_1,\ell_2,\ell_3)$ is given by 

\begin{equation}
    B^{{\rm kSZ}^2 \times \delta_g,\mathrm{1h}}_{\ell_1,\ell_2,\ell_3} = \int  \mathrm{d}\eta \, \eta^2 \frac{v_\mathrm{rms}^2}{3c^2}\int \mathrm{d}M \frac{\mathrm{d}N}{\mathrm{d}M\mathrm{d}V} u_{\ell_1}^\tau u_{\ell_2}^\tau u_{\ell_3}^g,  \label{eq:clgg1h}
\end{equation}
where $\eta$ is the comoving distance, $v_\mathrm{rms}^2$ is the velocity dispersion  given in Eq.~\eqref{eq:vrms2},  $\frac{\mathrm{d}N}{\mathrm{d}M\mathrm{d}V}$ is the differential number of halos per unit mass and volume  determined by the halo mass function,  $u_\ell^{\tau}$ is the 2D Fourier transform of the halo optical depth profile, and $u_\ell^{g}$ is the 2D Fourier transform of the halo density profile populated with galaxies. The optical depth term depends on halo masses and redshift and can be written as
\begin{equation}
    u_\ell^\tau(M,z) = W_\tau (z) C_\tau(M,z)\phi_\ell(M,z),\label{eq:utau}
\end{equation}
where  the redshift dependent term is given by
 \begin{equation}
     W_\tau(z) = \frac{1 }{D_A(z)^2} \frac{\sigma_\mathrm{T}}{m_\mathrm{p}\mu_\mathrm{e}}f_\mathrm{b}f_\mathrm{free}\label{eq:w_tau}, 
 \end{equation}
where $D_A(z)$ is the angular diameter distance, $m_\mathrm{p}$ is the proton mass, $\mu_\mathrm{e}$ is the mean molecular weight per free electron. It is computed in terms of the primordial hydrogen mass fraction $X_\mathrm{H}$ as $\mu_\mathrm{e}=2/(1+X_\mathrm{H})\simeq1.14$ for the fiducial value $X_\mathrm{H}=0.76$. 
Assuming that the electron density within halos follows that of matter approximated by the Navarro-Frenk-White (NFW) profile \cite{Navarro_1997}, the term $C_\tau(M,z)$, which depends on mass and redshift but not on multipole, sets the normalization of the $\tau$-profile in terms of the  concentration $c_{_\mathrm{NFW}}$, i.e.,
 \begin{equation}
     C_\tau (M,z) = M \left[\ln(1+c_{_\mathrm{NFW}})-\frac{c_{_\mathrm{NFW}}}{1+c_{_\mathrm{NFW}}}\right]^{-1}.
 \end{equation}
The last term in Eq. \eqref{eq:utau} is the 2D Fourier transform of the 3D density profile projected along the LOS, written in a dimensionless fashion and with its normalisation factored out in $C_\tau$. With the NFW profile, it reads  as
 \begin{equation}
     \phi_\ell(M,z)= \int \mathrm{d}x \frac{x}{(1+x)^2} \mathrm{sinc}\left(\frac{\ell+\tfrac{1}{2}}{\ell_s}x\right) \label{eq:phiellnfw}
 \end{equation}
 with the radial variable $x=r/r_\mathrm{s}$,  
 where the characteristic multipole is $\ell_\mathrm{s}=D_A/r_\mathrm{s} $ and the  scale radius $r_\mathrm{s}= r_{_\mathrm{NFW}}/c_{_\mathrm{NFW}}$. Since  the integral in Eq.~\eqref{eq:phiellnfw} is divergent, it is carried out from the center $x=0$ (or $r=0$) to a limiting  $x_\mathrm{out}=\Delta c_{_\mathrm{NFW}}$ (or equivalently $r=\Delta r_{_\mathrm{NFW}}$, where $\Delta$ is a numerical factor). To choose the outer cut-off radius we  compare the prediction of the lensing power spectrum computed within the halo-model and within Halofit, as in \cite{Hill_2014}. We perform the comparison numerically using the public code \verb|class_sz|\footnote{\href{https://github.com/borisbolliet/class_sz}{https://github.com/borisbolliet/class\_sz}} \cite{Bolliet:2017lha}. In our settings, with fiducial cosmological parameters, halo mass function from \cite{Tinker_2010} and concentration from \cite{Duffy_2008} both defined with respect to the overdensity masses $M_{200m}$ (hence,  $r_{_\mathrm{NFW}}=r_{200m}$ and $c_{_\mathrm{NFW}}=c_{200m}$) we find that $\Delta=1$ ensures a nearly perfect match between both methods, over all multipoles of interest. Setting an upper bound to the the radial integration is equivalent to taking the Fourier transform of the NFW profile truncated at $r_{_\mathrm{trunc}}=\Delta r_{_\mathrm{NFW}}$. Fortunately, this Fourier transform has an analytical expression \citep[see, e.g., Eq. 18 of][]{Koukoufilippas:2019ilu} which can be used to speed up the numerical computation of the bispectrum   without loss of accuracy. Nevertheless, note that   assuming a rescaled NFW profile for the electron density is a strong assumption. Indeed, it  already appears to be invalidated by recent kSZ measurements~\cite{schaan2020act,Amodeo2020}. But it provides a useful starting point and    further work will investigate the validity of this hypothesis, for instance by considering electron density profiles directly calibrated on simulations~\cite{Battaglia_2016}, or by measuring them from data~\cite{battaglia2019probing}. 
 
The galaxy term entering the bispectrum  can be written as
\begin{equation}
    u_\ell^g(M,z) = W_g(z)\left[U_\ell^g(M,z)^2 + 2U_\ell^g(M,z)\right]^{1/2},\label{eq:ulg}
\end{equation}
with 
\begin{equation}
    U_\ell^g(M,z) = N_\mathrm{sat}(M,z)\phi_\ell^g(M,z),\label{eq:Ulg}
\end{equation}
where $N_\mathrm{sat}(M,z)$ is determined by a specific HOD and $\phi_\ell^g(M,z)$ is the Fourier transform of a truncated NFW profile. The profile is truncated at the same radius as before, namely $r_{_\mathrm{trunc}}=\Delta r_{_\mathrm{NFW}}$ with $r_{_\mathrm{NFW}}=r_{200m}$ and $
\Delta=1$ in our case.  The redshift dependent term is given by
\begin{equation}
        W_g(z)=\frac{H}{c\eta^2}\frac{\varphi_g^\prime(z)}{\bar{n}_g(z)}\label{eq:wgz}
\end{equation}
where $\varphi_g^\prime(z)$ is the normalized differential galaxy distribution of the studied sample, i.e., 
\begin{equation}
    \varphi_g^\prime(z) = \frac{1}{N_g^\mathrm{tot}}\frac{\mathrm{d}N_g}{\mathrm{d}z},\quad \mathrm{with}\quad N_g^\mathrm{tot}=\int \mathrm{d}z\frac{\mathrm{d}N_g}{\mathrm{d}z}, \label{eq:varphig}
\end{equation} the total number of galaxies in the sample and $\bar{n}_g(z)$ is the galaxy number density predicted within the HOD model in terms of central and satellite galaxy populations, namely
\begin{equation}
    \bar{n}_g(z) =\int\mathrm{d}M \frac{\mathrm{d}N}{\mathrm{d}M\mathrm{d}V}\left\{N_\mathrm{cent}(M,z)+N_\mathrm{sat}(M,z)\right\}.\label{eq:ngbar}
\end{equation}
For central and satellite galaxy populations we use slightly modified version of Eq.~(B1) and Eq.~(B2) from \cite{Alex}. In particular, we find that replacing\footnote{$\log_{10}M_\mathrm{cut}$ sets the host halo  mass threshold for central and and satellite galaxies. We use formulas obtained as by-products of the analysis of \cite{Alex}, and available in \texttt{class\_sz}.} $\log_{10}M_\mathrm{cut}$ by $1.03\log_{10}M_\mathrm{cut}$, and using 1.3 instead of 0.1 in the numerator of Eq.~(B2) provides a good fit to the galaxy-galaxy and CMB lensing-galaxy auto and cross power spectra measurements from \cite{Alex}. (Note that these numbers are relevant to the green sample of $unWISE$ galaxies and may differ slightly for the red and blue sample.)

\begin{figure}
    \includegraphics[width=1.\columnwidth]{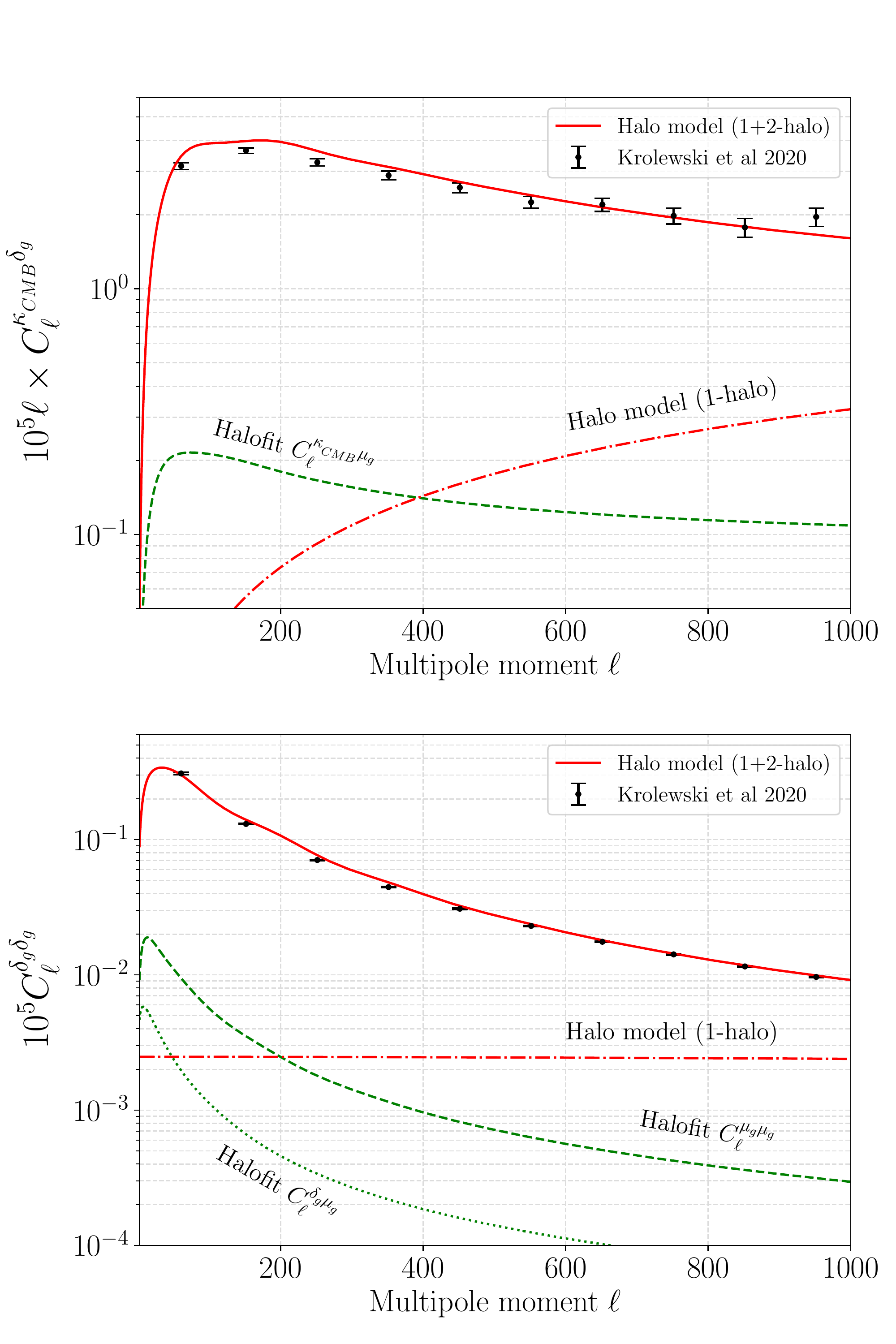}
    \caption{Dimensionless CMB lensing-galaxy cross power spectrum (top panel) and galaxy auto power spectrum (bottom panel). Measurements from \cite{Alex} are shown as the black data points with error bars. The halo model prediction (1+2-halo) is shown as the solid red lines. The 1-halo term contribution is shown as the dash-dotted red lines. The other curves (in green) show the contribution from lensing magnification bias and are computed according to Eq.~6.4 and 6.5 of \cite{Alex}, using the matter power spectrum computed with Halofit. For this figure we use the green sample of \emph{unWISE} galaxies and carry out all the Halofit and halo model computations with \texttt{class\_sz}.} 
    \label{fig:halo-model-gg-gk}
\end{figure}

\begin{figure}
    \includegraphics[width=1.\columnwidth]{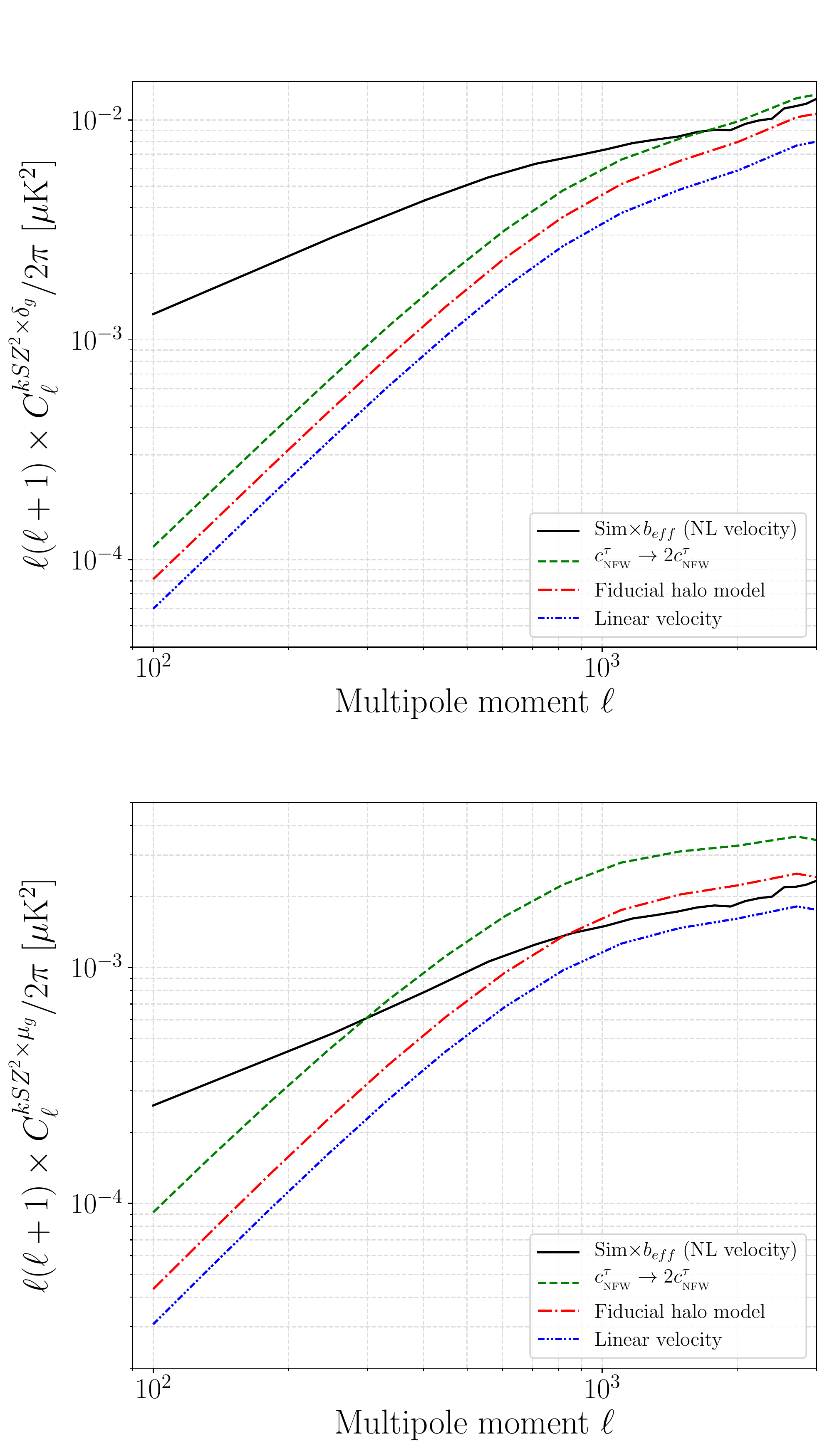}
    \caption{ Halo model predictions for the one-halo term contribution to $C_\ell^{{\rm kSZ}^2\times \delta_g}$ (top panel) and $C_\ell^{{\rm kSZ}^2\times \mu_g}$ (bottom panel) under three different assumptions: (i) fiducial halo model (dash-dotted red lines); (ii) velocity dispersion computed without non-linear corrections in the matter power spectrum, i.e., using the linear matter power spectrum rather than Halofit in Eq.~\ref{eq:vrms2} (dot-dot-dashed blue lines); (iii) a $\tau$-profile concentration twice as high as the fiducial model (green dashed lines). Note that as we change the concentration we keep the total mass fixed (see text). The solid black lines depict the templates used in the data analysis in the main part of this paper (see Fig.~\ref{Fiducial}). For this figure we use the green sample of \emph{unWISE} galaxies and carry out all the halo model computations with \texttt{class\_sz}. In future work we will compute the other halo-model contributions to these signals, beyond the one-halo term.} 
    \label{fig:halo-model-cvir}
\end{figure}

In Fig.~\ref{fig:halo-model-gg-gk}, we compare the halo model prediction for this HOD (red curves) to the data (black data points).  We show results only for the \emph{unWISE} green sample, as this was the sample that yielded results most discrepant with our fiducial theoretical model in the main analysis, and thus we want to validate the theory calculation.  Here our comparison is tentative as our goal is to set the HOD parameter values approximately rather than precisely, since this is enough to check the consistency between our fiducial template for $C_{\ell}^{{\rm kSZ}^2 \times \delta_g}$ and the halo model predictions. But in principle, one could perform a maximum likelihood analysis to set the HOD parameters based on such measurements, eventually enabling constraints on the parameters of the cluster optical depth profile with measurements of $C_{\ell}^{{\rm kSZ}^2 \times \delta_g}$. This point can be directly appreciated by looking at Fig.~\ref{fig:halo-model-cvir}. There, once the HOD parameters have been fixed thanks to the comparison shown in  Fig.~\ref{fig:halo-model-gg-gk}, we plotted the 1-halo term of the kSZ$^2$-galaxy (top panel) and kSZ$^2$-lensing magnification (bottom panel) cross power spectra under three different assumptions for the kSZ effect. These are compared with the template for $C_{\ell}^{{\rm kSZ}^2 \times \delta_g}$ and $C_{\ell}^{{\rm kSZ}^2 \times \mu_g}$ (solid black lines) used in the data analysis presented in the main part of this paper (see, e.g.,  Fig.~\ref{Fiducial}). Our fiducial halo model (dash-dotted red curves) is  consistent with the templates at small scales: at $\ell\approx3000$, they agree within $5-10\%$. Furthermore, the 1-halo term seems to be the main contribution to $C_{\ell}^{{\rm kSZ}^2 \times \delta_g}$ and $C_{\ell}^{{\rm kSZ}^2 \times \mu_g}$ for $\ell\gtrsim 10^{3}$. By comparing the fiducial halo model (dash-dotted red lines) with the lines labeled as `Linear velocity', we see that non-linear correction to the velocity dispersion (i.e., using the Halofit power spectrum rather than the linear matter power in Eq.~\ref{eq:vrms2}) is responsible for $20-30\%$ of the power over all scales. Finally, we find that when setting the concentration to the $\tau$-profile twice as high as for the fiducial halo model, the amplitudes of both power spectra increase significantly. Note that as we change $c_{_\mathrm{NFW}}$ in the $\tau$-profile, we impose that $M_{_\mathrm{NFW}}$, the total mass enclosed within $r_{_\mathrm{NFW}}$, remains the same so that on large angular scales $u_\ell^\tau\propto M_{_\mathrm{NFW}}$ also remains the same. In this case the main effect of changing the concentration is via its effect on the characteristic multipole $\ell_\mathrm{s}$ that enters Eq.\ \eqref{eq:phiellnfw}. It is important to emphasize that we do not expect the large scale  power of $C_\ell^{{\rm kSZ}^2\times \delta_g}$ to remain the same here, even though we impose constant $M_\mathrm{NFW}$.  Indeed, this observable is an integral over a bispectrum and the small scales of the bispectrum (so-called squeezed triangles) also contribute to the large scale amplitude of $C_\ell^{{\rm kSZ}^2\times \delta_g}$  (see Eq.~\ref{eq.Clintegral}) and $C_\ell^{{\rm kSZ}^2\times \mu_g}$.

\section{Validation plots}
\label{sec:append_valid_plots}
\begin{figure*}[t]
    \centering
    \includegraphics[width=0.45\textwidth]{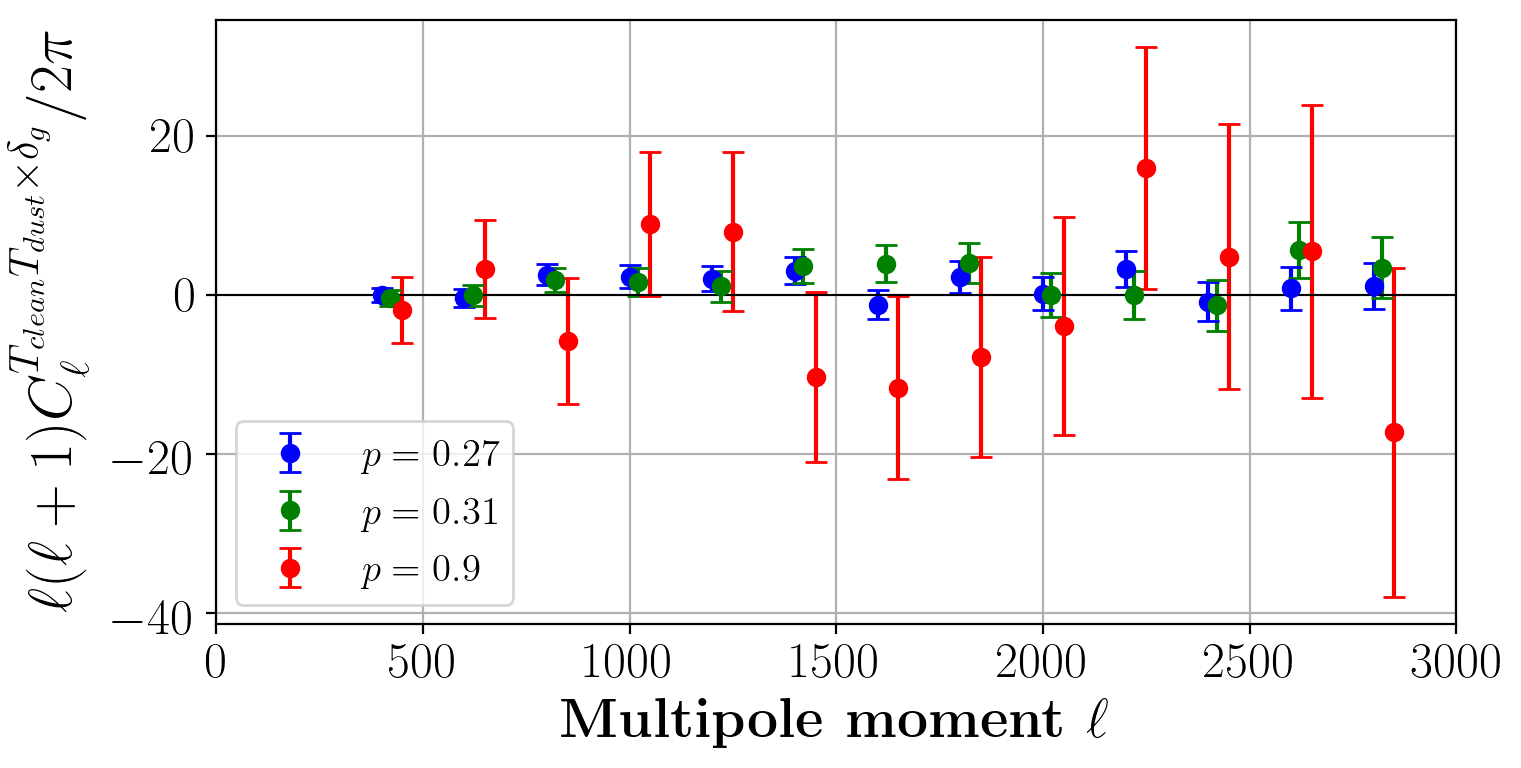}
    \includegraphics[width=0.45\textwidth]{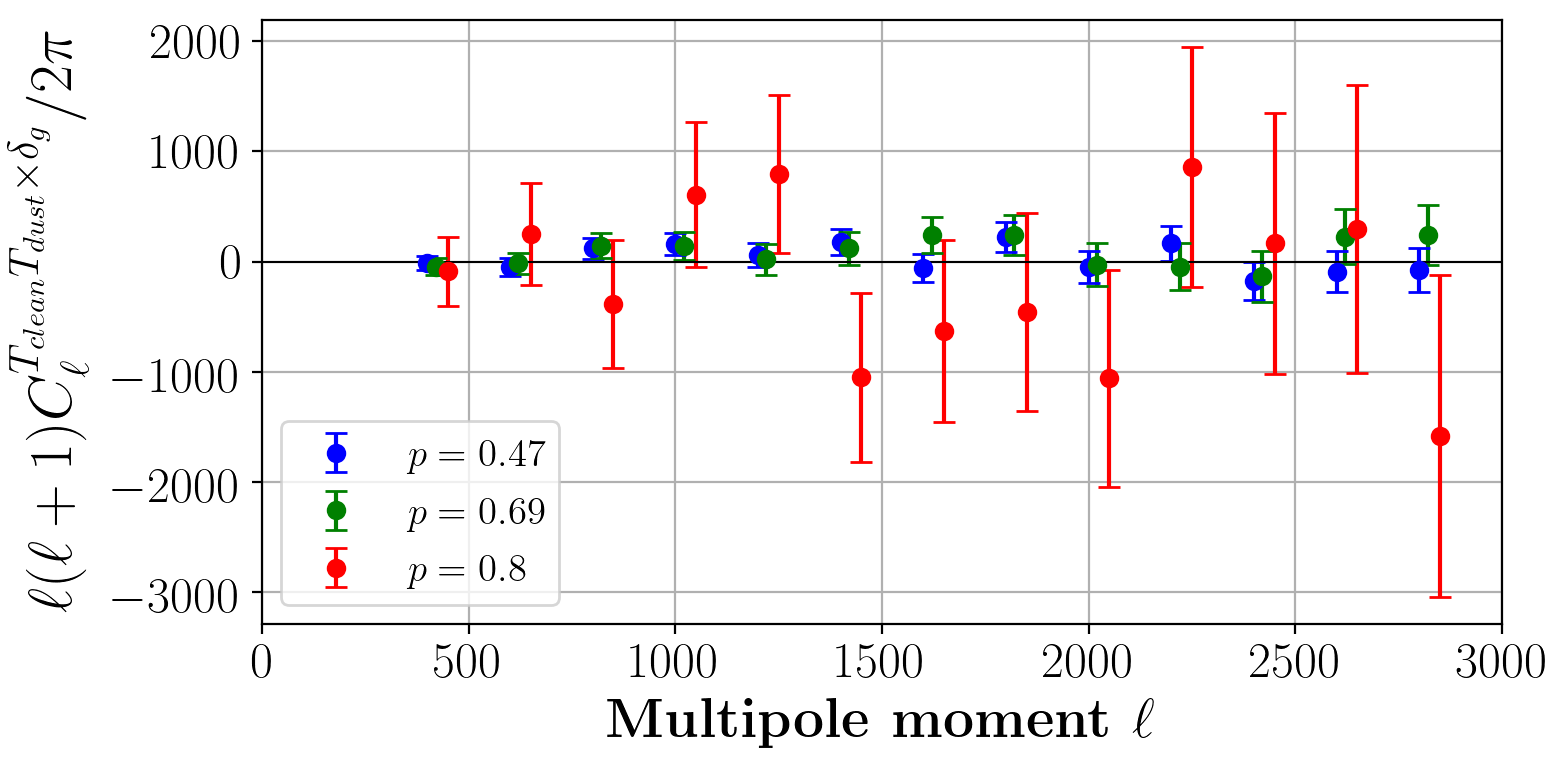}
    \caption{Dust null tests on the $\alpha$-cleaned SMICA-noSZ map, analogous to those shown in Fig.~\ref{fig_tests} for the $\alpha$-cleaned LGMCA map.
    We show the dust null tests for the $\alpha$-cleaned LGMCA and $\alpha$-cleaned SMICA-noSZ maps, and not for the original SMICA maps, since the latter has non-negligible tSZ residuals and thus only the former two maps can be used as primary maps in our estimator.  All are color coded based on the \emph{unWISE} subsample colors: blue, red, and green, and the legends give the probabilities-to-exceed for fits to null.  The green sample points are offset by $\ell=20$, and the red by $\ell=50$ with respect to the true multipole moment values, for visual purposes. Left: Cross-correlation of $(T_{\rm clean}T_{\rm dust})$ and \emph{unWISE}, where $T_{\rm dust}$ is the \emph{Planck} 545 GHz map and $(T_{\rm clean}$ is the clean SMICA-noSZ map. Right: Cross-correlation of $(T_{\rm clean}T_{\rm dust})$ and \emph{unWISE}, where $T_{\rm dust}$ is the \emph{Planck} 857 GHz map and $(T_{\rm clean}$ is the clean SMICA-noSZ map. Both are consistent with null.} 
    \label{fig_tests_smicanosz}
\end{figure*}

\begin{figure}[htbp!]
    \includegraphics[width=1.\columnwidth ]{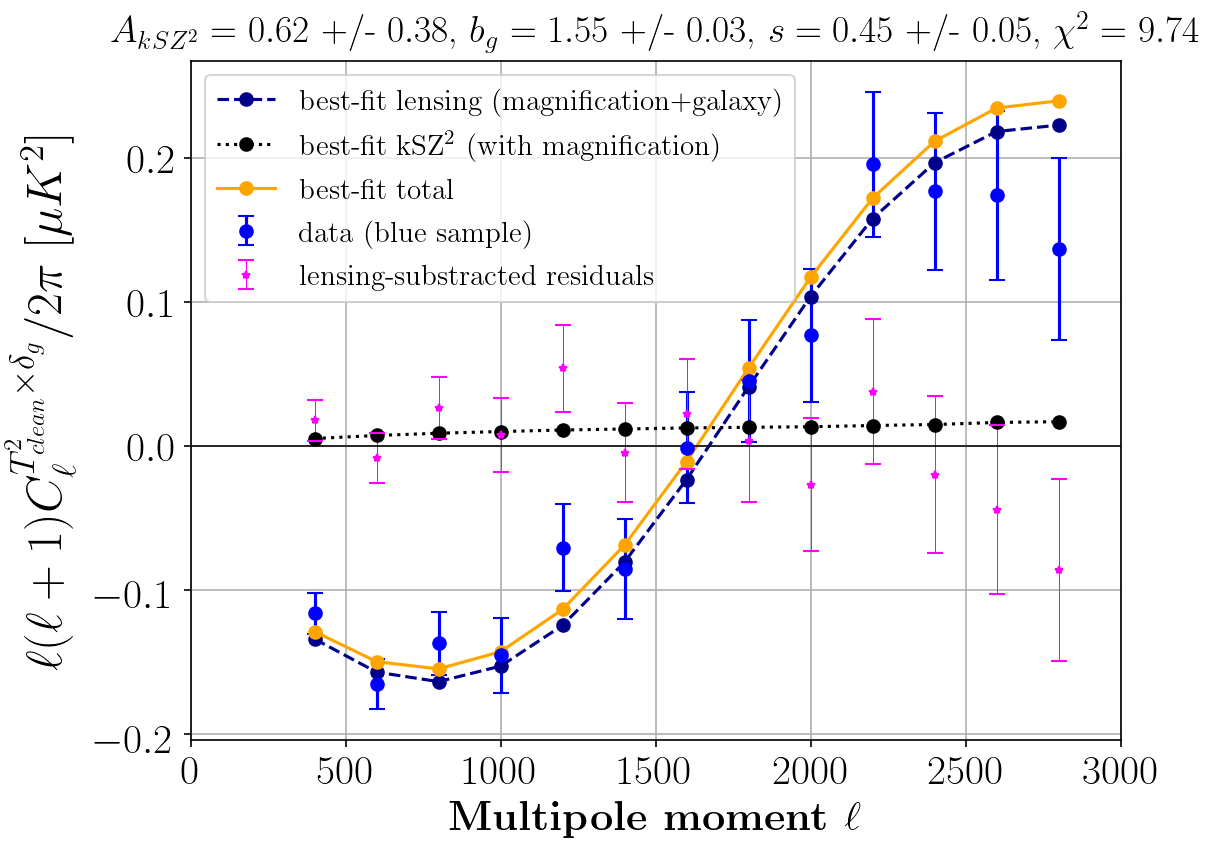}
    \includegraphics[width=1.\columnwidth ]{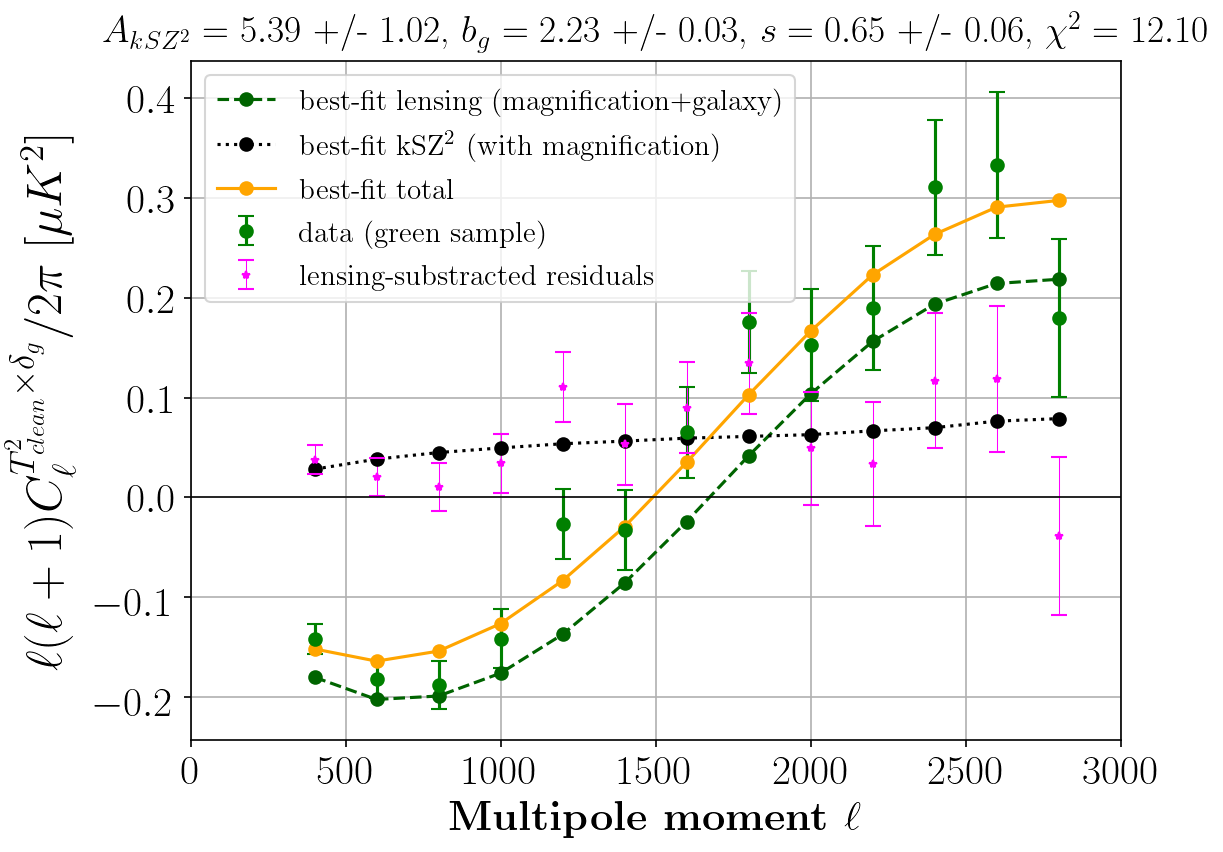}
    \includegraphics[width=1.\columnwidth ]{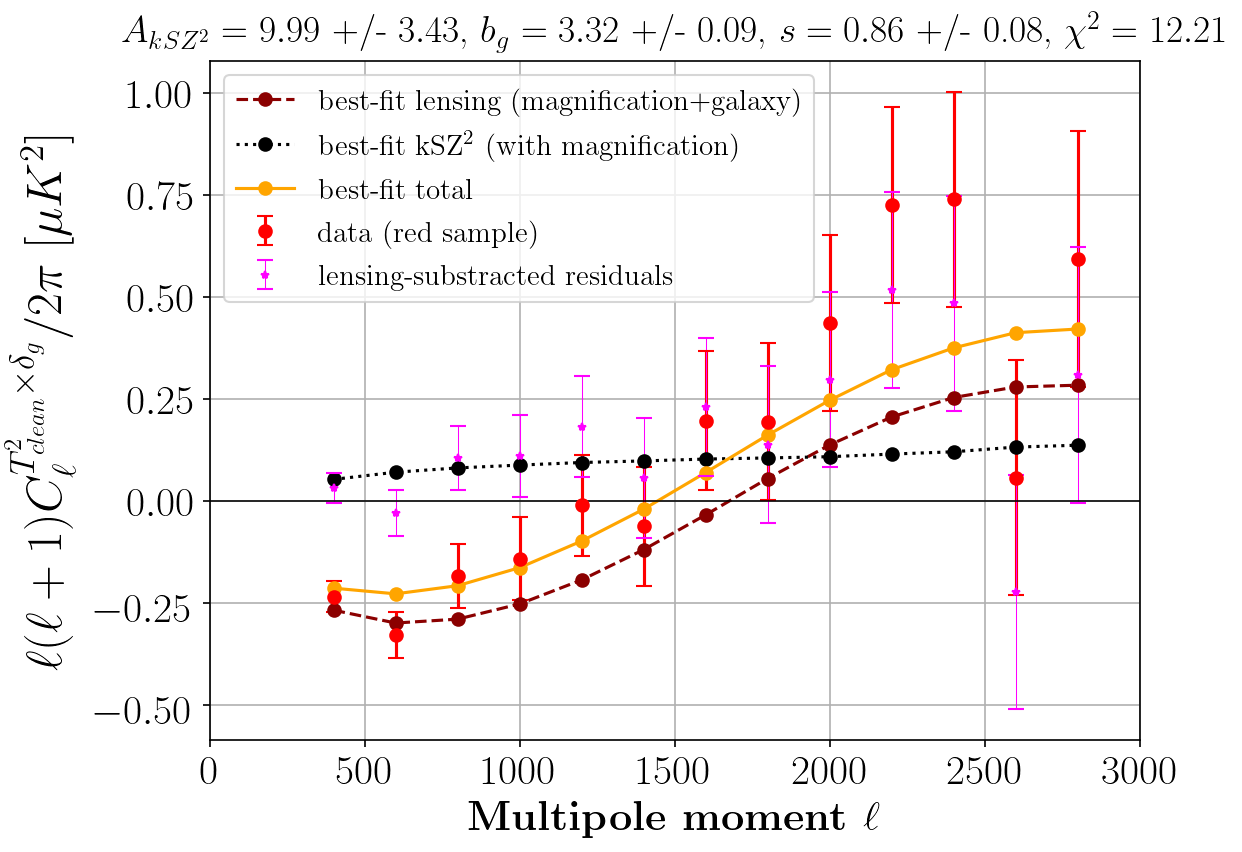}
    \caption{Cross-power spectra of the real-space product of the cleaned, filtered SMICA-noSZ map and the filtered SMICA map with each of the \emph{unWISE} galaxy maps: blue, green, and red (data points in respective colors), analogous to our main analysis in Fig.~\ref{fig:1}. The thin dashed curves show the best-fit lensing contribution, the black dotted are kSZ  and the pink stars the residuals. The yellow solid curves in each plot show the total best-fit curves. The best-fit values for the free parameters (the kSZ$^2$ amplitude $A_{{\rm kSZ}^2}$, galaxy bias $b_g$, and the magnification response $s$) are presented in the plot titles. These results validate our main analysis. } 
    \label{fits_smicanosz_smica}
\end{figure}

\begin{figure}[htbp!]
    \includegraphics[width=1.\columnwidth ]{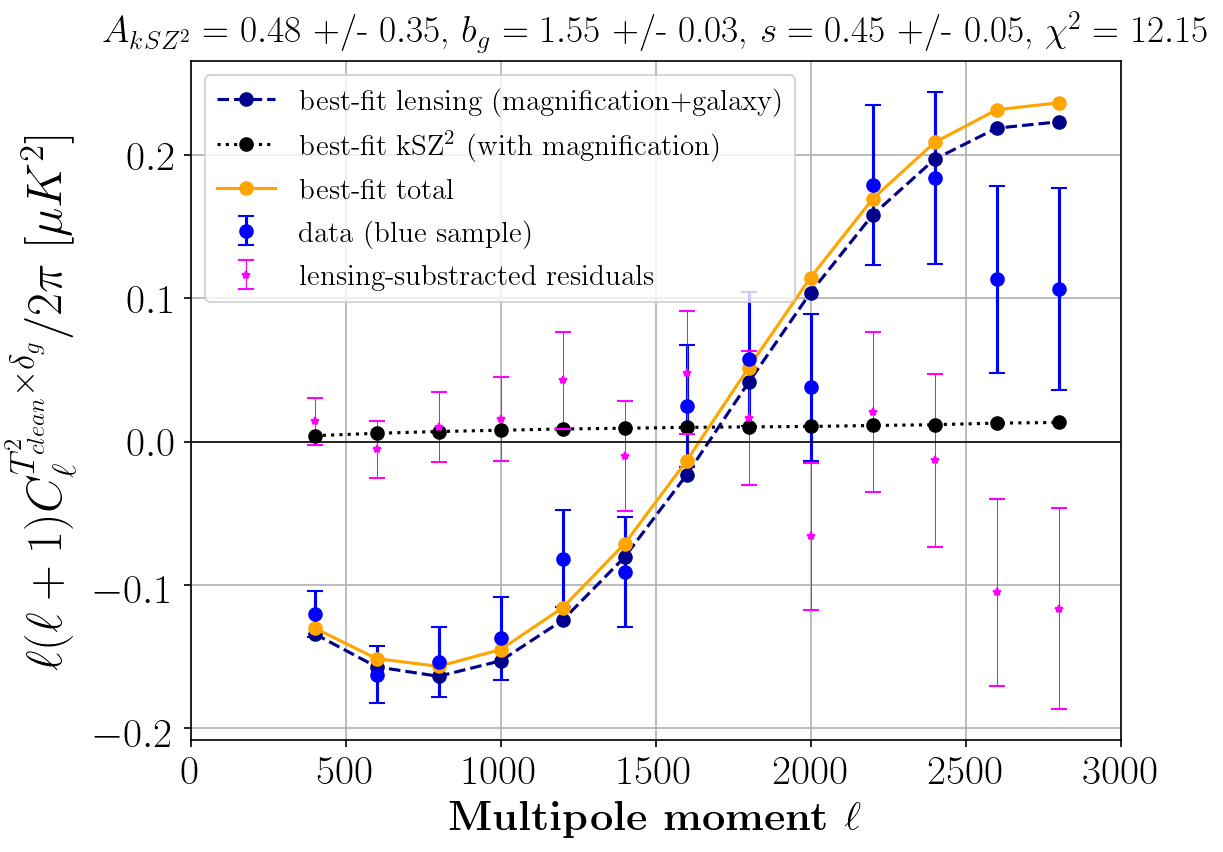}
    \includegraphics[width=1.\columnwidth ]{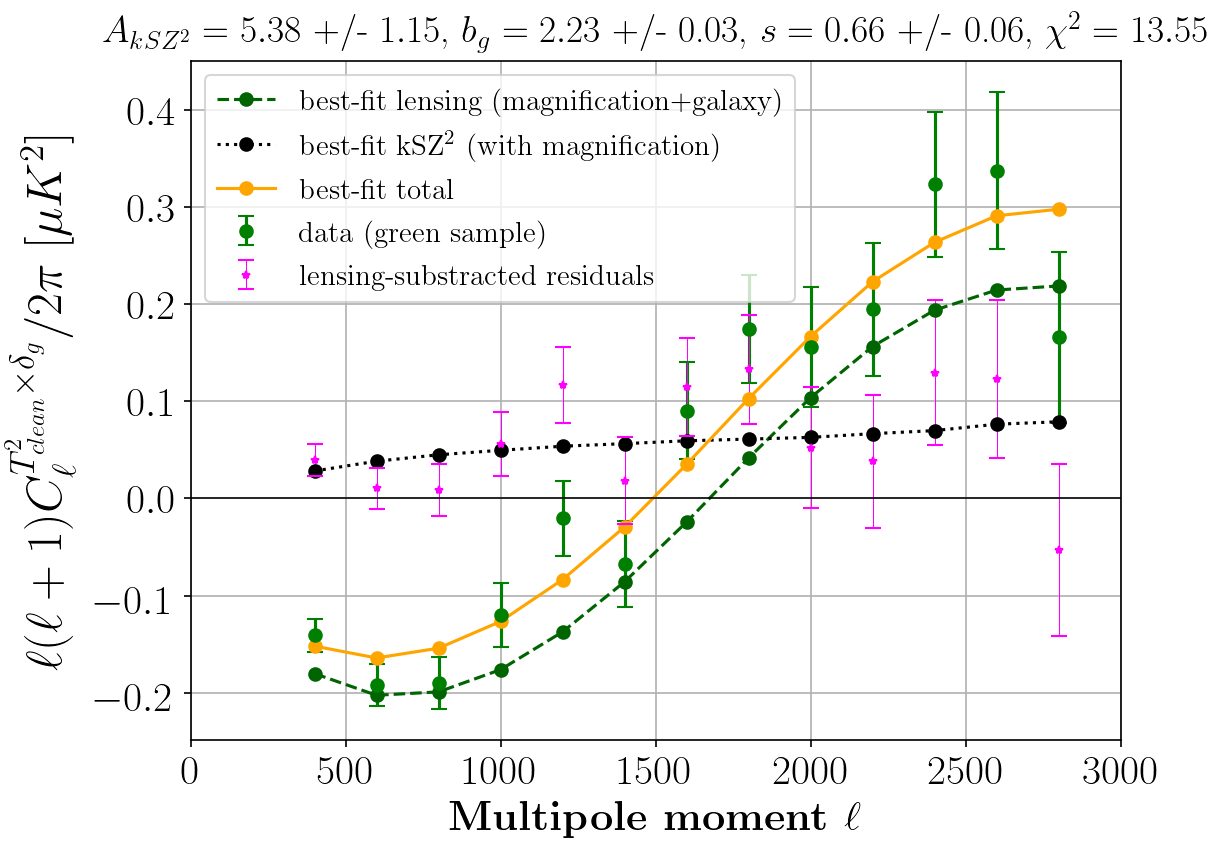}
    \includegraphics[width=1.\columnwidth ]{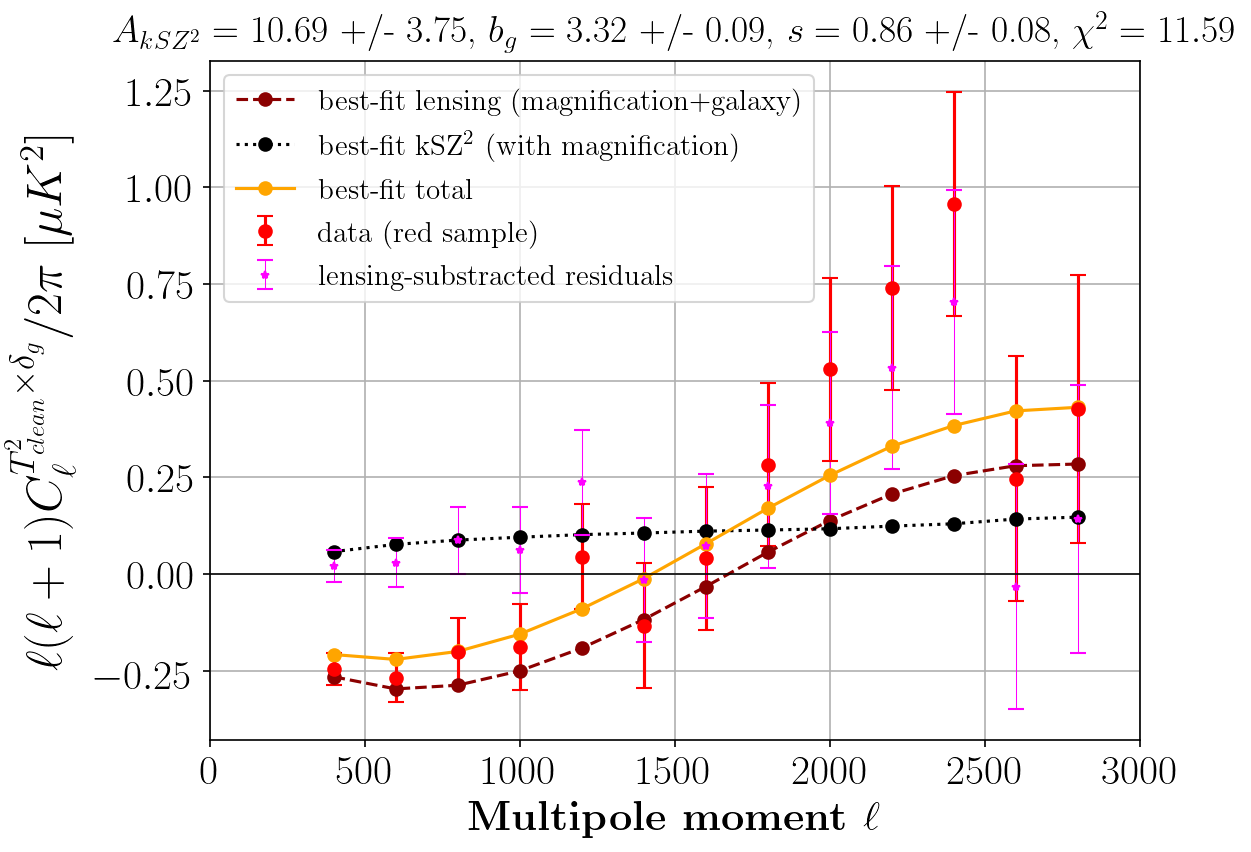}
    \caption{Cross-power spectra of the real-space square of the cleaned, filtered LGMCA map with each of the \emph{unWISE} galaxy maps: blue, green, and red (data points in respective colors), analogous to our main analysis in Fig.~\ref{fig:1}. The thin dashed curves show the best-fit lensing contribution, the black dotted the kSZ and the pink stars the residuals. The yellow solid curves in each plot show the total best-fit curves. The best-fit values for the free parameters (the kSZ$^2$ amplitude $A_{{\rm kSZ}^2}$, galaxy bias $b_g$, and the magnification response $s$) are presented in the plot titles. These results validate our main analysis, and also demonstrate that the ``asymmetric'' estimator employed in our fiducial analysis is robust to foregrounds, while generally yielding smaller error bars than the approach used in these plots (which matches that in~\citetalias{Hill2016}). } 
    \label{LGMCA^2}
\end{figure}

\begin{figure}[htbp!]
    \includegraphics[width=1.\columnwidth ]{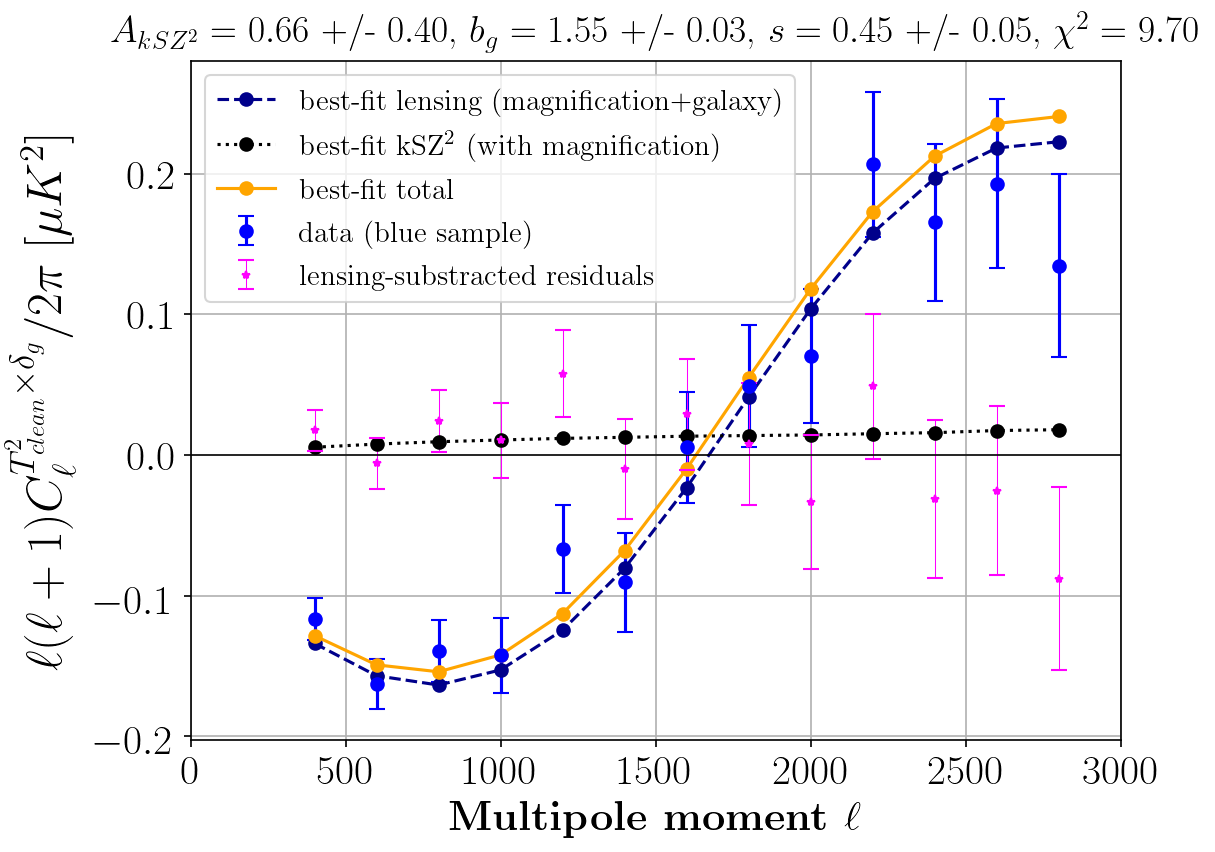}
    \includegraphics[width=1.\columnwidth ]{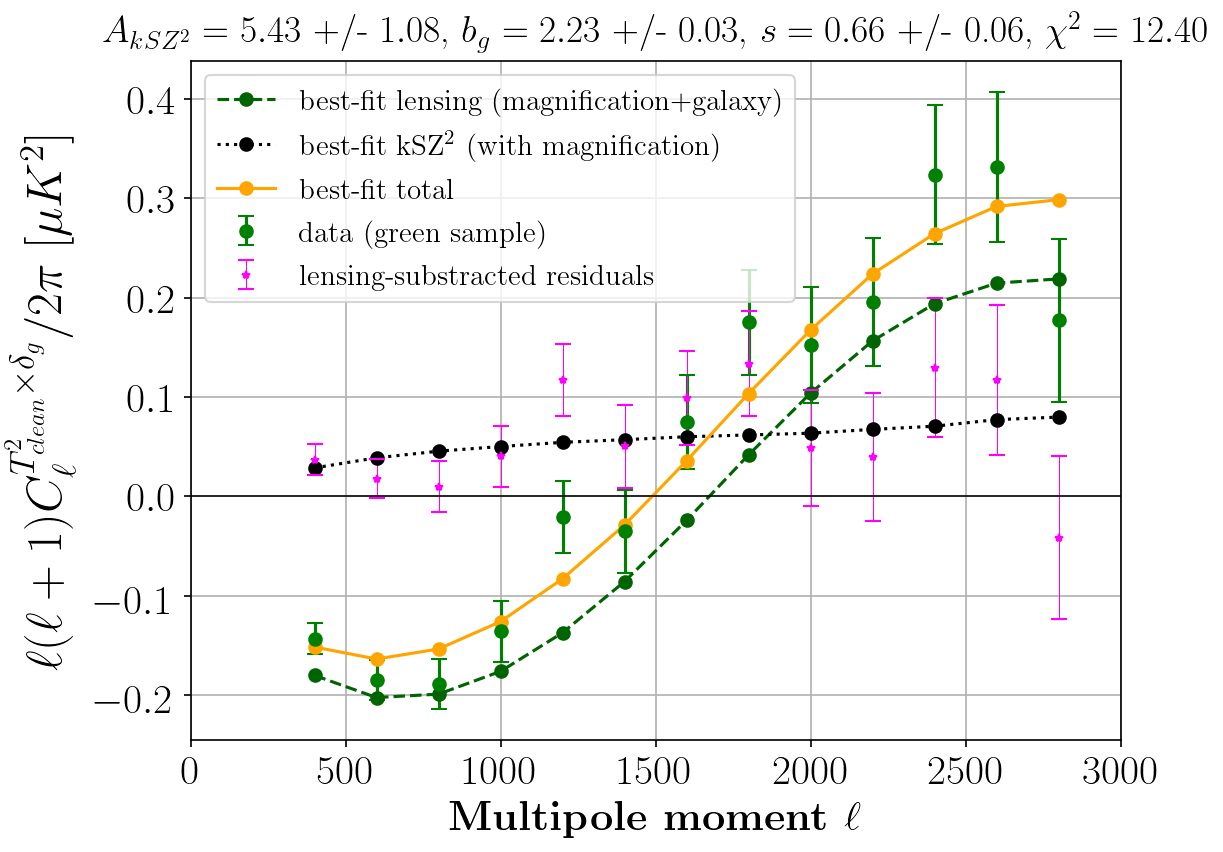}
    \includegraphics[width=1.\columnwidth ]{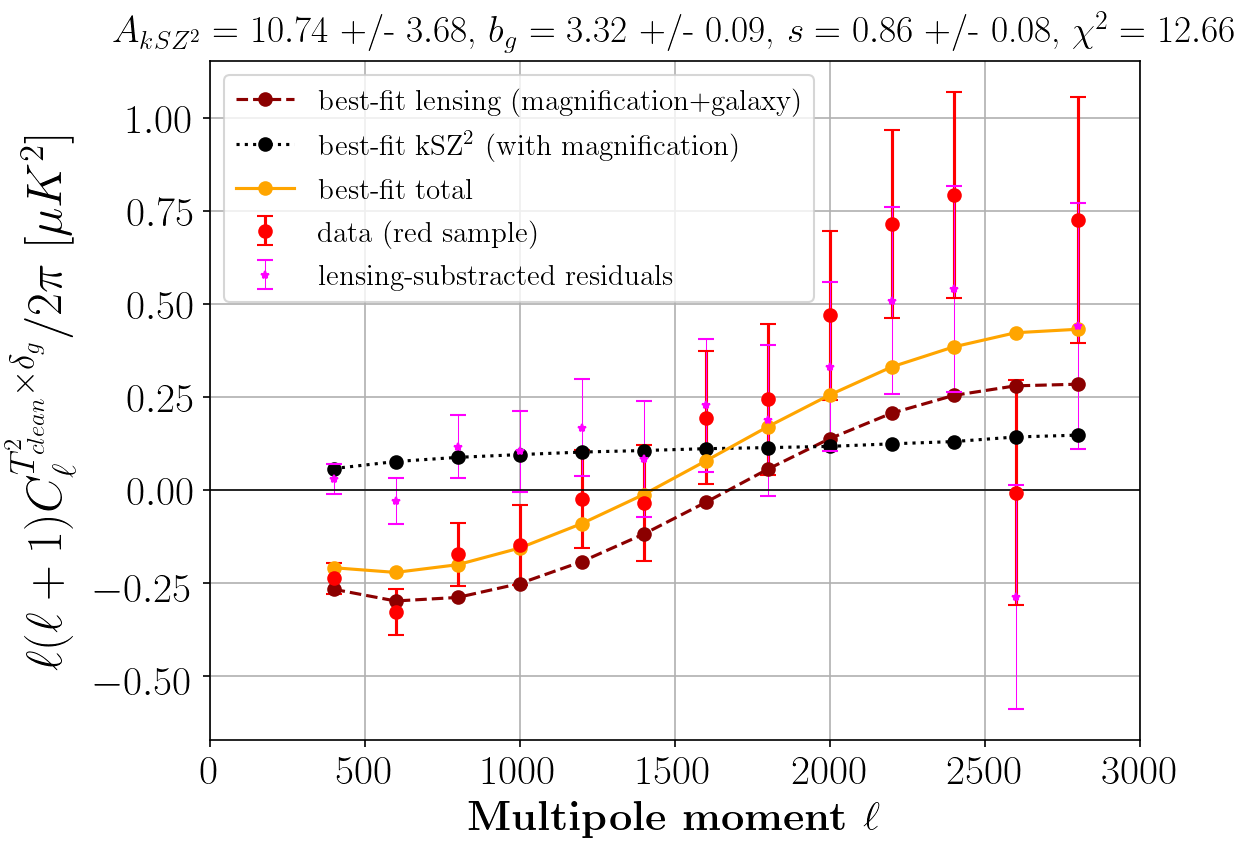}
    \caption{Cross-power spectra of the real-space square of the cleaned, filtered SMICA-noSZ map with each of the \emph{unWISE} galaxy maps: blue, green, and red (data points in respective colors), analogous to our main analysis in Fig.~\ref{fig:1}.  The thin dashed curves show the best-fit lensing contribution, the black dotted the kSZ and the pink stars the residuals. The yellow solid curves in each plot show the total best-fit curves. The best-fit values for the free parameters (the kSZ$^2$ amplitude $A_{{\rm kSZ}^2}$, galaxy bias $b_g$, and the magnification response $s$) are presented in the plot titles. These results validate our main analysis, and again also verify the robustness of the asymmetric estimator used in our fiducial analysis. } 
    \label{SMICAnoSZ^2}
\end{figure}

\begin{figure*}[htbp!]
\includegraphics[width=1.\columnwidth]{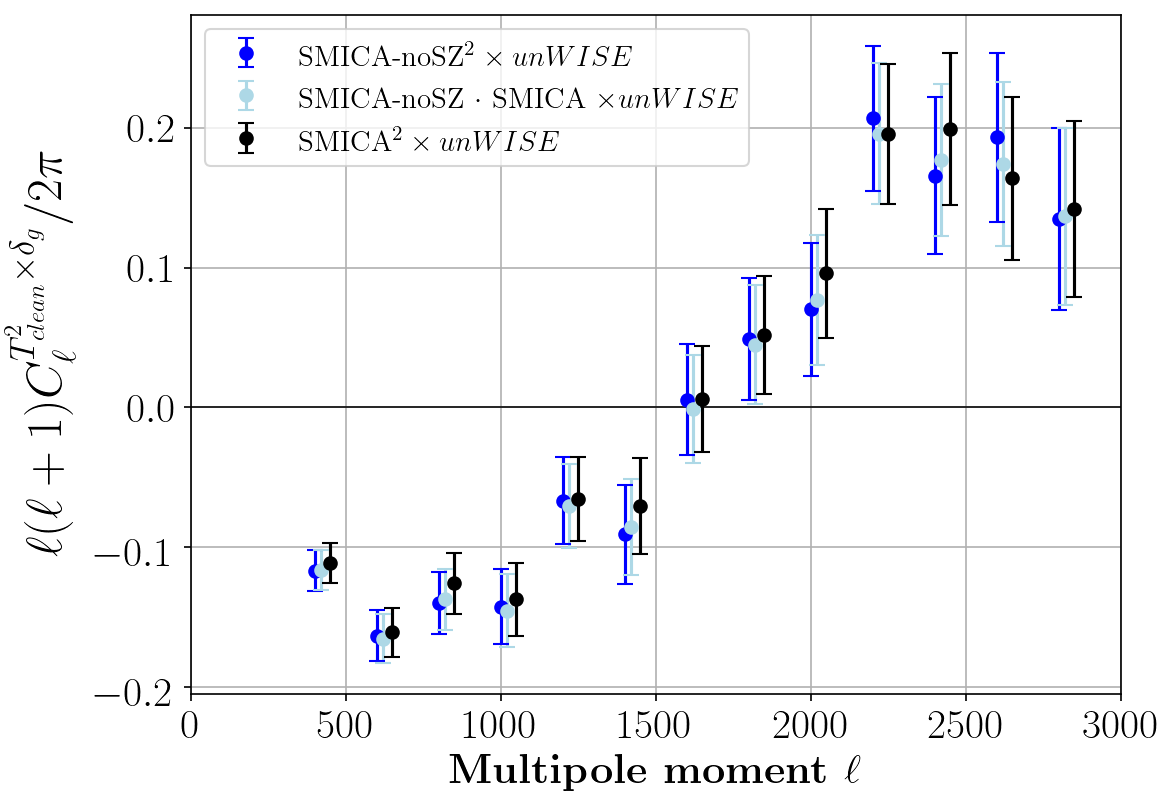}
\includegraphics[width=1.\columnwidth]{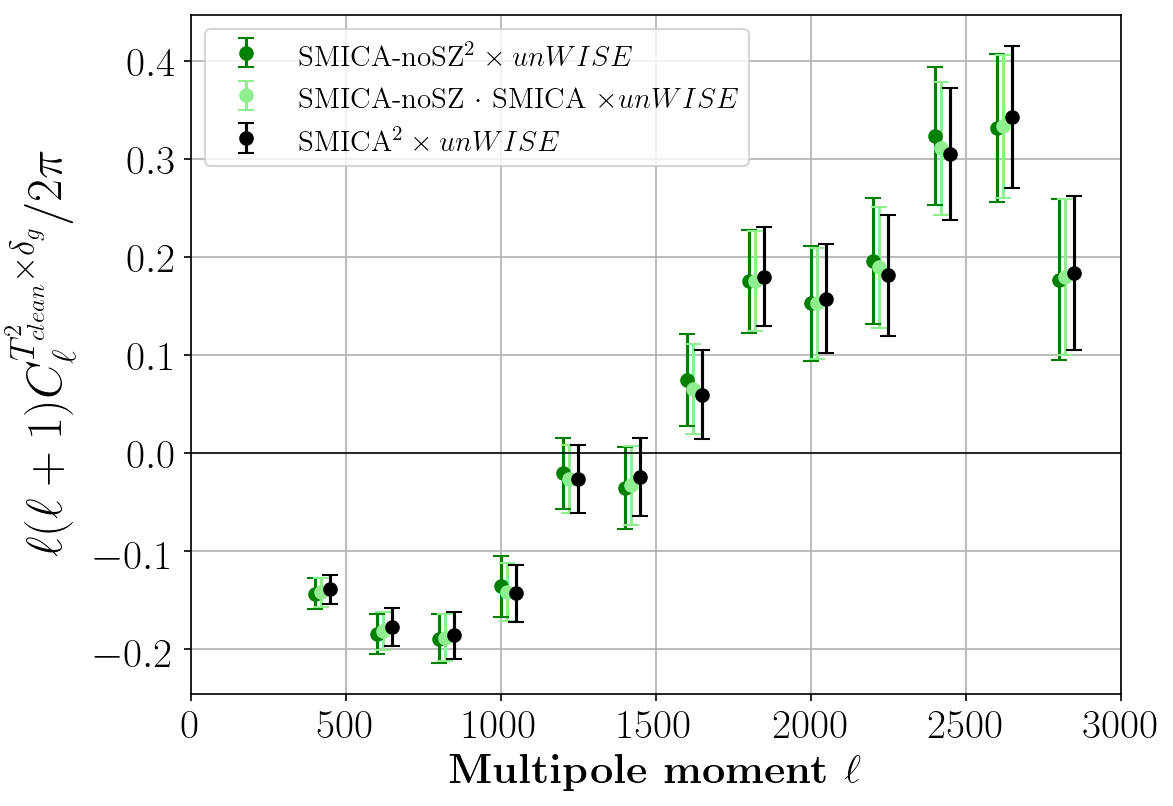}
\includegraphics[width=1.\columnwidth]{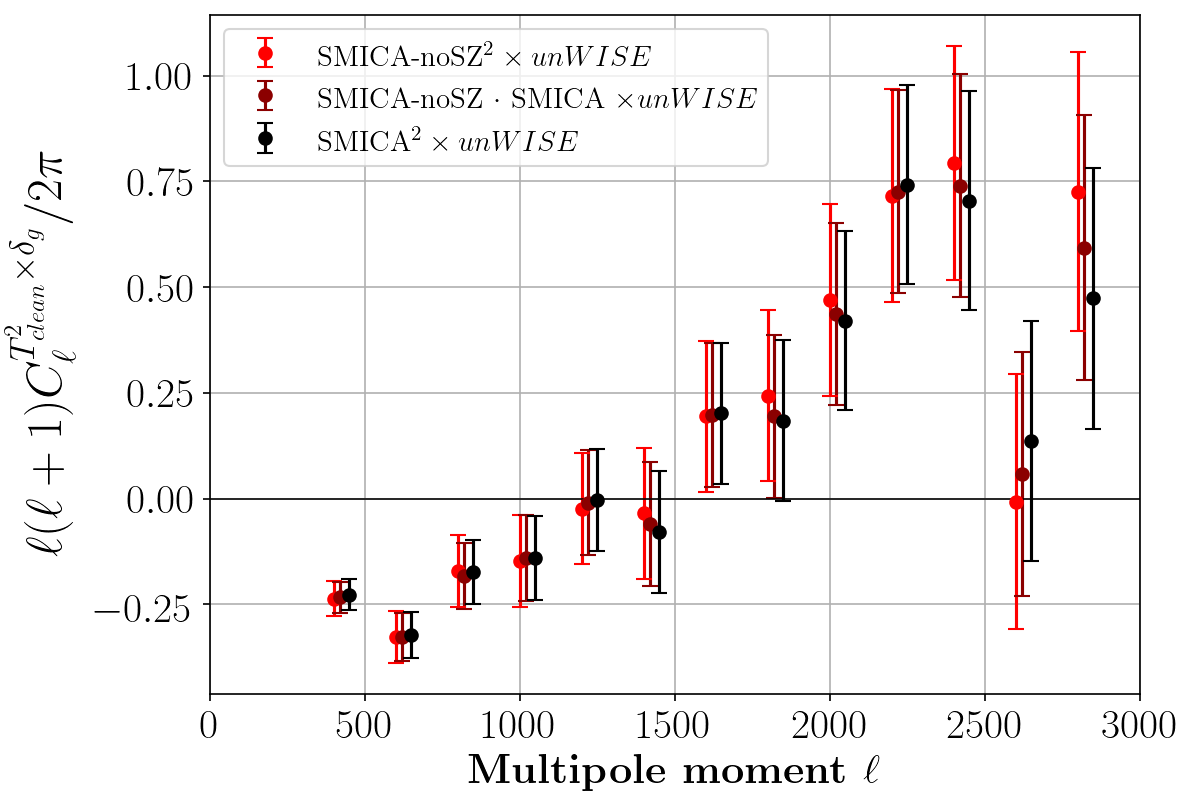}
\caption{A comparison of all the combinations of SMICA and SMICA-noSZ maps cross-correlated with the \emph{unWISE} maps (blue, green, and red, color-coded). The (SMICA-noSZ$_{\rm clean}\cdot$SMICA) $\times$ \emph{unWISE} points are offset by $\ell=20$, and the SMICA$_{\rm clean}^2$ $\times$ \emph{unWISE} points are offset by $\ell=50$ with respect to the true multipole moment values, for visual purposes.  All analyses yield consistent results with one another, and with our main analysis in Fig.~\ref{fig:1}, which validates the robustness of our analysis.}
\label{smicas_all_comparison}
\end{figure*}

\begin{figure*}[htbp!]
\includegraphics[width=1.\columnwidth]{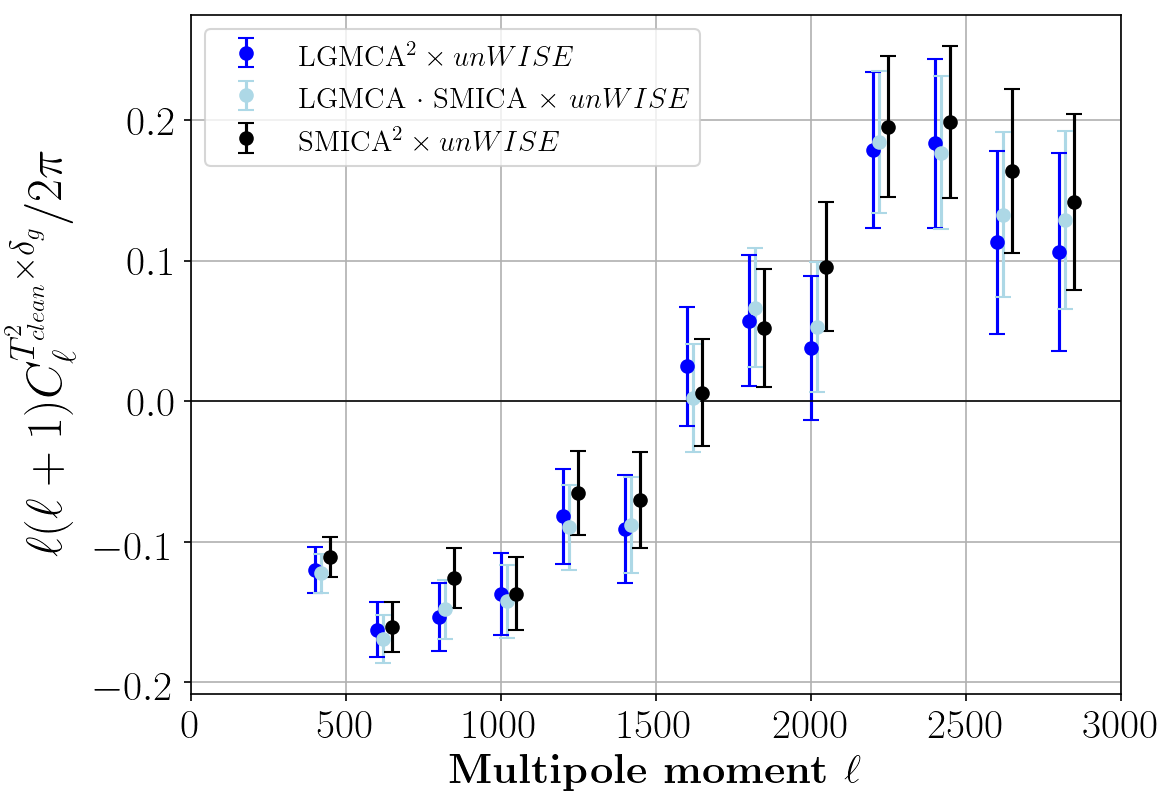}
\includegraphics[width=1.\columnwidth]{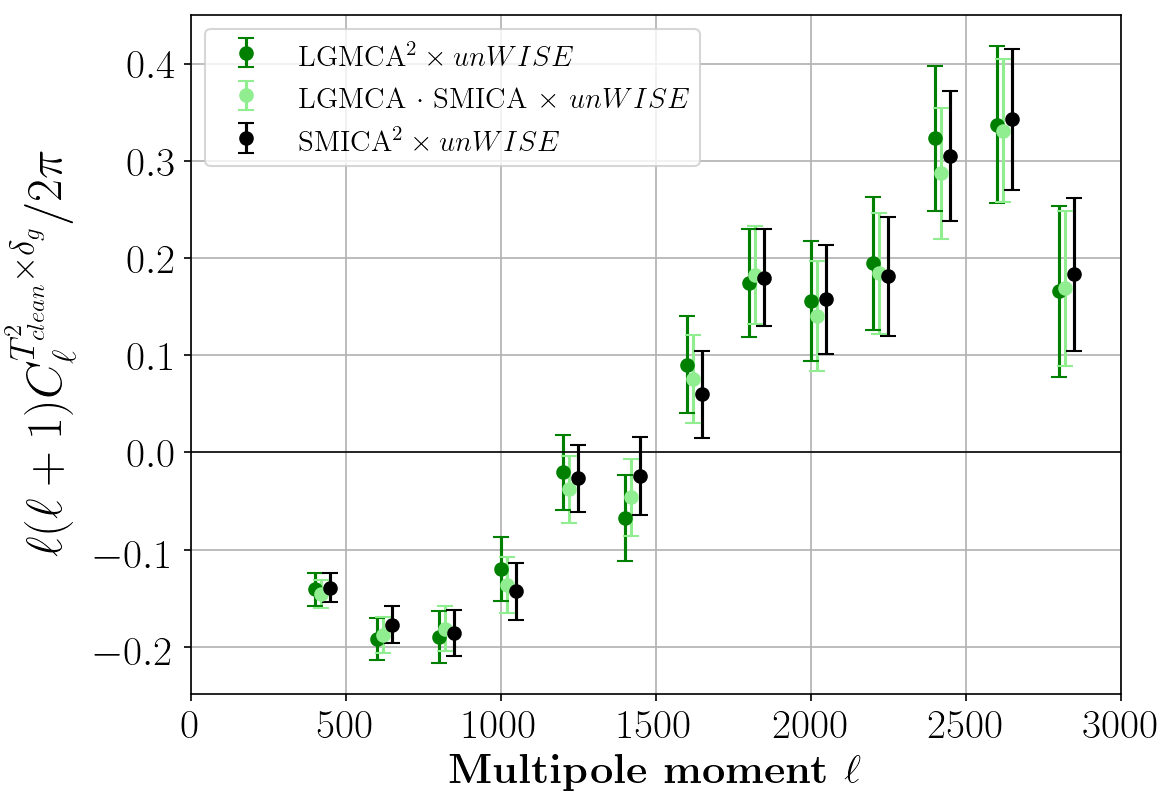}
\includegraphics[width=1.\columnwidth]{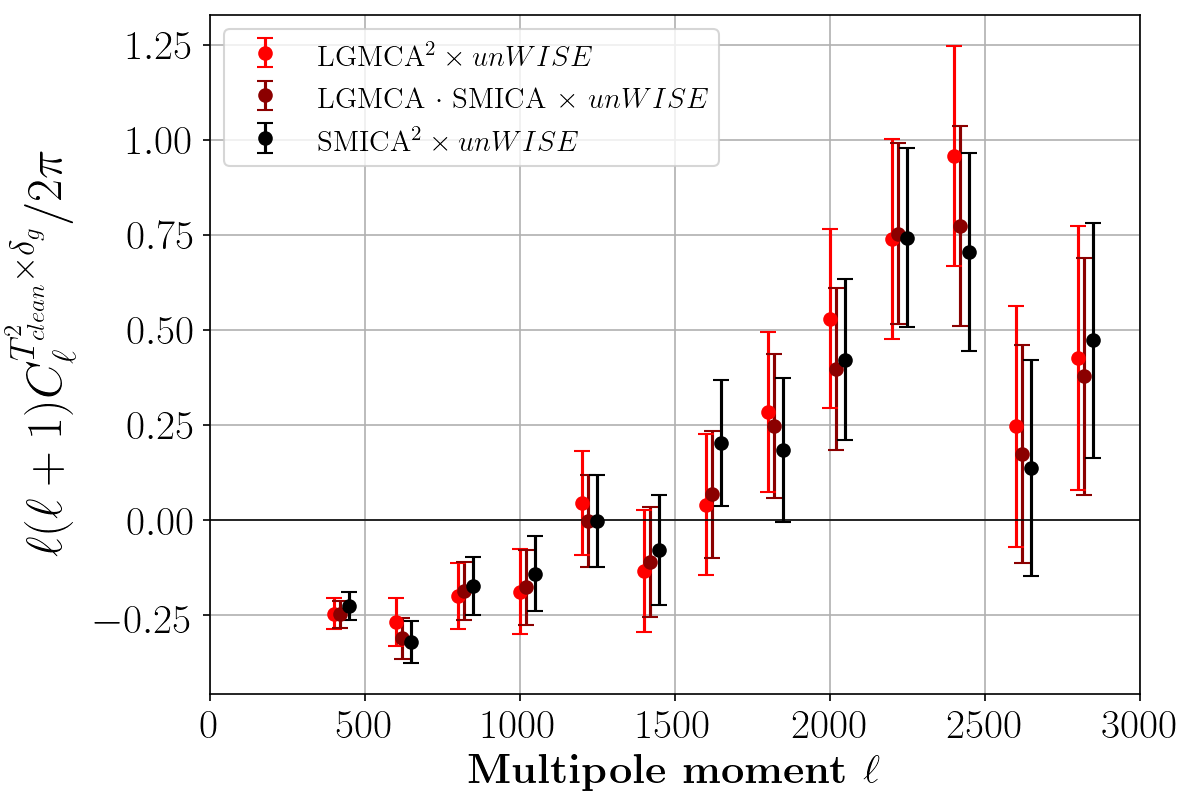}
\caption{A comparison of all the combinations of SMICA and LGMCA maps cross-correlated with the \emph{unWISE} maps (blue, green, and red, color-coded). The (LGMCA$_{\rm clean}\cdot$SMICA) $\times$ \emph{unWISE} points are offset by $\ell=20$, and the SMICA$_{\rm clean}^2$ $\times$ \emph{unWISE} points are offset by $\ell=50$ with respect to the true multipole moment values, for visual purposes. All analyses yield consistent results with one another, although the SMICA$_{\rm clean}^2$ results show some small differences, which are likely due to the tSZ residuals present in this map (hence why we do not use this map for our main analysis).  Note that our main results from Fig.~\ref{fig:1} are also shown here as the cyan, light green, and maroon points, and their consistency with the LGMCA$_{\rm clean}^2$ points validates the robustness of our analysis.}
\label{smica_lgmca_comparison}
\end{figure*}

We present dust tests results for the $\alpha$-cleaned SMICA-noSZ map (see Sec. \ref{subsec:cleaning} for an analogous cleaning with LGMCA), which are consistent with null (Fig. \ref{fig_tests_smicanosz}). We also show validation plots to our main analysis, described in details in Sec. \ref{sec:validation}: (SMICA-noSZ$\cdot$SMICA) $\times$ \emph{unWISE} in Fig. \ref{fits_smicanosz_smica}, 
LGMCA$^2 \times$ \emph{unWISE} in Fig. \ref{LGMCA^2}, 
and SMICA-noSZ$^2 \times$ \emph{unWISE} in Fig. \ref{SMICAnoSZ^2}, along with their comparison in Figs. \ref{smicas_all_comparison} and \ref{smica_lgmca_comparison}.

\bibliographystyle{apsrev}
\bibliography{unWISE}

\end{document}